\documentclass[10pt,preprintnumbers,nofootinbib]{revtex4-2}%
\usepackage{float} 
\usepackage{cancel}
\usepackage[utf8]{inputenc}
\usepackage{amsmath}
\usepackage{amssymb}
\usepackage{dcolumn}
\usepackage{graphicx}
\usepackage[colorlinks=true, pdfstartview=FitV, linkcolor=blue, citecolor=red, urlcolor=magenta]{hyperref}
\usepackage{amsfonts}
\usepackage{subfigure}
\usepackage{soul}
\usepackage{cleveref}
\usepackage{listings}
\usepackage{xcolor}
\usepackage{array}
\usepackage{url}
\usepackage{color}%
\usepackage{tikz}
\usepackage{pgfplots}
\usepackage{multirow}
\usepackage[outdir=./Figures]{epstopdf}

\usepackage{fontawesome5}

\usepackage{booktabs} 
\newcommand{\qbar}{\raisebox{-1.5ex}[1ex][0pt]{$\mathchar'26$}\mkern-9mu q}

\usepackage{xcolor}

\usepackage{subcaption}

\usepackage[colorinlistoftodos, textsize=small]{todonotes}

\graphicspath{{Figures/}}

\begin{document}

\title{Imprints of $\qbar$uantum space in the ringdown phase of Kerr black holes}

\author{Nikola Herceg}
\email{nherceg@irb.hr}
\affiliation{  Faculty of Electrical Engineering and Computing, University of Zagreb, 10000 Zagreb, Croatia \\
Rudjer Bo\v{s}kovi\'c Institute, Bijeni\v cka  c.54, HR-10002 Zagreb, Croatia}

\author{Tajron Juri\'c}
\email{tjuric@irb.hr}

\author{A. Naveena Kumara }
\email{nathith@irb.hr}

\author{Andjelo Samsarov}
\email{asamsarov@irb.hr}
\affiliation{Rudjer Bo\v{s}kovi\'c Institute, Bijeni\v cka  c.54, HR-10002 Zagreb, Croatia}

\author{Ivica Smoli\'c}
\email{ismolic@phy.hr}
\affiliation{Department of Physics, Faculty of Science, University of Zagreb, Bijeni\v cka cesta 32, 10000 Zagreb, Croatia}

\author{Kumar S. Gupta}
\email{kumars.gupta@gmail.com}
\affiliation{13 Mahendra Road, Kolkata 700025, India}

\date{\today}

\begin{abstract}

We investigate $\qbar$uantum-space corrections to the gravitational quasinormal modes (QNMs) of Kerr black holes within the framework of noncommutative geometry. The analysis employs a pseudo-semi-Killing Drinfeld twist, which preserves the classical Kerr geometry as an exact background solution while introducing nontrivial $\qbar$uantum corrections to linearized gravitational perturbations. Working to first order in the black hole spin and the $\qbar$uantum deformation parameter, including their mixed contribution, we derive a Schr\"odinger-like master equation for axial gravitational perturbations with a frequency- and spin-dependent effective potential. We compute the fundamental QNM frequencies using the WKB approximation and Leaver's continued-fraction method. Our results reveal corrections to the QNM oscillation frequencies and damping rates arising from rotation, pure $\qbar$uantum effects, and their interplay. To the best of our knowledge, this is the first systematic analysis of gravitational metric perturbations in a noncommutative Kerr black hole spacetime using a Drinfeld twist, extending prior work beyond the Schwarzschild case. Our findings demonstrate that $\qbar$uantum deformation effects alter both the oscillation frequencies and damping rates of quasinormal modes, with spin–quantum coupling introducing additional spin-dependent corrections to the Kerr QNM spectrum. Using a Kerr-anchored phenomenological prescription, we estimate the fractional frequency and damping-time shifts for selected merger remnants from the GWTC-3 and GWTC-5.0 catalogs. These estimates illustrate the potential signatures of $\qbar$uantum corrections in gravitational-wave ringdown.

\end{abstract}

\maketitle

\section{Introduction}
 \label{sec:intro}
The direct detection of gravitational waves (GWs) from the binary black hole coalescence event GW150914 by the LIGO Collaboration~\cite{LIGOScientific:2016wyt} inaugurated the era of observational strong-field gravity. Over the subsequent observing runs of the LIGO-Virgo-KAGRA (LVK) network, spanning GWTC-3~\cite{KAGRA:2021vkt}, GWTC-4~\cite{LIGOScientific:2025wao, LIGOScientific:2025cmm, LIGOScientific:2026wpt}, and the recently released GWTC-5 catalog~\cite{LIGOScientific:2026wfs, LIGOScientific:2026qqv, LIGOScientific:2026oim}, gravitational-wave astronomy has transitioned from individual discovery events to high-precision, population-level tests of General Relativity (GR) in the highly dynamical, nonlinear regime. A central pillar of this observational program is black hole spectroscopy~\cite{Dreyer:2003bv, Berti:2005ys, Berti:2018vdi, Franchini:2023eda, Carullo:2025oms, Berti:2025hly}. Following the merger of  compact binary black holes, the newly formed remnant black hole relaxes toward a stationary equilibrium state by radiating gravitational waves in its ringdown phase. At linear order, this relaxation is governed by a superposition of damped sinusoids known as quasinormal modes (QNMs)~\cite{Vishveshwara:1970zz, Kokkotas:1999bd, Nollert:1999ji, Berti:2009kk}. These modes are purely ingoing at the horizon and purely outgoing at spatial infinity and they are characterized by complex frequencies $\omega_{\ell m n} = \omega_R + i\omega_I$, where the real part determines the oscillation frequency and the imaginary part fixes the inverse damping time.

In classical GR, under the standard assumptions of black hole uniqueness theorems, stationary, asymptotically flat, electrically neutral vacuum black holes are described by the Kerr geometry~\cite{Kerr:1963ud, Chrusciel:2012jk}. The Kerr hypothesis assumes that the late-time exterior geometry of an isolated, electrically neutral astrophysical black hole is well approximated by the Kerr spacetime. Consequently, the QNM spectrum of a Kerr black hole depends solely on its mass $M$ and angular momentum $J$. Extracting multiple ringdown modes, whether through fundamental angular multipoles~\cite{Capano:2021etf, Siegel:2023lxl, Gennari:2023gmx} or overtones near the strain peak~\cite{London:2014cma, Giesler:2019uxc, Isi:2019aib, Cotesta:2022pci}, and constraining fractional deviations from the Kerr spectrum via full inspiral-merger-ringdown (IMR) and time-domain analyses~\cite{Brito:2018rfr, Ghosh:2021mrv, Maggio:2022hre, Carullo:2019flw, Torri:2025tmm} therefore provide a direct experimental test of the Kerr hypothesis. Recent high-signal-to-noise-ratio (SNR) detections in the fourth observing run (O4), such as GW230814\_230901~\cite{LIGOScientific:2025cmm}, GW231226\_101520~\cite{LIGOScientific:2026wpt}, GW250114~\cite{Kumar:2026yhu}, and GW250207~\cite{LIGOScientific:2026qqv}, alongside comprehensive ringdown tests of GR~\cite{LIGOScientific:2026oim, Kumar:2026rcv, Lemos:2025duv, Song:2025dbf, Gong:2023ghh}, have enabled increasingly stringent constraints on deviations from the Kerr QNM spectrum. Next-generation gravitational-wave detectors, such as the Einstein Telescope, Cosmic Explorer, and LISA, are expected to further improve the precision of black hole spectroscopy~\cite{ET:2025xjr, Evans:2021gyd, LISA:2024hlh}.

Despite the remarkable empirical successes of classical GR, the corresponding quantum theory of gravity continues to be one of the most significant challenges. There have been many attempts to reconcile gravity with quantum physics, including string theory~\cite{Green:2012oqa, Witten:1995ex, Polchinski:1998rq, Becker:2006dvp}, loop quantum gravity (LQG) and spin foam models~\cite{Rovelli:2004tv, Thiemann:2007pyv, Ashtekar:2021kfp}, asymptotically safe quantum gravity~\cite{Reuter:1996cp, Percacci:2017fkn}, causal dynamical triangulations~\cite{Ambjorn:2012jv}, and causal set theory~\cite{Bombelli:1987aa}. While these approaches differ significantly in their microscopic degrees of freedom and ultraviolet completions, most of them share a recurring physical feature: at length scales approaching the Planck scale $\ell_P = \sqrt{\hbar G/c^3}$ (or a characteristic mesoscopic quantum-gravity scale), the classical notion of a smooth spacetime continuum breaks down. In this regime, spacetime acquires a granular, fuzzy, or nonlocal structure that precludes the localization of physical events with arbitrary precision. In a significant development, Doplicher et al. ~\cite{Doplicher:1994zv, Doplicher:1994tu} showed that GR and Heisenberg's uncertainty principle together lead to spacetime noncommutativity, henceforth we shall use this noncommutative (NC) geometry as a model of the quantum spacetime.

Connecting such quantum-spacetime frameworks to gravitational-wave observations has become a major frontier of contemporary theoretical physics~\cite{Barack:2018yly, Barausse:2020rsu, Addazi:2021xuf}. Because the ringdown phase probes the immediate vicinity of the photon sphere and the event horizon, black hole QNMs serve as natural observables for testing quantum modifications of gravity, effective field theories of gravity~\cite{Endlich:2017tqa, Cano:2019ore, Cardoso:2019mqo}, horizon-scale quantum effects~\cite{Cardoso:2016rao, Cardoso:2016oxy, Agullo:2020hxe}, and modified dispersion or ringdown signatures~\cite{Gong:2023ghh, Lemos:2025duv, Song:2025dbf, Kumar:2026rcv}. Within NC gravity, a variety of approaches to gravitational dynamics and black hole solutions have been explored. Early gauge-theoretic and Seiberg-Witten-map constructions examined deformed gravitational actions and corrections to Schwarzschild, charged, and cosmological black hole geometries~\cite{Chamseddine:2000si, Chaichian:2007we, Chaichian:2007dr, Calmet:2005qm, Calmet:2006iz, Kobakhidze:2007jn}. Twist-based formulations provided further NC gravitational solutions, including Schwarzschild-like and exact black hole configurations~\cite{Aschieri:2005yw, Aschieri:2005zs, Aschieri:2009qh, Ohl:2009pv, Schupp:2009pt, Schenkel:2011biz}. Other lines of investigation modeled noncommutativity through smeared matter sources~\cite{Nicolini:2005vd} or examined scalar, fermionic, and gauge-field perturbations on prescribed NC black hole backgrounds~
\cite{Gupta:2015uga, Gupta:2017lwk, DimitrijevicCiric:2017rmz, DimitrijevicCiric:2019hqq, Juric:2022bnm}. However, probing the genuine gravitational ringdown requires analyzing metric perturbations governed by NC-deformed Einstein field equations. 

For a long time, this program faced a structural obstacle: if the Drinfeld twist is constructed solely from the Killing vector fields of a highly symmetric background (a Killing twist), the background metric along with the resulting modes that are consistent with these isometries remain undeformed at linear order, pushing the leading NC corrections to quadratic order in the metric perturbation \cite{Aschieri:2005zs, Aschieri:2005yw, Aschieri:2009qh}. To overcome this limitation, a novel framework for NC gravitational perturbations was recently established in a series of works~\cite{Herceg:2023zlk, Herceg:2023pmc, Herceg:2024vwc, Herceg:2024upt, Herceg:2025rll}. In \cite{Herceg:2023zlk, Herceg:2023pmc}, a noncommutative (i.e. $\qbar$-deformed\footnote{$\qbar$ is the NC deformation parameter.}) Einstein-manifold condition $\hat{\rm R}_{\mu\nu}(g)=0$, formulated in terms of the $\mathcal{R}$-symmetrized NC Ricci tensor, was introduced alongside the class of pseudo-semi-Killing Abelian twists. By pairing a Killing vector field of the background geometry with a commuting vector field that is not Killing (such as the radial derivative $\partial_r$), the pseudo-semi-Killing twist ensures that the classical black hole metric remains an exact vacuum solution of the $\qbar$-deformed Einstein equations while simultaneously inducing nontrivial $\qbar$uantum corrections in the linearized metric perturbation equations at $\mathcal{O}(\qbar)$. Applying this construction to the static Schwarzschild background in Ref.~\cite{Herceg:2023zlk}, the $\qbar$-deformed Regge-Wheeler equation was derived for axial gravitational perturbations, while in Refs.~\cite{Herceg:2023pmc, Herceg:2024upt, Herceg:2025rll} the resulting QNM spectrum was computed and there it was demonstrated how a near-horizon radial coordinate translation regularizes
the NC boundary behavior. Subsequently, Ref.~\cite{Herceg:2024vwc} extended the formalism to polar gravitational perturbations of the Schwarzschild black hole, deriving the $\qbar$-deformed Zerilli equation and demonstrating that quantum-spacetime noncommutativity breaks the classical isospectrality between axial and polar modes.

While the analyses carried out in Refs.~\cite{Herceg:2023zlk, Herceg:2023pmc, Herceg:2024vwc, Herceg:2024upt, Herceg:2025rll} revealed the essential properties of $\qbar$-deformed gravitational
perturbations, it turns out that the setting in which it was applied is a rather limited one, involving only a Schwarzschild-type of background.  On the other hand, the astrophysical black holes formed in binary mergers inevitably possess a nonvanishing spin and  thus they are not
adequately taken care of in these studies. It therefore appears that confronting
the $\qbar$-deformed Einstein-manifold proposal with realistic gravitational-wave ringdown signals requires extending the noncommutative perturbation theory to rotating Kerr black holes. In classical GR, perturbations of the Kerr spacetime are famously analyzed via the Teukolsky formalism~\cite{Teukolsky:1972my,Teukolsky:1973ha}, which employs the Newman-Penrose formalism to decouple and separate the equations governing the Weyl scalars, enabling high-precision numerical evaluations of Kerr QNMs~\cite{Leaver:1985ax,Cook:2014cta}. Beyond GR, however, modified field equations generally spoil the separability of the perturbation equations or lack a decoupled Newman-Penrose master variable (which is also the problem in NC framework \cite{Herceg:2023zlk, Herceg:2023pmc, Herceg:2024vwc, Herceg:2024upt, Herceg:2025rll}). Over the past decade, several powerful techniques have been developed to compute QNMs of rotating black holes in such nonseparable systems, including the modified Teukolsky formalism~\cite{Li:2022pcy} (see also Refs.~\cite{Hussain:2022ins, Cano:2023tmv}), direct metric-perturbation partial differential equation (PDE) and spectral methods~\cite{Chung:2023zdq, Chung:2023wkd, Blazquez-Salcedo:2023hwg}, and slow-rotation perturbation theory~\cite{Pani:2012vp, Pani:2012bp, Pani:2013ija, Pani:2013wsa}.

Slow-rotation perturbation theory provides a systematic and analytically transparent framework for studying perturbations of rotating compact objects through a perturbative expansion in the dimensionless spin parameter $\chi=J/M^2$. Originally developed in the context of slowly rotating relativistic stars~\cite{Kojima:1992ie, 1993ApJ...414..247K, 1993PThPh..90..977K}, the formalism was subsequently extended to black hole perturbations~\cite{Pani:2012vp, Pani:2012bp, Pani:2013pma, Franchini:2023xhd} and successfully employed in the analysis of coupled gravitoelectromagnetic perturbations of Kerr-Newman black holes~\cite{Pani:2013ija, Pani:2013wsa}. By decomposing the metric perturbations directly into tensor spherical harmonics and treating rotation perturbatively without relying on the Newman-Penrose formalism, this approach allows one to reduce nonseparable systems to an effective radial problem. Crucially, at first order in rotation, the couplings between modes of opposite parity with multipole indices $\ell \pm 1$ do not contribute to the QNM eigenfrequency shifts~\cite{Pani:2012bp, Pani:2013pma}, allowing the axial and polar sectors to be analyzed independently. The method has proved particularly fruitful in modified theories of gravity, where exact rotating solutions or separable perturbation equations are often unavailable~\cite{Cano:2021myl, Pierini:2021jxd, Wagle:2021tam, Srivastava:2021imr, Pierini:2022eim}.

Motivated by these developments, in the present work we employ the slow-rotation approximation to investigate $\qbar$uantum corrections to axial gravitational perturbations of rotating Kerr black holes within the NC geometric setting. This work presents the first systematic study of metric perturbations of a noncommutative Kerr black hole, extending previous investigations restricted to Schwarzschild backgrounds. Our results show that $\qbar$uantum deformation modifies both the oscillation frequencies and damping rates, with the mixed spin-$\qbar$uantum contribution introducing additional spin-dependent corrections to the Kerr QNM spectrum. Finally, we explore the phenomenological implications of these corrections for black hole ringdown and estimate their potential signatures in gravitational-wave observations of binary black hole merger remnants.

The remainder of this paper is organized as follows. In Sec.~\ref{sec:nc_geometry}, we review the Drinfeld-twist formulation of NC differential geometry, the $\qbar$-deformed vacuum Einstein equations, the symmetry structure of the Kerr background, and the construction of the pseudo-semi-Killing $t$-$r$ twist. In Sec.~\ref{sec:nc_slow_rotation}, we apply the slow-rotation formalism to axial metric perturbations of the Kerr spacetime, project the $\qbar$-deformed field equations onto tensor spherical harmonics, and derive the Schrödinger-like axial master equation and its effective potential. In Sec.~\ref{sec:nc_qnms}, we present  two independent semi-analytic and numerical techniques used to extract the QNM frequencies: the WKB method (Sec.~\ref{sec:wkb_method}) and the Leaver continued-fraction method (Sec.~\ref{sec:leaver_tr}). In Sec.~\ref{sec:results}, we present our numerical results for the QNM spectrum, quantify the validity domain of the slow-rotation expansion against the exact Kerr spectrum, and analyze the interplay between spin and noncommutativity in the complex-frequency plane. In Sec.~\ref{sec:ringdown_phenomenology}, we review recent advances in observational ringdown spectroscopy, apply our Kerr-anchored prescription to estimate $\qbar$uantum shifts for 11 LVK events from O3b and O4b, and present an illustrative comparison with GW250114. Finally, we summarize our main findings and discuss future directions in Sec.~\ref{sec:discussion}. Technical details, including spherical-harmonic derivative identities and the explicit commutative and $\qbar$-dependent radial expansion coefficients, are collected in Appendices~\ref{identities} and~\ref{coefficient_list}.

\section{Noncommutative geometry}
\label{sec:nc_geometry}

Noncommutative geometry offers a mathematically rigorous and an effective framework to capture the quantum nature of spacetime without requiring a full resolution of the ultraviolet degrees of freedom~\cite{Connes:1994yd, Chamseddine:1996zu, Majid:1996kd}. The physical motivation for spacetime noncommutativity is formalized by Doplicher, Fredenhagen, and Roberts (DFR)~\cite{Doplicher:1994zv, Doplicher:1994tu} where the interplay between the Heisenberg uncertainty principle and GR  lead to a conclusion where one has to replace the classical commuting coordinates with noncommuting operators satisfying $[x^\mu, x^\nu] = i\theta^{\mu\nu}$, leading to spacetime uncertainty relations. 
From an algebraic point of view, the transition from classical to such quantum geometry is rooted in Gel'fand-Naimark duality, which establishes an equivalence between the category of smooth manifolds $\mathcal{M}$ and the commutative algebra of smooth functions $\mathcal{A} = C^\infty(\mathcal{M})$ defined upon them~\cite{Connes:1994yd}. Rather than modifying the point set of the underlying manifold directly, one deforms the commutative pointwise multiplication of functions into a NC, associative $\star$-product. Within the Hopf-algebraic formulation of NC differential geometry developed by Wess, Aschieri and collaborators~\cite{Aschieri:2005yw, Aschieri:2005zs, Aschieri:2009qh} (building on twisted symmetries~\cite{Drinfeld:1989st, Chaichian:2004za, Wess:2003da}), this deformation is systematically generated by a Drinfeld twist $\mathcal{F}$ constructed from elements of the universal enveloping algebra of diffeomorphisms $U(\Xi)$. The twist deforms not only the algebra of functions into $\mathcal{A}_\star$, but also the entire tensor algebra, the exterior calculus, the $\star$-covariant derivative, and the Riemann curvature tensor in a manner that preserves twisted diffeomorphism covariance. In this way, classical differential geometry is elevated to a NC-deformed gravitational theory.\\

\emph{Noncommutative manifold:} First, we briefly summarize the NC gravitational framework employed in the present work.  The construction is based on Drinfeld-twist deformation of differential geometry, in which the underlying differentiable manifold is retained while the algebra of functions, and consequently the tensor algebra, is deformed through a NC $\star$-product~\cite{Aschieri:2005yw, Aschieri:2005zs, Aschieri:2009qh}. The formulation used for gravitational perturbations follows Refs.~\cite{Herceg:2023zlk,Herceg:2023pmc, Herceg:2024vwc}, where a NC analogue of the Einstein-manifold condition was introduced and applied to black hole perturbation theory.

Let $\mathcal A=C^\infty(\mathcal M)$ denote the algebra of smooth functions on a spacetime manifold $\mathcal M$.  A deformation of this algebra is specified by a Drinfeld twist
\begin{equation}
    \mathcal F
    =
    f^A\otimes f_A,
    \qquad
    \mathcal F^{-1}
    =
    \bar f^{A}\otimes \bar f_{A},
    \label{eq:general_twist}
\end{equation}
where summation over the index $A$ is understood.  The ordinary pointwise product $\mu(f\otimes g)=fg$ is then replaced by the $\star$-product defined by
\begin{equation}
    f\star g
    \equiv
    \mu\!\left[
       \mathcal F^{-1}(f\otimes g)
    \right]
    =
    \bar f^{A}(f)\,
    \bar f_{A}(g).
    \label{eq:star_product_general}
\end{equation}
Thus, the deformation does not introduce a new set of spacetime points; rather, it modifies the algebraic relations among fields defined on the manifold.

For an Abelian twist generated by two mutually commuting vector fields
$X_1$ and $X_2$, we take
\begin{equation}
    \mathcal F
    =
    \exp\left[
       -\frac{i \qbar}{2}
       \left(
          X_1\otimes X_2-X_2\otimes X_1
       \right)
    \right],
    \qquad [X_1,X_2]=0 ,
    \label{eq:abelian_twist}
\end{equation}
where $\qbar$ denotes the NC deformation parameter~\footnote{Hereafter, we refer to the NC deformation as the $\qbar$-deformation and use the adjective $\qbar$uantum.}. Since the twist generators used below involve derivatives with respect to spacetime coordinates, $\qbar$ has dimensions of length squared in geometrized units.  To first order in $\qbar$, the corresponding $\star$-product becomes
\begin{equation}
    f\star g
    =
    fg
    +\frac{i \qbar}{2}
      \left[
        X_1(f)X_2(g)-X_2(f)X_1(g)
      \right]
    +\mathcal O(\qbar^2).
    \label{eq:star_product_linear}
\end{equation}
All calculations in the present work are performed consistently to first order in the deformation parameter.

The deformation of the product induces corresponding modifications of the geometrical quantities.  The $\qbar$-deformed connection is defined through the $\star$-covariant derivative $\widehat{\nabla}^{\star}$, whose coefficients in a natural basis are given by
\begin{equation}
    \widehat{\nabla}^{\star}_{\partial_\mu}\partial_\nu
    =
    \widehat{\Gamma}^{\rho}{}_{\mu\nu}
    \star \partial_\rho .
    \label{eq:nc_connection}
\end{equation}
In a natural basis adapted to the Abelian twist, the $\qbar$-deformed, metric compatible and torsion free connection coefficients are given by \cite{Herceg:2023zlk, Herceg:2023pmc}
\begin{equation}
    \widehat{\Gamma}^{\rho}{}_{\mu\nu}
    =
    \frac{1}{2}
    \widehat{g}^{\rho\sigma}
    \star
    \left(
       \partial_\mu g_{\nu\sigma}
       +\partial_\nu g_{\mu\sigma}
       -\partial_\sigma g_{\mu\nu}
    \right)
    \label{eq:nc_christoffel}
\end{equation}
where the inverse metric is replaced by its $\star$-inverse $g_{\star}^{\mu\nu}\equiv \widehat{g}^{\mu\nu}$, defined through
\begin{equation}
    \widehat{g}^{\mu\rho}\star g_{\rho\nu}
    =
    g_{\nu\rho}\star \widehat{g}^{\rho\mu}
    =
    \delta^\mu{}_\nu .
    \label{eq:star_inverse}
\end{equation}
Notice that, in general, $\widehat{g}^{\mu\nu}$ is not simply the ordinary matrix inverse $g^{\mu\nu}$ once the metric itself carries coordinate dependence on which the twist acts. The hats $\string^$ will be used to distinguish $\qbar$-deformed geometrical quantities from their commutative counterparts.  The corresponding curvature tensor is
\begin{align}
    \widehat R^{\rho}{}_{\sigma\mu\nu}
    ={}&
    \partial_\mu
       \widehat{\Gamma}^{\rho}{}_{\nu\sigma}
    -
    \partial_\nu
       \widehat{\Gamma}^{\rho}{}_{\mu\sigma}
    +
    \widehat{\Gamma}^{\lambda}{}_{\nu\sigma}
       \star
    \widehat{\Gamma}^{\rho}{}_{\mu\lambda}
    -
    \widehat{\Gamma}^{\lambda}{}_{\mu\sigma}
       \star
    \widehat{\Gamma}^{\rho}{}_{\nu\lambda}.
    \label{eq:nc_riemann}
\end{align}
The corresponding $\qbar$-deformed Ricci tensor is obtained by contraction,
\begin{equation}
    \widehat R_{\mu\nu}
    \equiv
    \widehat R^{\rho}{}_{\mu\rho\nu}.
    \label{eq:nc_ricci}
\end{equation}

Because the product is noncommutative, contractions and index ordering that are equivalent in the commutative theory need not remain equivalent after deformation.  The appropriate $\qbar$-deformed Ricci flat condition employed here is therefore formulated in terms of the $\mathcal R$-symmetrized Ricci tensor~\cite{Herceg:2023zlk,Herceg:2023pmc, Herceg:2024vwc}.  For the class of Abelian twists and the natural basis used below, it reduces to usual symmetrization
\begin{equation}
    \widehat{\rm R}_{\mu\nu}
    =
    \frac{1}{2}
    \left(
       \widehat R_{\mu\nu}
       +
       \widehat R_{\nu\mu}
    \right),
    \label{eq:r_sym_ricci}
\end{equation}
and the $\qbar$-deformed vacuum Einstein condition is
\begin{equation}
    \widehat{\rm R}_{\mu\nu}=0 .
    \label{eq:nc_einstein}
\end{equation}
In the commutative limit $\qbar\rightarrow0$, the $\star$-product reduces to the ordinary product and Eq.~\eqref{eq:nc_einstein} reduces to the usual commutative vacuum Einstein equation $R_{\mu\nu}=0$.\\

\emph{Kerr black hole and its properties:} The commutative geometry underlying our construction is the Kerr spacetime, which describes an asymptotically flat, stationary and axisymmetric rotating black hole in vacuum~\cite{Kerr:1963ud, Boyer:1966qh, Chandrasekhar:1985kt}. The Kerr metric in Boyer-Lindquist coordinates $(t,r,\theta,\varphi)$ can be written as
\begin{align}
ds^2_{\rm Kerr}
={}&
-\left(1-\frac{2Mr}{\Sigma}\right)dt^2
-\frac{4M^2\chi r\sin^2\theta}{\Sigma}\,
dt\,d\varphi
+\frac{\Sigma}{\Delta}\,dr^2
+\Sigma\,d\theta^2
+
\left(
r^2+\chi^2M^2
+\frac{2\chi^2M^3r\sin^2\theta}{\Sigma}
\right)
\sin^2\theta\,d\varphi^2 ,
\label{eq:full_kerr_metric}
\end{align}
where
\begin{equation}
    \Sigma
    =
    r^2+\chi^2M^2\cos^2\theta,
    \qquad
    \Delta
    =
    r^2-2Mr+\chi^2M^2,
    \label{eq:kerr_Sigma_Delta}
\end{equation}
and
\begin{equation}
    \chi
    \equiv
    \frac{J}{M^2}
\label{eq:dimensionless_spin}
\end{equation}
is the dimensionless spin parameter. The Kerr spacetime is a Petrov type D vacuum solution of the Einstein field equations,
\begin{equation}
    R_{\mu\nu}=0.
    \label{eq:kerr_vacuum}
\end{equation}

The black hole mass $M$ sets the overall length scale of the geometry, while $J=\chi M^2$ denotes its angular momentum. The function $\Sigma$ controls the angular dependence of the geometry and vanishes only at
$r=0$, $\theta=\pi/2$, corresponding to the Kerr ring singularity. The zeros of $\Delta$ determine the two characteristic horizon radii,
\begin{equation}
    r_{\pm}
    =
    M\left(
       1\pm\sqrt{1-\chi^2}
    \right),
    \label{eq:kerr_horizons}
\end{equation}
where $r_+$ is the event horizon and $r_-$ is the inner Cauchy horizon. A black hole solution therefore requires
\begin{equation}
    |\chi|\leq1 .
\end{equation}
The metric \eqref{eq:full_kerr_metric} reduces to the Schwarzschild geometry when $\chi\to0$.

The Kerr spacetime is stationary and axially symmetric, with the associated commuting Killing vector fields $\xi_{(t)} = \partial_t$ and $\xi_{(\varphi)} = \partial_\varphi$, such that
satisfying
\begin{equation}
    \mathcal L_{\xi_{(t)}}g_{\mu\nu}=0,
    \qquad
    \mathcal L_{\xi_{(\varphi)}}g_{\mu\nu}=0,
    \qquad
    [\xi_{(t)},\xi_{(\varphi)}]=0 .
\label{eq:kerr_killing_properties}
\end{equation}
 Consequently, any linear combination
\begin{equation}
    K
    =
    \alpha\,\xi_{(t)}
    +
    \beta\,\xi_{(\varphi)}
    =
    \alpha\,\partial_t
    +
    \beta\,\partial_\varphi
    \qquad
    \alpha,\beta\in\mathbb R
\label{eq:kerr_combined_killing}
\end{equation}
is also a Killing vector.  This observation is particularly important for the twist deformation considered below, since precisely such a linear combination is used as one of the generators of the Drinfeld twist.

The event horizon is itself generated by a particular linear combination
of the stationary and axial Killing vectors,
\begin{equation}
    \xi_H
    =
    \partial_t
    +
    \Omega_H\,\partial_\varphi ,
\label{eq:horizon_generator}
\end{equation}
where the constant
\begin{equation}
    \Omega_H
    =
    \frac{\chi M}{r_+^2+\chi^2M^2}
    =
    \frac{\chi}{2r_+}
\label{eq:horizon_angular_velocity}
\end{equation}
is the angular velocity of the Kerr horizon. These symmetry properties underlie much of the special structure of black hole perturbation theory.\\

\emph{Pseudo-semi-Killing twist:} 
The symmetry structure of the Kerr metric naturally motivates the class of twists considered below. Since $\partial_t$ and $\partial_\varphi$ are commuting Killing vector fields, the combination $K=\alpha\partial_t+\beta\partial_\varphi$ is itself Killing. We combine it with the radial vector field $\partial_r$, which is not a Killing vector of the Kerr geometry. The resulting deformation belongs to the pseudo-semi-Killing class of Abelian twists~\footnote{The choice of twist is motivated by physical considerations and, in principle, should be constrained by observations. In the absence of such constraints, the symmetries of the background spacetime provide useful guidance. A twist generated exclusively by Killing vectors leaves the background geometry undeformed, and the leading $\qbar$uantum
contributions to the gravitational field equations appear only at quadratic order in metric perturbations. Consequently, these effects are not captured within linear perturbation theory. The pseudo-semi-Killing construction offers an alternative by combining a Killing vector with a commuting vector field that is not necessarily Killing. This choice preserves the classical background as a solution while allowing $\qbar$uantum modifications to the gravitational perturbation equations already at linear order~\cite{Herceg:2023zlk,Herceg:2023pmc,Herceg:2024vwc}.}. They are generated by
\begin{equation}
   X_1\equiv K
    =
    \alpha\,\partial_t
    +
    \beta\,\partial_\varphi,
    \qquad
    X_2=\partial_r,
    \qquad
    \alpha,\beta\in\mathbb R ,
    \label{eq:semi_pseudo_generators}
\end{equation}
so that
\begin{equation}
    \mathcal F
    =
    \exp\left[
       -\frac{i \qbar}{2}
       \left(
          K\otimes\partial_r
          -
          \partial_r\otimes K
       \right)
    \right].
    \label{eq:semi_pseudo_twist}
\end{equation}
The resulting coordinate algebra is given by
\begin{equation}
    [t,r]_\star=i \qbar \alpha,
    \qquad
    [\varphi,r]_\star=i \qbar \beta,
    \label{eq:nc_coordinate_relations}
\end{equation}
while the remaining coordinate commutators vanish.

For a perturbation mode with harmonic dependence
\begin{equation}
    h_{\mu\nu}
    \propto
    e^{-i\omega t+i m\varphi},
    \label{eq:harmonic_dependence_nc}
\end{equation}
the action of $K$ is diagonal,
\begin{equation}
    \mathcal L_K h_{\mu\nu}
    =
    i\lambda\,h_{\mu\nu},
    \qquad
    \lambda
    =
    -\alpha\omega+\beta m ,
    \label{eq:lambda_definition}
\end{equation}
where \(\omega\) is the complex mode frequency and \(m\) the azimuthal quantum number. The relation \eqref{eq:lambda_definition} is useful because it converts the action of the Killing generator appearing in the twist into an algebraic factor at the level of the separated perturbation equations, which will be used later on.

In the present work we specialize to the $t$-$r$ deformation,
\begin{equation}
    X_1\equiv K=\partial_t,
    \qquad
    X_2=\partial_r,
    \label{eq:tr_twist_choice}
\end{equation}
corresponding to $\alpha=1$ and $\beta=0$, so that $\lambda=-\omega$. Consequently, the first-order $\qbar$-dependent corrections generated by the twist involve the combination $\qbar \lambda=-\qbar \omega$. To first order in $\qbar$,
\begin{equation}
    f\star g
    =
    fg
    +
    \frac{i \qbar }{2}
    \left(
       \partial_t f\,\partial_r g
       -
       \partial_r f\,\partial_t g
    \right)
    +\mathcal O(\qbar^2).
    \label{eq:tr_star_product}
\end{equation}
For the $t$-$r$ deformation, the dimension of the parameter $\qbar$ is $L^2$, and we introduce the dimensionless $q$-deformation parameter~\footnote{In SI units, the dimensionful deformation parameter is $\qbar=\hat\qbar(GM/c^2)^2$, where $G$ and $c$ are the gravitational constant and the speed of light, respectively. For a black hole of one solar mass, $GM_\odot/c^2 \simeq 1.477\,\mathrm{km}$, giving $\qbar\simeq 2.18\times10^6\,\hat\qbar\,\mathrm{m}^2$. For example, $\hat\qbar=0.1$ corresponds to $\qbar\simeq 2.18\times10^5\,\mathrm{m}^2$.}
\begin{equation}
    \hat \qbar
    \equiv
    \frac{\qbar}{M^2}.
    \label{eq:dimensionless_nc_parameter}
\end{equation}
The perturbative parameters used in the analysis are therefore $\chi$ and $\hat\qbar$, together with the perturbation $h$. Equations~\eqref{eq:nc_einstein} and \eqref{eq:tr_star_product} constitute the starting point for the slow-rotation perturbation analysis developed below.

\section{$\hat\qbar$-deformed metric perturbations}
\label{sec:nc_slow_rotation}

Although the pseudo-semi-Killing twist preserves the classical Kerr background, the $\qbar$-deformation modifies the gravitational field equations through the deformed connection and curvature. Consequently, the conventional Newman--Penrose/Teukolsky formalism cannot be applied directly without establishing the corresponding decoupling of the deformed perturbation equations. We therefore adopt the slow-rotation formalism and work directly with metric perturbations~\cite{Pani:2012vp,Pani:2012bp,Pani:2013pma,Franchini:2023xhd}.

 A general stationary axisymmetric spacetime can be written as ~\cite{Chandrasekhar:1985kt}
\begin{equation}
 ds^2_0
 =
 -\mathcal{H}^2 dt^2
 +\mathcal{Q}^2 dr^2
 +r^2\mathcal{K}^2
 \left[
 d\theta^2
 +\sin^2\theta
 \left(d\varphi-\mathcal{L}dt\right)^2
 \right],
 \label{eq:stationary_axisymmetric_metric}
\end{equation}
where $\mathcal{H}$, $\mathcal{Q}$, $\mathcal{K}$, and $\mathcal{L}$ are functions of $r$ and $\theta$ only. For a slowly rotating object, the metric can be expanded perturbatively in the dimensionless spin parameter $\chi$. Retaining terms through first order in $\chi$, Eq.~\eqref{eq:stationary_axisymmetric_metric} reduces to
\begin{equation}
 ds^2_0
 =
 -F(r)\,dt^2
 +B(r)^{-1}dr^2
 -2\varpi(r)\sin^2\theta\,d\varphi\,dt
 +r^2d\Omega^2
 +\mathcal O(\chi^2),
 \label{eq:slow_rotating_metric}
\end{equation}
where
\begin{equation}
 d\Omega^2=d\theta^2+\sin^2\theta\,d\varphi^2 .
\end{equation}
For the slowly rotating Kerr black hole,
\begin{equation}
 F(r)=B(r)=1-\frac{2M}{r},
 \qquad
 \varpi(r)=\frac{2M^2\chi}{r}.
 \label{Kerr1st}
\end{equation}
Equation~\eqref{Kerr1st} is the linear (in $\chi$) approximation of the Kerr geometry \eqref{eq:full_kerr_metric} introduced in Sec.~\ref{sec:nc_geometry}. We work to linear order in the metric perturbation $h$ and retain the terms through first order in rotation. The NC contributions are included below.

To study the odd-parity, or axial, perturbation equations, we follow the standard harmonic decomposition of the metric in tensor spherical harmonics,
\begin{equation}
\label{oddpart}
\delta g_{\mu\nu}^{\rm odd}
=
 \begin{pmatrix}
  0 & 0 & h_0^\ell S_\theta^{\ell}
        & h_0^\ell S_\varphi^{\ell} \\
  0 & 0 & h_1^\ell S_\theta^{\ell}
        & h_1^\ell S_\varphi^{\ell} \\
  \textit{sym} & \textit{sym}
        & -h_2^\ell \dfrac{X^\ell}{\sin\theta}
        & h_2^\ell\sin\theta\,W^\ell \\
  \textit{sym} & \textit{sym}
        & \textit{sym}
        & h_2^\ell\sin\theta\,X^\ell
 \end{pmatrix},
\end{equation}
where we use the conventions
\begin{alignat}{2}
S_\theta^\ell
&= -\frac{\partial_\varphi Y^\ell}{\sin\theta},
\qquad &
S_\varphi^\ell
&= \sin\theta\,\partial_\theta Y^\ell,
\\
X^\ell
&= 2\left(
\partial_\theta\partial_\varphi Y^\ell
-\cot\theta\,\partial_\varphi Y^\ell
\right),
\qquad &
W^\ell
&= \partial_\theta^2 Y^\ell
-\cot\theta\,\partial_\theta Y^\ell
-\frac{\partial_\varphi^2 Y^\ell}{\sin^2\theta}.
\label{XW}
\end{alignat}
We use the abbreviation $Y^\ell\equiv Y_{\ell m}(\theta,\varphi)$, with the azimuthal index $m$ suppressed when
no ambiguity arises. Under the parity transformation $(\theta,\varphi)\rightarrow
(\pi-\theta,\varphi+\pi)$, the axial (odd-parity) and polar (even-parity) tensor spherical harmonics
have parity eigenvalues $(-1)^{\ell+1}$ and $(-1)^\ell$, respectively~\cite{Pani:2013pma}. We adopt the Regge-Wheeler gauge, in which $ h_2^\ell=0.$ The axial sector is therefore described by the two functions $h_0^\ell(t,r)$ and $h_1^\ell(t,r)$. For the mode analysis we use the harmonic time dependence
\begin{equation}
 h_i^\ell(t,r)=h_i^\ell(r)e^{-i\omega t},
 \qquad i=0,1.
 \label{eq:fourier_axial}
\end{equation}

The separation of the angular dependence makes repeated use of
\begin{equation}
 \partial_\varphi Y_{\ell m}=imY_{\ell m},
 \qquad
 \left[
 \partial_\theta^2+\cot\theta\,\partial_\theta
 +\frac{1}{\sin^2\theta}\partial_\varphi^2
 \right]Y_{\ell m}
 =
 -\ell(\ell+1)Y_{\ell m},
 \label{eq:spherical_harmonic_identities}
\end{equation}
together with the orthogonality relation
\begin{equation}
 \int d\Omega\,
 Y_{\ell m}^{*}Y_{\ell' m'}
 =
 \delta_{\ell\ell'}\delta_{mm'}.
 \label{eq:Y_orthogonality}
\end{equation}
For the axial vector harmonics one similarly has
\begin{equation}
 \int d\Omega\,
 \gamma^{ab}
 (S_a^{\ell m})^{*}S_b^{\ell'm'}
 =
 \ell(\ell+1)
 \delta_{\ell\ell'}\delta_{mm'},
 \label{eq:axial_vector_orthogonality}
\end{equation}
where $\gamma_{ab}=\operatorname{diag}(1,\sin^2\theta)$ is the metric on the unit two-sphere. Further angular identities used in the calculation are collected in Appendix~\ref{identities}.

We now apply the $\hat\qbar$-deformed Einstein equations of Sec.~\ref{sec:nc_geometry} to axial perturbations. For the $t$-$r$ twist we take
\begin{equation}
    X_1=\partial_t,
    \qquad
    X_2=\partial_r.
\end{equation}
The $\hat\qbar$-deformed Einstein equations \eqref{eq:nc_einstein} are expanded consistently to first order in the metric perturbation, the dimensionless spin $\chi$, and the deformation parameter $\hat\qbar$. Thus, the perturbation equations retain the sectors
\begin{equation}
    \mathcal O(h),\qquad
    \mathcal O(\chi h),\qquad
    \mathcal O(\hat\qbar h),\qquad
    \mathcal O(\hat\qbar\chi h),
    \label{eq:nc_bookkeeping}
\end{equation}
while terms of order $\mathcal O(h^2)$, $\mathcal O(\chi^2 h)$, $\mathcal O(\hat\qbar^2 h)$, and higher orders are discarded.

For the $t$-$r$ twist, the deformation acts only along the temporal and radial directions. Consequently, after applying the same spherical-harmonic identities as in the commutative case, the angular dependence of the $\hat\qbar$-deformed Einstein equations \eqref{eq:nc_einstein} can be organized in the same harmonic basis. The $(t,\theta)$, $(r,\theta)$, and $(\theta,\varphi)$ components therefore take the form
\begin{align}
\widehat{\rm R}_{t\theta}: \qquad
&\left(
\alpha^{(t)}_{\ell}
+\widetilde{\alpha}^{(t)}_{\ell}\cos\theta
\right)
\partial_{\theta}Y^{\ell}
-\left(
\beta^{(t)}_{\ell}
+\widetilde{\beta}^{(t)}_{\ell}\cos\theta
\right)
\frac{\partial_{\varphi}Y^{\ell}}{\sin\theta}
+\eta^{(t)}_{\ell}\sin\theta\,Y^{\ell}
+\xi^{(t)}_{\ell}X^{\ell}
+\zeta^{(t)}_{\ell}\sin\theta\,W^{\ell}
=0,
\label{e13nc}
\\[1ex]
\widehat{\rm R}_{r\theta}: \qquad
&\left(
\alpha^{(r)}_{\ell}
+\widetilde{\alpha}^{(r)}_{\ell}\cos\theta
\right)
\partial_{\theta}Y^{\ell}
-\left(
\beta^{(r)}_{\ell}
+\widetilde{\beta}^{(r)}_{\ell}\cos\theta
\right)
\frac{\partial_{\varphi}Y^{\ell}}{\sin\theta}
+\eta^{(r)}_{\ell}\sin\theta\,Y^{\ell}
+\xi^{(r)}_{\ell}X^{\ell}
+\zeta^{(r)}_{\ell}\sin\theta\,W^{\ell}
=0,
\label{e23nc}
\\[1ex]
\widehat{\rm R}_{\theta\varphi}: \qquad
&f_{\ell}\sin\theta\,\partial_{\theta}Y^{\ell}
+g_{\ell}\partial_{\varphi}Y^{\ell}
+s_{\ell}\frac{X^{\ell}}{\sin\theta}
+t_{\ell}W^{\ell}
=0.
\label{e34nc}
\end{align}

To distinguish the $\qbar$uantum corrections from their commutative counterparts, we write each radial coefficient schematically as
\begin{equation}
    \mathcal C_{\ell}^{(i)}
    =
    \mathcal C_{\ell,\mathrm{GR}}^{(i)}
    +
    \hat\qbar M^2\,\mathcal C_{\ell,\qbar}^{(i)},
    \qquad
    \mathcal C_{\ell}^{(i)}
    \in
    \{
    \alpha,\widetilde{\alpha},
    \beta,\widetilde{\beta},
    \eta,\xi,\zeta
    \},
    \label{eq:nc_coeff_decomposition}
\end{equation}
with an analogous decomposition for $f_\ell$, $g_\ell$, $s_\ell$, and $t_\ell$. Both the GR and $\qbar$uantum coefficients are retained only through first order in $\chi$. The commutative contributions are listed in Appendix~\ref{app:comm_coefficients}, and the explicit $\hat\qbar$-dependent corrections in Appendix~\ref{app:nc_coefficients}.

Thus, after reducing higher angular derivatives using the spherical-harmonic identities, the entire angular dependence is expressed in the basis $Y^\ell$, $\partial_\theta Y^\ell$, $\partial_\varphi Y^\ell/\sin\theta$, $X^\ell$, and $W^\ell$. Projection onto this basis using the corresponding orthogonality relations yields a system of radial equations. In a rotating axisymmetric background, modes with different values of $m$ remain decoupled, whereas different multipoles $\ell$ can in general mix. At first order in $\chi$, a mode with multipolar index $\ell$ couples to opposite-parity modes with indices $\ell\pm1$. However, these couplings do not contribute to the QNM eigenfrequencies at first order in $\chi$ ~\cite{Pani:2012bp,Pani:2013pma}. They can therefore be neglected for the first-order eigenvalue problem, allowing the axial and polar sectors to be treated separately at this order.

Projecting onto this harmonic basis using the orthogonality relations above gives
\begin{align}
0={}&
\ell(\ell+1)\beta^{(t)}_{\ell}
+i m
\left[
(\ell-1)(\ell+2)\zeta^{(t)}_{\ell}
+\widetilde{\alpha}^{(t)}_{\ell}
+\eta^{(t)}_{\ell}
\right],
\label{eq13nc}
\\
0={}&
\ell(\ell+1)\beta^{(r)}_{\ell}
+i m
\left[
(\ell-1)(\ell+2)\zeta^{(r)}_{\ell}
+\widetilde{\alpha}^{(r)}_{\ell}
+\eta^{(r)}_{\ell}
\right],
\label{eq14nc}
\\
0={}&
\ell(\ell+1)t_{\ell}
+i m g_{\ell}.
\label{eq15nc}
\end{align}
Only two of these three equations are independent. Using the harmonic time dependence $e^{-i\omega t}$ defined above, we solve Eq.~\eqref{eq14nc} for $h_0'$ and substitute the result into Eq.~\eqref{eq15nc}. This yields a second-order radial equation for $h_1^\ell(r)$,
\begin{equation}
\left(
A_{\rm GR}+ \hat\qbar M^2 A_{\qbar}
\right)h_1''
+
\left(
B_{\rm GR}+\hat\qbar M^2 B_{\qbar}
\right)h_1'
+
\left(
C_{\rm GR}+\hat\qbar M^2 C_{\qbar}
\right)h_1
=0,
\label{eq:nc_second_order_h1}
\end{equation}
where the explicit coefficients are given in Appendix~\ref{app:nc_radial_coefficients}.

Equation~\eqref{eq:nc_second_order_h1} can be reduced to Schr\"odinger form by a field redefinition and a change of radial coordinate. Writing
\begin{equation}
    h_1^\ell(r)
    =
    \mathcal P(r)\,\psi^\ell(r),
    \qquad
    n(r)\equiv\frac{dr_*}{dr},
    \label{eq:nc_field_redefinition}
\end{equation}
the absence of a first derivative of $\psi^\ell$ fixes $\mathcal P(r)$, up to an overall normalization, through
\begin{equation}
    \frac{\mathcal P'}{\mathcal P}
    =
    -\frac{1}{2}
    \left(
    \frac{B}{A}
    +
    \frac{n'}{n}
    \right),
    \qquad
    A=A_{\rm GR}+\hat\qbar M^2 A_{\qbar},
    \quad
    B=B_{\rm GR}+\hat\qbar M^2 B_{\qbar}.
    \label{eq:field_redefinition_condition}
\end{equation}
The transformation of Eq.~\eqref{eq:nc_second_order_h1} into Schr\"odinger form involves a redefinition of the perturbation variable and the introduction of a generalized tortoise coordinate, following the procedure outlined in Appendix C of Ref.~\cite{Herceg:2023pmc}. The tortoise coordinate is defined by
\begin{equation}
    \frac{dr_*}{dr}
    =
    \frac{r}{r-2M}
    +
    \hat\qbar \,\frac{M^3\omega}{(r-2M)^2},
    \label{eq:nc_tortoise_adopted}
\end{equation}
or, equivalently,
\begin{equation}
    r_*
    =
    r
    +
    2M\ln\left(\frac{r-2M}{2M}\right)
    -
    \hat\qbar \,\frac{M^3\omega}{r-2M}
    +\mathcal O(\hat\qbar ^2).
    \label{eq:nc_tortoise_integrated}
\end{equation}
To first order in \(\hat\qbar\), the corresponding radial derivative operator is
\begin{equation}
    \frac{d}{dr_*}
    =
    \left(
    F(r)
    -
    \hat\qbar \,\frac{M^3\omega}{r^2}
    \right)
    \frac{d}{dr}
    +\mathcal O(\hat\qbar ^2).
    \label{eq:nc_inverse_tortoise}
\end{equation}
With this choice, the axial master equation becomes
\begin{equation}
 \frac{d^2\psi^\ell}{dr_*^2}
 +
 \left[
 \omega^2
 -
 \frac{4m\chi M^2\omega}{r^3}
 -
 V_{\rm eff}(r,\omega)
 \right]
 \psi^\ell
 =0,
 \label{eq:nc_axial_master}
\end{equation}
where the effective potential is decomposed as
\begin{equation}
    V_{\rm eff}(r,\omega)
    =
    V_{\rm axial}(r,\omega)
    +
    \hat\qbar M^2\,V_{\qbar}(r,\omega).
    \label{eq:nc_effective_potential}
\end{equation}
The commutative slow-rotation contribution is
\begin{equation}
 V_{\rm axial}(r,\omega)
 =
 F(r)
 \left[
 \frac{\ell(\ell+1)}{r^2}
 -\frac{6M}{r^3}
 +\frac{
 24m\chi M^2(3r-7M)
 }{
 \ell(\ell+1)\omega r^6
 }
 \right].
 \label{vaxial}
\end{equation}
For compactness, defining
\begin{equation}
    L\equiv\ell(\ell+1),
\end{equation}
the $\qbar$uantum correction is
\begin{align}
V_{\qbar}(r,\omega)
={}&
\omega^2
\left(
\frac{6m\chi M^2}{r^4}
-
\frac{3m\chi M^2}{Lr^4}
\right)
+
\frac{\omega}{2r^5}
\left[
2Lr^2
-
2(3L+5)Mr
+
32M^2
\right]
\nonumber\\
&+
\frac{3m\chi M^2}{4Lr^9}
\left[
(22L-408)r^3
-
4(23L-465)Mr^2
+
4(24L-779)M^2r
+
2120M^3
\right].
\label{eq:nc_potential_adopted}
\end{align}
The term proportional to $\omega$ survives in the nonrotating limit and represents the pure $\qbar$uantum correction, whereas the remaining terms are of mixed order $\mathcal O(\hat\qbar \chi)$. Rotation therefore introduces an additional $\omega^2$ dependence in the $\qbar$uantum part of the effective potential.

Several limits provide useful consistency checks. For $\hat\qbar\rightarrow0$, Eqs.~\eqref{eq:nc_tortoise_adopted} and \eqref{eq:nc_axial_master} reduce to the commutative slow-rotation equation with the potential $V_{\rm axial}$ given in Eq.~\eqref{vaxial}. For $\chi\rightarrow0$, they reduce to the corresponding $t$-$r$ $\qbar$uantum Schwarzschild axial equation~\cite{Herceg:2023zlk,Herceg:2023pmc, Herceg:2024vwc}. Finally, in the simultaneous limit $\hat\qbar \rightarrow0$ and $\chi\rightarrow0$, the standard Regge-Wheeler equation is recovered~\cite{Regge:1957td}.

\section{$\hat\qbar$-deformed quasinormal modes}
\label{sec:nc_qnms}

In this section we calculate the $\hat\qbar$-deformed quasinormal-mode frequencies of a slowly rotating Kerr black hole using two methods, namely the semianalytic WKB method and the continued-fraction method.

\subsection{WKB method}
\label{sec:wkb_method}

To apply the WKB method, we write the axial master equation \eqref{eq:nc_axial_master} in the compact form
\begin{equation}
 \frac{d^2\psi^\ell}{dr_*^2}
 +Q(r,\omega)\psi^\ell=0,
 \qquad
 Q(r,\omega)
 =
 \omega^2
 -\frac{4m\chi M^2\omega}{r^3}
 -V_{\rm eff}(r,\omega),
 \label{eq:wkb_master}
\end{equation}
where the effective potential $V_{\rm eff}$ and the corresponding tortoise coordinate are given in Eqs.~\eqref{eq:nc_effective_potential}-\eqref{eq:nc_potential_adopted} and Eq.~\eqref{eq:nc_tortoise_adopted}, respectively.

Unlike the Schwarzschild Regge-Wheeler potential, the effective potential in the present slow-rotation $\hat\qbar$-deformed problem depends explicitly on the frequency $\omega$. Following the WKB construction of
Refs.~\cite{Schutz:1985km,Iyer:1986np} and its extension to frequency-dependent rotating potentials~\cite{Seidel:1989bp,Tang:2025qaq} (also see Refs.~\cite{Kokkotas:1991vz,Franzin:2022iai}), we define
\begin{equation}
 Q_j\equiv
 \left.
 \left(\frac{d}{dr_*}\right)^jQ
 \right|_{r=\bar r},
 \qquad
 Q_1(\bar r,\omega)=0.
 \label{eq:qj_peak}
\end{equation}
Because $Q$ depends on the complex frequency, the stationary point $\bar r$ is generally complex. Therefore, $Q_1=0$ and the WKB quantization condition must be solved simultaneously for $(\bar r,\omega)$.

For $\nu=n+1/2$, the first-order condition is
\begin{equation}
 F_1\equiv
 Q_0+i\nu\sqrt{2Q_2}=0.
 \label{eq:wkb1_condition}
\end{equation}
Since $Q_2$ generally becomes complex in the rotating, $\hat\qbar$-deformed problem, the branch of the square root must be specified. We choose the positive square root in the Schwarzschild limit ($\chi=\hat\qbar=0$) and continuously extend this branch to nonzero values of $\chi$ and $\hat\qbar$.\footnote{In the Schwarzschild limit, $Q=\omega^2-V_{\rm RW}(r)$. At the maximum of the Regge-Wheeler potential, $d^2V_{\rm RW}/dr_*^2<0$, so that $Q_2=-d^2V_{\rm RW}/dr_*^2>0$. Consequently, $\sqrt{2Q_2}$ is real, and we choose its positive value. With the time dependence $e^{-i\omega t}$, this choice yields the usual damped QNM solution with $\omega _I<0$.}

At third order in the WKB approximation (WKB3), we use
\begin{equation}
 F_3\equiv
 Q_0-\mathcal A_2
 +i\nu\sqrt{2Q_2}\left(1+\mathcal B_3\right)=0,
 \label{eq:wkb3_condition}
\end{equation}
where
\begin{equation}
 \mathcal A_2=
 \frac18\frac{Q_4}{Q_2}
 \left(\frac14+\nu^2\right)
 -\frac1{288}
 \left(\frac{Q_3}{Q_2}\right)^2
 \left(7+60\nu^2\right),
 \label{eq:wkb_a2}
\end{equation}
and
\begin{align}
 \mathcal B_3=
 \frac{1}{2Q_2}\Bigg[
 &\frac5{6912}
 \left(\frac{Q_3}{Q_2}\right)^4
 (77+188\nu^2)
 -\frac1{384}
 \frac{Q_3^2Q_4}{Q_2^3}
 (51+100\nu^2)
 \nonumber\\
 &+\frac1{2304}
 \left(\frac{Q_4}{Q_2}\right)^2
 (67+68\nu^2)
 +\frac1{288}
 \frac{Q_3Q_5}{Q_2^2}
 (19+28\nu^2)
 -\frac1{288}
 \frac{Q_6}{Q_2}
 (5+4\nu^2)
 \Bigg].
 \label{eq:wkb_b3}
\end{align}
Second-order WKB follows from Eq.~\eqref{eq:wkb3_condition} by setting $\mathcal B_3=0$, whereas first order follows by also setting $\mathcal A_2=0$. These expressions are equivalent to the usual Iyer-Will form written in terms of derivatives of the barrier potential~\cite{Iyer:1986np}.

To implement the perturbative expansion, we decompose $Q$ into contributions of different orders in $\chi$ and $\hat\qbar$. We denote by $\mathcal{Q}_{ab}$ the coefficient of $\chi^a\hat\qbar^{\,b}$, where $a,b\in\{0,1\}$. Thus, $\mathcal{Q}_{00}$ represents the Schwarzschild contribution, $\mathcal{Q}_{10}$ and $\mathcal{Q}_{01}$ describe the first-order rotational and $\qbar$uantum corrections, respectively, and $\mathcal{Q}_{11}$ contains their mixed contribution. We write
\begin{equation}
 Q
 =
 \mathcal{Q}_{00}
 +\chi \mathcal{Q}_{10}
 +\hat \qbar\,\mathcal{Q}_{01}
 +\chi\hat \qbar\,\mathcal{Q}_{11},
 \qquad
 \frac{d}{dr_*}
 =
 \left(F+\hat \qbar\,g_{\qbar} \right)\frac{d}{dr},
 \qquad
 g_{\qbar}=-\frac{M^3\omega}{r^2}.
 \label{eq:sector_decomposition}
\end{equation}
Similarly, $\mathcal{Q}_{ab}^{(j)}$ denotes the coefficient of $\chi^a\hat\qbar^{\,b}$ in $Q_j$, with
$\mathcal{Q}_{ab}^{(0)}=\mathcal{Q}_{ab}$. The tortoise-derivative recursion is therefore
\begin{align}
 \mathcal{Q}_{00}^{(j+1)}
 &=F\,\partial_r \mathcal{Q}_{00}^{(j)},
 &
 \mathcal{Q}_{10}^{(j+1)}
 &=F\,\partial_r \mathcal{Q}_{10}^{(j)},
 \nonumber\\
 \mathcal{Q}_{01}^{(j+1)}
 &=F\,\partial_r \mathcal{Q}_{01}^{(j)}
 +g_{\qbar }\partial_r \mathcal{Q}_{00}^{(j)},
 &
 \mathcal{Q}_{11}^{(j+1)}
 &=F\,\partial_r \mathcal{Q}_{11}^{(j)}
 +g_{\qbar} \partial_r \mathcal{Q}_{10}^{(j)}.
 \label{eq:wkb_sector_recursion}
\end{align}
This construction prevents spurious $\chi^2$, $\hat \qbar^2$, and higher-order sectors from entering $Q_2,\ldots,Q_6$.

Finally, because the master equation itself was derived only through $\mathcal O(\chi)$, $\mathcal O(\hat \qbar)$, and $\mathcal O(\chi\hat \qbar)$, we employ a WKB calculation that is strictly perturbative in these parameters. The formally consistent QNM frequency is therefore written as
\begin{equation}
M\omega
=
M\omega_{00}
+\chi\,M\omega_{10}
+\hat\qbar\,M\omega_{01}
+\chi\hat\qbar\,M\omega_{11}
+\mathcal O(\chi^2,\hat\qbar^2).
 \label{eq:omega_strict_expansion}
\end{equation}
The coefficients in Eq.~\eqref{eq:omega_strict_expansion} are obtained from the third-order WKB approximation by solving the stationary-point and WKB conditions simultaneously and evaluating symmetric finite differences around the Schwarzschild limit $(\chi,\hat\qbar)=(0,0)$. The resulting frequencies are evaluated using the perturbative expansion in Eq.~\eqref{eq:omega_strict_expansion}, retaining only terms linear in $\chi$ and $\hat\qbar$, including their mixed contribution. The WKB results provide an independent consistency check of the continued-fraction calculations, which serve as our primary numerical reference.


\subsection{Continued-fraction method}
\label{sec:leaver_tr}

We compute the QNM frequencies using a Frobenius continued-fraction construction adapted from Leaver's method~\cite{Leaver:1985ax}. For convenience, we define
\begin{equation}
 R\equiv2M.
 \label{eq:leaver_R_definition}
\end{equation}
The quantities $L=\ell(\ell+1)$, $\chi=J/M^2$, and $\hat\qbar$ are those introduced in the preceding sections.

Using the axial master equation~\eqref{eq:nc_axial_master}, we define the total barrier
\begin{equation}
 \mathcal V(r,\omega)
 =
 \frac{4m\chi M^2\omega}{r^3}
 +V_{\rm eff}(r,\omega),
 \label{eq:total_barrier_leaver}
\end{equation}
so that
\begin{equation}
 \frac{d^2\psi^\ell}{dr_*^2}
 +\left[\omega^2-\mathcal V(r,\omega)\right]\psi^\ell=0.
 \label{eq:leaver_master_dimensional}
\end{equation}
Here $V_{\rm eff}$ is the effective potential defined in Eq.~\eqref{eq:nc_effective_potential}. Writing the tortoise coordinate~\eqref{eq:nc_tortoise_adopted} in terms of $R=2M$ gives
\begin{equation}
 \frac{dr_*}{dr}
 =
 \frac{r}{r-R}
 +\frac{\hat\qbar M^2\omega R}{2(r-R)^2}.
 \label{eq:second_tortoise_leaver}
\end{equation}

\subsubsection{Shifted-coordinate formulation and boundary conditions}

To implement Leaver's method, it is convenient to introduce a shifted radial coordinate $\rho$, together with the dimensionless variables $x=\rho/R$, $w=R\omega$, and $x_*=r_*/R$. The radial equation then takes the form
\begin{equation}
 \frac{d^2\psi^\ell}{dx_*^2}
 +\left[w^2-\mathcal U(x,w)\right]\psi^\ell=0,
 \qquad
 \frac{dx_*}{dx}=\frac{x}{x-1}.
 \label{eq:dimensionless_regular_equation}
\end{equation}
Here, the shifted coordinate is defined by~\footnote{A similar radial-coordinate shift was employed in Ref.~\cite{Herceg:2024upt} in the study of noncommutative Schwarzschild perturbations. The underlying radial translation was subsequently investigated from a nonperturbative perspective in Ref.~\cite{Herceg:2025rll}.}
\begin{equation}
 \rho
 =
 r-\frac{\hat\qbar M^2\omega}{2},
 \label{eq:rho_shift}
\end{equation}
which brings Eq.~\eqref{eq:second_tortoise_leaver} into the form
\begin{equation}
 \frac{dr_*}{d\rho}
 =
 \frac{\rho}{\rho-R}
 +\mathcal O(\hat\qbar^2).
 \label{eq:regular_tortoise}
\end{equation}
The potential is expressed in the same shifted coordinate, consistently retaining terms linear in $\chi$ and $\hat\qbar$, including their mixed contribution. Defining $j=m\chi/4$, we write
\begin{equation}
 \mathcal U(x,w)
  =
 \sum_{k=2}^{9}\frac{u_k}{x^k},
 \label{eq:regularized_U_expansion}
\end{equation}
with coefficients
\begin{align}
 u_2={}&L,\nonumber\\
 u_3={}&4jw-(L+3),\nonumber\\
 u_4={}&3+\frac{\hat \qbar}{4}
 \left(2w-\frac{3jw^2}{L}\right),\nonumber\\
 u_5={}&\frac{72j}{Lw}-\frac{\hat \qbar w}{2},\nonumber\\
 u_6={}&-\frac{156j}{Lw}
 +\frac{j\hat \qbar}{4}
 \left(\frac{33}{2}-\frac{486}{L}\right),
 \nonumber\\
 u_7={}&\frac{84j}{Lw}
 +\frac{3j\hat \qbar(777-23L)}{8L},\nonumber\\
 u_8={}&\frac{3j\hat \qbar(24L-1171)}{16L},\nonumber\\
 u_9={}&\frac{795j\hat \qbar}{16L}.
 \label{eq:regularized_U_coefficients}
\end{align}

At the characteristic horizon $x=1$, the squared radial wave number is
\begin{equation}
 \kappa_H^2
 =
 w^2-\mathcal U(1,w)
 =
 w^2-4jw+\frac{3j\hat \qbar w^2}{4L}.
 \label{eq:kappa_squared}
\end{equation}
The branch continuously connected to $\kappa_H=w$ at $(\chi,\hat \qbar)=(0,0)$ is, to the retained order,
\begin{equation}
 \kappa_H
 =
 w-2j+\frac{3j\hat \qbar w}{8L}.
 \label{eq:kappa_strict}
\end{equation}
The commutative part satisfies
\begin{equation}
 \frac{\kappa_H}{R}
 =
 \omega-m\Omega_H+\mathcal O(\chi^2),
 \qquad
 \Omega_H=\frac{\chi}{4M}.
 \label{eq:kappa_commutative_limit}
\end{equation}
With the convention $e^{-i\omega t}$ for the time dependent part of the perturbation (see \eqref{eq:fourier_axial}), the QNM boundary conditions take the standard ingoing and outgoing forms, with the horizon wave number $\kappa_H$,
\begin{align}
 \psi^\ell
 &\sim e^{-i\kappa_Hx_*}
 \sim (x-1)^{-i\kappa_H},
 &&x\rightarrow1,
 \label{eq:ingoing_bc_leaver}\\
 \psi^\ell
 &\sim e^{+iwx_*}
 \sim e^{iwx}x^{iw},
 &&x\rightarrow\infty.
 \label{eq:outgoing_bc_leaver}
\end{align}
The first condition is purely ingoing at  horizon, whereas the second is purely outgoing at spatial infinity. The $\qbar$uantum corrections decay sufficiently rapidly that the asymptotic outgoing exponential is unchanged.

Equation~\eqref{eq:dimensionless_regular_equation} has a regular singular point at $x=1$ and an irregular singular point at infinity. Because the noncommutative barrier contains terms through $x^{-9}$, $x=0$ is an irregular interior singularity. It does not obstruct the exterior series below because it is mapped outside its convergence disk. There are no additional finite exterior radial singularities in the shifted coordinate. The factors $1/w$ in Eq.~\eqref{eq:regularized_U_coefficients} make $w=0$ a parameter singularity, not a radial singular point.

\subsubsection{Frobenius ansatz and recurrence relation}

Let us define
\begin{equation}
 z=\frac{x-1}{x},
 \qquad
 p=-i\kappa_H.
 \label{eq:leaver_z_p}
\end{equation}
Thus the horizon, spatial infinity, and the interior singularity $x=0$ map to $z=0$, $z=1$, and $z=\infty$, respectively. An ansatz which incorporates both QNM boundary conditions, Eqs.~\eqref{eq:ingoing_bc_leaver} and \eqref{eq:outgoing_bc_leaver}, is
\begin{equation}
 \psi^\ell(x)
 =
 e^{iwx}(x-1)^p x^{iw-p}
 \sum_{n=0}^{\infty}b_n z^n.
 \label{eq:tr_leaver_ansatz}
\end{equation}
Near $x=1$, this has the factor $(x-1)^{-i\kappa_H}$. At infinity, $(x-1)^p x^{iw-p}\sim x^{iw}$, giving the outgoing behavior in Eq.~\eqref{eq:outgoing_bc_leaver}.
Substitution of Eq.~\eqref{eq:tr_leaver_ansatz} into Eq.~\eqref{eq:dimensionless_regular_equation}, followed by consistent removal of $\chi^2$, $\hat \qbar^2$, and higher-order sectors, gives
\begin{align}
 0={}&
 \alpha_n b_{n+1}
 +\beta_n b_n
 +\gamma_n b_{n-1}
 +\delta_n b_{n-2}
 +\epsilon_n b_{n-3}
 +\zeta_n b_{n-4}
 +\eta_n b_{n-5}
 +\xi_n b_{n-6},
 \qquad n\geq0,
 \label{eq:eight_term_recurrence}
\end{align}
where coefficients multiplying $b_k$ with $k<0$ are set to zero and
\begin{align}
 \alpha_n={}&(n+1)(n+2p+1),\nonumber\\
 \beta_n={}&-2n^2+(-4p+4iw-2)n+c_0,\nonumber\\
 \gamma_n={}&(n-1)(n+2p-2iw+1)+c_1,\nonumber\\
 \delta_n={}&c_2,\qquad
 \epsilon_n=c_3,\qquad
 \zeta_n=c_4,\qquad
 \eta_n=c_5,\qquad
 \xi_n=c_6,
 \label{eq:eight_recurrence_coefficients}
\end{align}
with
\begin{align}
 c_0={}&
 -L+3+4w^2+2iw+4ipw-2p
 +j\left(-4w+\frac{12}{Lw}\right)
 -\frac{\hat \qbar w}{2}
 +\frac{j\hat \qbar}{4}
 \left(\frac32+\frac{21}{4L}\right),
 \nonumber\\
 c_1={}&
 -3-2w^2-2iw-2ipw+2p
 +j\left(4w-\frac{120}{Lw}\right)
 +\hat \qbar w
 +\frac{j\hat \qbar}{4}
 \left(-24+\frac{261}{L}\right),
 \nonumber\\
 c_2={}&
 \frac{288j}{Lw}
 -\frac{\hat \qbar w}{2}
 +\frac{j\hat \qbar}{4}
 \left(81-\frac{3591}{4L}\right),
 \nonumber\\
 c_3={}&
 -\frac{264j}{Lw}
 +\frac{j\hat \qbar}{4}
 \left(-114+\frac{876}{L}\right),
 \nonumber\\
 c_4={}&
 \frac{84j}{Lw}
 +\frac{j\hat \qbar}{4}
 \left(\frac{147}{2}+\frac{279}{4L}\right),
 \nonumber\\
 c_5={}&
 \frac{j\hat \qbar}{4}
 \left(-18-\frac{513}{L}\right),
 \nonumber\\
 c_6={}&
 \frac{795j\hat \qbar}{16L}.
 \label{eq:c0_to_c6}
\end{align}
The eight-term recurrence~\eqref{eq:eight_term_recurrence} is reduced to a three-term relation by five Gaussian eliminations (see Appendix~A of Ref.~\cite{Hegde:2026vyd}),
\begin{equation}
 \widehat\alpha_n b_{n+1}
 +\widehat\beta_n b_n
 +\widehat\gamma_n b_{n-1}
 =0.
 \label{eq:reduced_three_term}
\end{equation}
For completeness, if $C^{(s)}_{n,k}$ denotes a row of an $s$-term recurrence, one elimination step is
\begin{align}
 F_n^{(s)}
 &=
 \frac{C^{(s)}_{n,s-1}}
 {C^{(s-1)}_{n-1,s-2}},
 \nonumber\\
 C^{(s-1)}_{n,0}
 &=
 C^{(s)}_{n,0},
 \nonumber\\
 C^{(s-1)}_{n,k}
 &=
 C^{(s)}_{n,k}
 -F_n^{(s)}C^{(s-1)}_{n-1,k-1},
 \qquad k=1,\ldots,s-2.
 \label{eq:gaussian_recurrence_reduction}
\end{align}
The desired minimal solution satisfies the characteristic continued fraction
\begin{equation}
 0
 =
 \widehat\beta_0
 -
 \frac{\widehat\alpha_0\widehat\gamma_1}
 {\widehat\beta_1-\displaystyle
  \frac{\widehat\alpha_1\widehat\gamma_2}
  {\widehat\beta_2-\displaystyle
   \frac{\widehat\alpha_2\widehat\gamma_3}
   {\widehat\beta_3-\cdots}}}.
 \label{eq:leaver_characteristic_cf}
\end{equation}
This minimal-solution condition implements the outgoing boundary condition
at $z=1$~\cite{Leaver:1985ax}.

The eight-term recurrence is reduced to a three-term recurrence by five successive Gaussian eliminations. At each evaluation, the characteristic continued fraction in Eq.~\eqref{eq:leaver_characteristic_cf} is truncated at $N_{\max}\sim300$ and evaluated by backward recursion.
The coefficients in Eq.~\eqref{eq:omega_strict_expansion} are determined from the continued-fraction roots on the branch continuously connected to the Schwarzschild solution. As a numerical check, the calculation reproduces the Schwarzschild axial fundamental mode~\cite{Chandrasekhar:1975zza, Leaver:1985ax, Herceg:2024vwc}
\begin{equation}
 M\omega_{220}
 =
 0.373672-0.088962\,i.
 \label{eq:schwarzschild_leaver_check}
\end{equation}

\section{QNM spectrum and numerical results}
\label{sec:results}

We now present the fundamental $(\ell,m,n)=(2,2,0)$ axial gravitational QNMs obtained from the WKB3 and Leaver methods. Throughout the analytical calculations, we retain the explicit dependence on the black hole mass $M$, while all numerical evaluations are performed with $M=1$. The resulting frequencies are reported in terms of the dimensionless quantity $M\omega$, using the dimensionless deformation parameter $\hat\qbar$ defined in Eq.~\eqref{eq:dimensionless_nc_parameter}.

All frequencies reported below are perturbative results of the form
\begin{equation}
 M\omega_{\ell m}
 =
 \omega_{00}^{(\ell)}
 +m\chi\,A_{\ell}
 +\hat\qbar \,B_{\ell}
 +m\chi \hat\qbar \,C_{\ell}
 +\mathcal O(\chi^2,\hat\qbar ^2).
 \label{eq:strict_spectrum_results}
\end{equation}
The coefficients $\omega_{00}^{(\ell)}$, $A_\ell$, $B_\ell$, and $C_\ell$ are independent of $m$, with the azimuthal dependence entering through the factors $m\chi$ and $m\chi\hat\qbar$. Here, $\omega_{00}^{(\ell)}=M \omega_{00}$ is the Schwarzschild frequency, $B_\ell$ represents the pure $\qbar$uantum correction, and $A_\ell$ and $C_\ell$ characterize the rotational and mixed corrections, respectively.
For $m\neq0$, $A_\ell=M\omega_{10}/m$, $B_\ell=M\omega_{01}$, and $C_\ell=M\omega_{11}/m$; these coefficients are determined at $m=1$ and apply to any allowed $m$ by factorization.

The complex coefficients $\omega_{00}^{(\ell)}$, $A_\ell$, $B_\ell$, and $C_\ell$ obtained with the  Leaver continued-fraction method are listed in Table~\ref{tab:strict_nc_coefficients}. The coefficients specify, respectively, the Schwarzschild limit, the rotational correction, the pure $\qbar$uantum correction, and the mixed rotation-$\qbar$uantum correction. 
\begin{table}[H]
\centering
\small

\begin{tabular}{ccrrrr}
\toprule
$\ell$ & Part & $\omega_{00}^{(\ell)}$ & $A_\ell$ & $B_\ell$ & $C_\ell$ \\
\midrule
\multirow{2}{*}{$2$}
& $\mathrm{Re}$ & $0.373672$ & $0.062883$ & $0.006834$ & $-0.004048$ \\
& $\mathrm{Im}$ & $-0.088962$ & $0.000998$ & $-0.001714$ & $0.000057$ \\
\addlinespace[3pt]
\multirow{2}{*}{$3$}
& $\mathrm{Re}$ & $0.599443$ & $0.067355$ & $0.007578$ & $0.000142$ \\
& $\mathrm{Im}$ & $-0.092703$ & $0.000652$ & $-0.001467$ & $-0.000428$ \\
\addlinespace[3pt]
\multirow{2}{*}{$4$}
& $\mathrm{Re}$ & $0.809178$ & $0.069814$ & $0.007855$ & $0.001071$ \\
& $\mathrm{Im}$ & $-0.094164$ & $0.000340$ & $-0.001183$ & $-0.000386$ \\
\bottomrule
\end{tabular}

\caption{Complex coefficients of the fundamental $(\ell,m,n)=(2,2,0)$ axial gravitational QNM spectrum, obtained from the Leaver continued-fraction method at $m=1$. The coefficients determine $M\omega_{\ell m}$ through Eq.~\eqref{eq:strict_spectrum_results} and are independent of $m$, with the azimuthal dependence factored out of $A_\ell$ and $C_\ell$.}
\label{tab:strict_nc_coefficients}
\end{table}
Only Leaver results are used for this coefficient table; the third-order WKB calculation is retained as an independent comparison for the QNM frequencies in Tables~\ref{tab:qnm_wkb_leaver_chi0}-\ref{tab:qnm_wkb_leaver_chi040506}.
\begin{table}[H]
\centering
\caption{Fundamental axial gravitational QNM frequencies $M\omega$ for the nonrotating ($\chi=0$) Schwarzschild background with $\qbar$uantum-spacetime deformation. For each $\hat\qbar$, the upper entry is WKB3 and the lower entry is Leaver; $\hat\qbar=0$ gives the ordinary Schwarzschild spectrum.}
\label{tab:qnm_wkb_leaver_chi0}
\scriptsize
\setlength{\tabcolsep}{10pt}
\renewcommand{\arraystretch}{0.88}
\begin{tabular}{ccccc}
\toprule
$\chi$ & $\hat\qbar$ & $\ell=m=2$ & $\ell=m=3$ & $\ell=m=4$ \\
& & \multicolumn{1}{c}{$M\omega$ (WKB3/Leaver)}
& \multicolumn{1}{c}{$M\omega$ (WKB3/Leaver)}
& \multicolumn{1}{c}{$M\omega$ (WKB3/Leaver)} \\
\midrule

\multirow{18}{*}{$0$}
& \multirow{2}{*}{$-0.2$} & $0.371802-0.088899\,i$ & $0.597749-0.092435\,i$ & $0.807527-0.093934\,i$ \\
& & $0.372305-0.088620\,i$ & $0.597928-0.092410\,i$ & $0.807607-0.093927\,i$ \\
\addlinespace[2.5pt]
& \multirow{2}{*}{$-0.1$} & $0.372482-0.089058\,i$ & $0.598507-0.092582\,i$ & $0.808312-0.094053\,i$ \\
& & $0.372988-0.088791\,i$ & $0.598685-0.092556\,i$ & $0.808393-0.094046\,i$ \\
\addlinespace[2.5pt]
& \multirow{2}{*}{$-0.01$} & $0.373094-0.089202\,i$ & $0.599189-0.092714\,i$ & $0.809019-0.094159\,i$ \\
& & $0.373603-0.088945\,i$ & $0.599368-0.092688\,i$ & $0.809100-0.094152\,i$ \\
\addlinespace[2.5pt]
& \multirow{2}{*}{$-0.001$} & $0.373155-0.089216\,i$ & $0.599258-0.092727\,i$ & $0.809090-0.094170\,i$ \\
& & $0.373665-0.088961\,i$ & $0.599436-0.092702\,i$ & $0.809171-0.094163\,i$ \\
\addlinespace[2.5pt]
& \multirow{2}{*}{$0$} & $0.373162-0.089217\,i$ & $0.599265-0.092728\,i$ & $0.809098-0.094171\,i$ \\
& & $0.373672-0.088962\,i$ & $0.599443-0.092703\,i$ & $0.809178-0.094164\,i$ \\
\addlinespace[2.5pt]
& \multirow{2}{*}{$0.001$} & $0.373169-0.089219\,i$ & $0.599273-0.092730\,i$ & $0.809106-0.094172\,i$ \\
& & $0.373679-0.088964\,i$ & $0.599451-0.092705\,i$ & $0.809186-0.094165\,i$ \\
\addlinespace[2.5pt]
& \multirow{2}{*}{$0.01$} & $0.373230-0.089233\,i$ & $0.599341-0.092743\,i$ & $0.809176-0.094183\,i$ \\
& & $0.373740-0.088979\,i$ & $0.599519-0.092718\,i$ & $0.809257-0.094176\,i$ \\
\addlinespace[2.5pt]
& \multirow{2}{*}{$0.1$} & $0.373842-0.089377\,i$ & $0.600023-0.092875\,i$ & $0.809883-0.094289\,i$ \\
& & $0.374355-0.089134\,i$ & $0.600201-0.092850\,i$ & $0.809964-0.094282\,i$ \\
\addlinespace[2.5pt]
& \multirow{2}{*}{$0.2$} & $0.374522-0.089536\,i$ & $0.600781-0.093021\,i$ & $0.810669-0.094408\,i$ \\
& & $0.375038-0.089305\,i$ & $0.600959-0.092996\,i$ & $0.810749-0.094401\,i$ \\
\bottomrule
\end{tabular}
\end{table}

Notice that, for every entry of the  WKB3 and Leaver frequencies  presented in Tables~\ref{tab:qnm_wkb_leaver_chi0}-\ref{tab:qnm_wkb_leaver_chi040506}, $\omega_I<0$; hence no unstable mode appears in the parameter range examined. At fixed spin, increasing $\hat\qbar$ moves the modes toward larger $M\omega_R$ and larger $|M\omega_I|$. For example, at $(\ell,m,\chi)=(2,2,0.1)$, changing $\hat\qbar$ from $-0.2$ to $0.2$ changes the Leaver result from
\begin{equation}
    M\omega=0.385043-0.088422i
\end{equation}
to
\begin{equation}
    M\omega=0.387453-0.089103i.
\end{equation}
The analogous trends persist for $\ell=3$ and $4$. For the representative configurations with $\chi=0.1,0.2$, the Leaver-WKB3 difference remains below approximately $0.22\%$ and falls to about $0.01\%$ for $\ell=4$.

At fixed $\chi$, Eq.~\eqref{eq:strict_spectrum_results} can be written in the affine form
\begin{equation}
    M\omega_{\ell m}
    =\mathcal A_{\ell m}(\chi)
    +\hat\qbar\,\mathcal B_{\ell m}(\chi),
    \label{eq:nc_affine_spectrum}
\end{equation}
where $\mathcal A_{\ell m}(\chi)=\omega_{00}^{(\ell)}+m\chi A_\ell$
and $\mathcal B_{\ell m}(\chi)=B_\ell+m\chi C_\ell$.
Consequently, the slopes with respect to the deformation and spin parameters are
\begin{align}
    \left.\frac{\partial(M\omega_{\ell m})}{\partial\hat\qbar}\right|_{\chi}
    &=B_\ell+m\chi C_\ell,
    \label{eq:nc_deformation_slope}\\
    \left.\frac{\partial(M\omega_{\ell m})}{\partial\chi}\right|_{\hat\qbar}
    &=m\left(A_\ell+\hat\qbar C_\ell\right),
    \label{eq:nc_spin_slope}\\
    \frac{\partial^2(M\omega_{\ell m})}{\partial\chi\,\partial\hat\qbar}
    &=m C_\ell.
    \label{eq:nc_mixed_slope}
\end{align}
As the coefficients are complex, these relations describe the slopes of both $M\omega_R$ and $M\omega_I$. Thus, the dependence on $\hat\qbar$ is linear at fixed $\chi$, but its slope varies with rotation through the mixed $\chi\hat\qbar$ contribution. Conversely, the slope with respect to $\chi$ depends on the deformation parameter. The coefficient $C_\ell$ therefore quantifies the interplay between the rotational and $\qbar$uantum corrections. For example, the Leaver coefficients for $(\ell,m)=(2,2)$ give $\partial_{\hat\qbar}\operatorname{Re}(M\omega_{220}) \simeq 0.006834-0.008095\chi$, showing that the deformation-induced real-frequency shift decreases with increasing black hole spin for the
co-rotating mode\footnote{For the harmonic dependence $e^{-i\omega t+im\varphi}$, co-rotating modes propagate in the same azimuthal direction as the black hole's rotation. For $\chi>0$ and $\omega_R>0$, these correspond to $m>0$, whereas $m<0$ describes counter-rotating modes~\cite{Berti:2009kk,Pani:2013pma}.}. At the retained perturbative order, the trajectories at fixed $\chi$ are straight lines in the complex-frequency plane, with spin-dependent slopes.

\begin{table}[H]
\centering
\caption{Fundamental axial gravitational QNM frequencies $M\omega$ for $\chi=0.1,0.2,0.3$. For each $\hat\qbar$, the upper entry is WKB3 and the lower entry is Leaver.}
\label{tab:qnm_wkb_leaver_chi010203}
\scriptsize
\setlength{\tabcolsep}{10pt}
\renewcommand{\arraystretch}{0.88}
\begin{tabular}{ccccc}
\toprule
$\chi$ & $\hat\qbar$ & $\ell=m=2$ & $\ell=m=3$ & $\ell=m=4$ \\
& & \multicolumn{1}{c}{$M\omega$ (WKB3/Leaver)}
& \multicolumn{1}{c}{$M\omega$ (WKB3/Leaver)}
& \multicolumn{1}{c}{$M\omega$ (WKB3/Leaver)} \\
\midrule

\multirow{18}{*}{$0.1$}
& \multirow{2}{*}{$-0.2$} & $0.384894-0.088987\,i$ & $0.617945-0.092225\,i$ & $0.835365-0.093768\,i$ \\
& & $0.385043-0.088422\,i$ & $0.618126-0.092188\,i$ & $0.835447-0.093760\,i$ \\
\addlinespace[2.5pt]
& \multirow{2}{*}{$-0.1$} & $0.385471-0.089130\,i$ & $0.618708-0.092384\,i$ & $0.836193-0.093902\,i$ \\
& & $0.385646-0.088593\,i$ & $0.618888-0.092348\,i$ & $0.836276-0.093894\,i$ \\
\addlinespace[2.5pt]
& \multirow{2}{*}{$-0.01$} & $0.385990-0.089258\,i$ & $0.619394-0.092527\,i$ & $0.836939-0.094023\,i$ \\
& & $0.386188-0.088746\,i$ & $0.619573-0.092491\,i$ & $0.837021-0.094015\,i$ \\
\addlinespace[2.5pt]
& \multirow{2}{*}{$-0.001$} & $0.386042-0.089271\,i$ & $0.619463-0.092541\,i$ & $0.837014-0.094035\,i$ \\
& & $0.386242-0.088761\,i$ & $0.619642-0.092506\,i$ & $0.837096-0.094027\,i$ \\
\addlinespace[2.5pt]
& \multirow{2}{*}{$0$} & $0.386047-0.089272\,i$ & $0.619470-0.092543\,i$ & $0.837022-0.094036\,i$ \\
& & $0.386248-0.088763\,i$ & $0.619650-0.092507\,i$ & $0.837104-0.094028\,i$ \\
\addlinespace[2.5pt]
& \multirow{2}{*}{$0.001$} & $0.386053-0.089274\,i$ & $0.619478-0.092545\,i$ & $0.837030-0.094038\,i$ \\
& & $0.386254-0.088764\,i$ & $0.619657-0.092509\,i$ & $0.837112-0.094029\,i$ \\
\addlinespace[2.5pt]
& \multirow{2}{*}{$0.01$} & $0.386105-0.089286\,i$ & $0.619547-0.092559\,i$ & $0.837105-0.094050\,i$ \\
& & $0.386309-0.088780\,i$ & $0.619726-0.092523\,i$ & $0.837187-0.094041\,i$ \\
\addlinespace[2.5pt]
& \multirow{2}{*}{$0.1$} & $0.386624-0.089415\,i$ & $0.620233-0.092702\,i$ & $0.837851-0.094170\,i$ \\
& & $0.386851-0.088933\,i$ & $0.620412-0.092667\,i$ & $0.837932-0.094162\,i$ \\
\addlinespace[2.5pt]
& \multirow{2}{*}{$0.2$} & $0.387201-0.089557\,i$ & $0.620996-0.092861\,i$ & $0.838679-0.094304\,i$ \\
& & $0.387453-0.089103\,i$ & $0.621174-0.092826\,i$ & $0.838761-0.094295\,i$ \\
\midrule

\multirow{18}{*}{$0.2$}
& \multirow{2}{*}{$-0.2$} & $0.397986-0.089076\,i$ & $0.638142-0.092015\,i$ & $0.863203-0.093602\,i$ \\
& & $0.397782-0.088225\,i$ & $0.638323-0.091967\,i$ & $0.863287-0.093593\,i$ \\
\addlinespace[2.5pt]
& \multirow{2}{*}{$-0.1$} & $0.398460-0.089201\,i$ & $0.638909-0.092186\,i$ & $0.864074-0.093752\,i$ \\
& & $0.398303-0.088394\,i$ & $0.639090-0.092139\,i$ & $0.864159-0.093743\,i$ \\
\addlinespace[2.5pt]
& \multirow{2}{*}{$-0.01$} & $0.398886-0.089314\,i$ & $0.639599-0.092340\,i$ & $0.864859-0.093887\,i$ \\
& & $0.398773-0.088546\,i$ & $0.639779-0.092295\,i$ & $0.864943-0.093877\,i$ \\
\addlinespace[2.5pt]
& \multirow{2}{*}{$-0.001$} & $0.398928-0.089326\,i$ & $0.639668-0.092356\,i$ & $0.864937-0.093900\,i$ \\
& & $0.398820-0.088561\,i$ & $0.639848-0.092310\,i$ & $0.865021-0.093890\,i$ \\
\addlinespace[2.5pt]
& \multirow{2}{*}{$0$} & $0.398933-0.089327\,i$ & $0.639676-0.092358\,i$ & $0.864946-0.093902\,i$ \\
& & $0.398825-0.088563\,i$ & $0.639856-0.092312\,i$ & $0.865030-0.093892\,i$ \\
\addlinespace[2.5pt]
& \multirow{2}{*}{$0.001$} & $0.398938-0.089328\,i$ & $0.639684-0.092359\,i$ & $0.864955-0.093903\,i$ \\
& & $0.398830-0.088565\,i$ & $0.639864-0.092314\,i$ & $0.865038-0.093893\,i$ \\
\addlinespace[2.5pt]
& \multirow{2}{*}{$0.01$} & $0.398980-0.089340\,i$ & $0.639753-0.092375\,i$ & $0.865033-0.093917\,i$ \\
& & $0.398877-0.088580\,i$ & $0.639933-0.092329\,i$ & $0.865117-0.093907\,i$ \\
\addlinespace[2.5pt]
& \multirow{2}{*}{$0.1$} & $0.399406-0.089452\,i$ & $0.640443-0.092529\,i$ & $0.865818-0.094051\,i$ \\
& & $0.399346-0.088732\,i$ & $0.640622-0.092484\,i$ & $0.865901-0.094041\,i$ \\
\addlinespace[2.5pt]
& \multirow{2}{*}{$0.2$} & $0.399879-0.089578\,i$ & $0.641210-0.092701\,i$ & $0.866690-0.094201\,i$ \\
& & $0.399868-0.088901\,i$ & $0.641389-0.092657\,i$ & $0.866772-0.094190\,i$ \\
\midrule

\multirow{18}{*}{$0.3$}
& \multirow{2}{*}{$-0.2$} & $0.411079-0.089165\,i$ & $0.658339-0.091804\,i$ & $0.891040-0.093436\,i$ \\
& & $0.410521-0.088028\,i$ & $0.658521-0.091746\,i$ & $0.891127-0.093426\,i$ \\
\addlinespace[2.5pt]
& \multirow{2}{*}{$-0.1$} & $0.411448-0.089273\,i$ & $0.659110-0.091988\,i$ & $0.891955-0.093602\,i$ \\
& & $0.410961-0.088196\,i$ & $0.659292-0.091931\,i$ & $0.892041-0.093591\,i$ \\
\addlinespace[2.5pt]
& \multirow{2}{*}{$-0.01$} & $0.411781-0.089371\,i$ & $0.659804-0.092154\,i$ & $0.892779-0.093750\,i$ \\
& & $0.411357-0.088347\,i$ & $0.659985-0.092098\,i$ & $0.892864-0.093739\,i$ \\
\addlinespace[2.5pt]
& \multirow{2}{*}{$-0.001$} & $0.411815-0.089381\,i$ & $0.659874-0.092170\,i$ & $0.892861-0.093765\,i$ \\
& & $0.411397-0.088362\,i$ & $0.660055-0.092114\,i$ & $0.892946-0.093754\,i$ \\
\addlinespace[2.5pt]
& \multirow{2}{*}{$0$} & $0.411818-0.089382\,i$ & $0.659881-0.092172\,i$ & $0.892870-0.093767\,i$ \\
& & $0.411402-0.088364\,i$ & $0.660063-0.092116\,i$ & $0.892955-0.093756\,i$ \\
\addlinespace[2.5pt]
& \multirow{2}{*}{$0.001$} & $0.411822-0.089383\,i$ & $0.659889-0.092174\,i$ & $0.892879-0.093768\,i$ \\
& & $0.411406-0.088365\,i$ & $0.660070-0.092118\,i$ & $0.892965-0.093757\,i$ \\
\addlinespace[2.5pt]
& \multirow{2}{*}{$0.01$} & $0.411855-0.089393\,i$ & $0.659958-0.092191\,i$ & $0.892962-0.093783\,i$ \\
& & $0.411446-0.088380\,i$ & $0.660140-0.092135\,i$ & $0.893047-0.093772\,i$ \\
\addlinespace[2.5pt]
& \multirow{2}{*}{$0.1$} & $0.412188-0.089490\,i$ & $0.660653-0.092356\,i$ & $0.893785-0.093932\,i$ \\
& & $0.411842-0.088531\,i$ & $0.660833-0.092301\,i$ & $0.893869-0.093920\,i$ \\
\addlinespace[2.5pt]
& \multirow{2}{*}{$0.2$} & $0.412558-0.089599\,i$ & $0.661424-0.092540\,i$ & $0.894700-0.094097\,i$ \\
& & $0.412283-0.088699\,i$ & $0.661604-0.092487\,i$ & $0.894783-0.094085\,i$ \\
\bottomrule
\end{tabular}
\end{table}

\begin{table}[H]
\centering
\caption{Fundamental axial gravitational QNM frequencies $M\omega$ for $\chi=0.4,0.5,0.6$. For each $\hat\qbar$, the upper entry is WKB3 and the lower entry is Leaver.}
\label{tab:qnm_wkb_leaver_chi040506}
\scriptsize
\setlength{\tabcolsep}{10pt}
\renewcommand{\arraystretch}{0.88}
\begin{tabular}{ccccc}
\toprule
$\chi$ & $\hat\qbar$ & $\ell=m=2$ & $\ell=m=3$ & $\ell=m=4$ \\
& & \multicolumn{1}{c}{$M\omega$ (WKB3/Leaver)}
& \multicolumn{1}{c}{$M\omega$ (WKB3/Leaver)}
& \multicolumn{1}{c}{$M\omega$ (WKB3/Leaver)} \\
\midrule

\multirow{18}{*}{$0.4$}
& \multirow{2}{*}{$-0.2$} & $0.424171-0.089253\,i$ & $0.678535-0.091594\,i$ & $0.918878-0.093270\,i$ \\
& & $0.423259-0.087830\,i$ & $0.678719-0.091525\,i$ & $0.918967-0.093260\,i$ \\
\addlinespace[2.5pt]
& \multirow{2}{*}{$-0.1$} & $0.424437-0.089345\,i$ & $0.679311-0.091790\,i$ & $0.919836-0.093451\,i$ \\
& & $0.423619-0.087997\,i$ & $0.679494-0.091723\,i$ & $0.919924-0.093440\,i$ \\
\addlinespace[2.5pt]
& \multirow{2}{*}{$-0.01$} & $0.424677-0.089427\,i$ & $0.680009-0.091967\,i$ & $0.920699-0.093614\,i$ \\
& & $0.423942-0.088147\,i$ & $0.680191-0.091901\,i$ & $0.920785-0.093602\,i$ \\
\addlinespace[2.5pt]
& \multirow{2}{*}{$-0.001$} & $0.424701-0.089436\,i$ & $0.680079-0.091985\,i$ & $0.920785-0.093630\,i$ \\
& & $0.423975-0.088162\,i$ & $0.680261-0.091919\,i$ & $0.920871-0.093618\,i$ \\
\addlinespace[2.5pt]
& \multirow{2}{*}{$0$} & $0.424704-0.089437\,i$ & $0.680087-0.091987\,i$ & $0.920794-0.093632\,i$ \\
& & $0.423978-0.088164\,i$ & $0.680269-0.091921\,i$ & $0.920881-0.093620\,i$ \\
\addlinespace[2.5pt]
& \multirow{2}{*}{$0.001$} & $0.424706-0.089437\,i$ & $0.680094-0.091989\,i$ & $0.920804-0.093634\,i$ \\
& & $0.423982-0.088166\,i$ & $0.680277-0.091923\,i$ & $0.920891-0.093621\,i$ \\
\addlinespace[2.5pt]
& \multirow{2}{*}{$0.01$} & $0.424730-0.089446\,i$ & $0.680164-0.092006\,i$ & $0.920890-0.093650\,i$ \\
& & $0.424014-0.088181\,i$ & $0.680346-0.091940\,i$ & $0.920977-0.093638\,i$ \\
\addlinespace[2.5pt]
& \multirow{2}{*}{$0.1$} & $0.424970-0.089528\,i$ & $0.680862-0.092183\,i$ & $0.921752-0.093813\,i$ \\
& & $0.424338-0.088331\,i$ & $0.681044-0.092119\,i$ & $0.921838-0.093800\,i$ \\
\addlinespace[2.5pt]
& \multirow{2}{*}{$0.2$} & $0.425236-0.089620\,i$ & $0.681638-0.092380\,i$ & $0.922710-0.093994\,i$ \\
& & $0.424697-0.088498\,i$ & $0.681819-0.092317\,i$ & $0.922795-0.093980\,i$ \\
\midrule

\multirow{18}{*}{$0.5$}
& \multirow{2}{*}{$-0.2$} & $0.437263-0.089342\,i$ & $0.698732-0.091383\,i$ & $0.946716-0.093105\,i$ \\
& & $0.435998-0.087633\,i$ & $0.698917-0.091303\,i$ & $0.946807-0.093093\,i$ \\
\addlinespace[2.5pt]
& \multirow{2}{*}{$-0.1$} & $0.437426-0.089417\,i$ & $0.699512-0.091592\,i$ & $0.947717-0.093301\,i$ \\
& & $0.436276-0.087799\,i$ & $0.699696-0.091514\,i$ & $0.947807-0.093288\,i$ \\
\addlinespace[2.5pt]
& \multirow{2}{*}{$-0.01$} & $0.437573-0.089484\,i$ & $0.700214-0.091781\,i$ & $0.948618-0.093478\,i$ \\
& & $0.436527-0.087948\,i$ & $0.700397-0.091704\,i$ & $0.948707-0.093464\,i$ \\
\addlinespace[2.5pt]
& \multirow{2}{*}{$-0.001$} & $0.437587-0.089491\,i$ & $0.700284-0.091799\,i$ & $0.948709-0.093495\,i$ \\
& & $0.436552-0.087963\,i$ & $0.700468-0.091723\,i$ & $0.948797-0.093482\,i$ \\
\addlinespace[2.5pt]
& \multirow{2}{*}{$0$} & $0.437589-0.089491\,i$ & $0.700292-0.091801\,i$ & $0.948719-0.093497\,i$ \\
& & $0.436555-0.087964\,i$ & $0.700475-0.091725\,i$ & $0.948807-0.093484\,i$ \\
\addlinespace[2.5pt]
& \multirow{2}{*}{$0.001$} & $0.437591-0.089492\,i$ & $0.700300-0.091804\,i$ & $0.948729-0.093499\,i$ \\
& & $0.436558-0.087966\,i$ & $0.700483-0.091727\,i$ & $0.948817-0.093486\,i$ \\
\addlinespace[2.5pt]
& \multirow{2}{*}{$0.01$} & $0.437605-0.089499\,i$ & $0.700370-0.091822\,i$ & $0.948819-0.093517\,i$ \\
& & $0.436583-0.087981\,i$ & $0.700553-0.091746\,i$ & $0.948907-0.093503\,i$ \\
\addlinespace[2.5pt]
& \multirow{2}{*}{$0.1$} & $0.437752-0.089566\,i$ & $0.701072-0.092011\,i$ & $0.949720-0.093694\,i$ \\
& & $0.436833-0.088130\,i$ & $0.701255-0.091936\,i$ & $0.949806-0.093679\,i$ \\
\addlinespace[2.5pt]
& \multirow{2}{*}{$0.2$} & $0.437915-0.089641\,i$ & $0.701852-0.092220\,i$ & $0.950721-0.093890\,i$ \\
& & $0.437112-0.088296\,i$ & $0.702034-0.092147\,i$ & $0.950806-0.093875\,i$ \\
\midrule

\multirow{18}{*}{$0.6$}
& \multirow{2}{*}{$-0.2$} & $0.450356-0.089431\,i$ & $0.718928-0.091173\,i$ & $0.974554-0.092939\,i$ \\
& & $0.448736-0.087436\,i$ & $0.719115-0.091082\,i$ & $0.974647-0.092926\,i$ \\
\addlinespace[2.5pt]
& \multirow{2}{*}{$-0.1$} & $0.450415-0.089488\,i$ & $0.719713-0.091394\,i$ & $0.975599-0.093151\,i$ \\
& & $0.448934-0.087600\,i$ & $0.719898-0.091306\,i$ & $0.975690-0.093137\,i$ \\
\addlinespace[2.5pt]
& \multirow{2}{*}{$-0.01$} & $0.450469-0.089540\,i$ & $0.720419-0.091594\,i$ & $0.976538-0.093341\,i$ \\
& & $0.449112-0.087748\,i$ & $0.720603-0.091507\,i$ & $0.976628-0.093326\,i$ \\
\addlinespace[2.5pt]
& \multirow{2}{*}{$-0.001$} & $0.450474-0.089545\,i$ & $0.720489-0.091614\,i$ & $0.976632-0.093360\,i$ \\
& & $0.449129-0.087763\,i$ & $0.720674-0.091527\,i$ & $0.976722-0.093345\,i$ \\
\addlinespace[2.5pt]
& \multirow{2}{*}{$0$} & $0.450475-0.089546\,i$ & $0.720497-0.091616\,i$ & $0.976643-0.093363\,i$ \\
& & $0.449131-0.087765\,i$ & $0.720682-0.091529\,i$ & $0.976732-0.093348\,i$ \\
\addlinespace[2.5pt]
& \multirow{2}{*}{$0.001$} & $0.450475-0.089547\,i$ & $0.720505-0.091618\,i$ & $0.976653-0.093365\,i$ \\
& & $0.449133-0.087766\,i$ & $0.720690-0.091532\,i$ & $0.976743-0.093350\,i$ \\
\addlinespace[2.5pt]
& \multirow{2}{*}{$0.01$} & $0.450480-0.089552\,i$ & $0.720576-0.091638\,i$ & $0.976747-0.093384\,i$ \\
& & $0.449151-0.087781\,i$ & $0.720760-0.091552\,i$ & $0.976837-0.093369\,i$ \\
\addlinespace[2.5pt]
& \multirow{2}{*}{$0.1$} & $0.450534-0.089604\,i$ & $0.721282-0.091838\,i$ & $0.977687-0.093575\,i$ \\
& & $0.449329-0.087929\,i$ & $0.721465-0.091753\,i$ & $0.977775-0.093558\,i$ \\
\addlinespace[2.5pt]
& \multirow{2}{*}{$0.2$} & $0.450593-0.089661\,i$ & $0.722066-0.092059\,i$ & $0.978731-0.093787\,i$ \\
& & $0.449527-0.088094\,i$ & $0.722249-0.091977\,i$ & $0.978817-0.093769\,i$ \\
\bottomrule
\end{tabular}
\end{table}

\begin{figure}[H]
\centering
\includegraphics[width=0.66\linewidth]{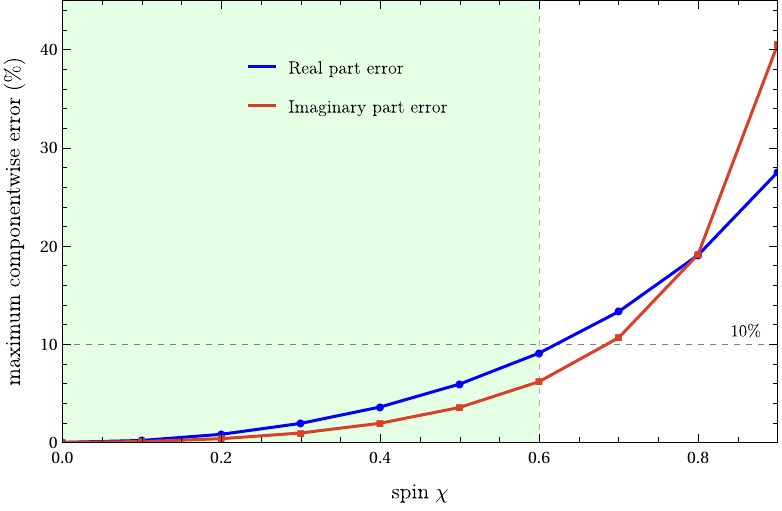}
\caption{Maximum componentwise relative difference between the commutative slow-rotation QNM spectrum obtained from our Leaver calculation and the full Kerr QNM spectrum computed using the \texttt{qnm} package~\cite{Stein:2019mop}. The shaded region indicates the spin range $0\leq\chi\leq0.6$, where the maximum componentwise relative difference remains below $10\%$ on the sampled spin grid.}
\label{fig:slow_rotation_validity}
\end{figure}

At $\chi=\hat\qbar=0$, the Leaver frequencies reproduce the established Schwarzschild spectrum~\cite{Berti:2009kk,Berti:2005ys,Stein:2019mop}. To assess the applicability of the first-order slow-rotation approximation, we compare the commutative Leaver frequencies with the full Kerr QNM spectrum obtained numerically using the \texttt{qnm} package~\cite{Stein:2019mop}, which employs Leaver's continued-fraction method together with a spectral treatment of the angular Teukolsky equation. The comparison is performed for $\ell=2,3,4$ and all allowed azimuthal numbers $m$. Adopting a maximum componentwise relative error of $10\%$ in both the real frequency and damping rate, we find that the slow-rotation approximation remains within this tolerance through $\chi=0.6$ on the sampled spin grid. At $\chi=0.6$, the maximum deviations in the real and imaginary parts are $9.091\%$ and $6.186\%$, respectively, whereas at $\chi=0.7$ they increase to $13.311\%$ and $10.650\%$. We therefore adopt $0\leq\chi\leq0.6$ as the working range under this accuracy criterion. The corresponding comparison is illustrated in Fig.~\ref{fig:slow_rotation_validity}.

The numerical QNM spectrum is evaluated over the parameter ranges
\begin{equation}\label{eq101}
    0\leq\chi\leq0.9,
    \qquad
    -1\leq \hat \qbar\leq1,
\end{equation}
with a step size of $0.01$ in both parameters. Although the numerical analysis covers this wider parameter space, we tabulate the QNM frequencies for the smaller deformation interval
\begin{equation}
    -0.2\leq \hat \qbar\leq0.2,
\end{equation}
with representative spin values up to $\chi=0.6$. The wider parameter range in \eqref{eq101} is used to explore the dependence of the QNM frequencies on spin and $\qbar$-deformation, but does not imply the validity of the perturbative approximation throughout this range. The restriction $\chi\leq0.6$ is motivated by the comparison with the full Kerr spectrum in the commutative limit, using the adopted $10\%$ accuracy criterion. However, no corresponding accuracy estimate is available for the NC corrections, including the mixed $\chi\hat\qbar$ contribution. Therefore, the chosen deformation interval $|\hat\qbar|\leq0.2$ should be regarded as illustrative rather than a rigorously established domain of perturbative validity.

\begin{figure}[H]
\centering
\includegraphics[width=0.8\textwidth]{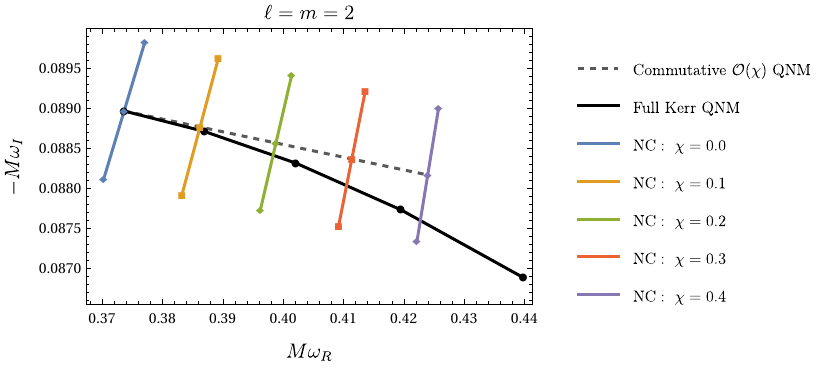}
\caption{Trajectories of the fundamental $(\ell,m,n)=(2,2,0)$ mode in the complex-frequency plane parametrized by $\chi$.
The dashed curve represents the commutative first-order slow-rotation Kerr spectrum, while the black solid curve corresponds to the full Kerr QNM frequencies obtained using the \texttt{qnm} package~\cite{Stein:2019mop}. Both commutative curves are evaluated at $\hat\qbar=0$ by varying the spin parameter over $0\leq\chi\leq0.4$. The colored lines represent the $\qbar$uantum-induced frequency shifts obtained by varying $-0.5\leq\hat\qbar\leq0.5$ at fixed $\chi=0,0.1,0.2,0.3$, and $0.4$. Each colored trajectory intersects the commutative slow-rotation curve at $\hat\qbar=0$. The separation between the dashed and solid curves illustrates the truncation error of the first-order slow-rotation approximation. As $\chi$ increases, this error becomes more significant and can mask the comparatively small $\qbar$uantum corrections. This motivates the Kerr-anchored prescription adopted in Sec.~\ref{sec:ringdown_phenomenology}, in which the perturbative $\qbar$uantum corrections are added to the full Kerr frequencies, effectively shifting the colored trajectories from the slow-rotation baseline to the full Kerr baseline at each fixed spin. The corresponding results for $\ell=m=3,4$ are presented in Appendix~\ref{app:higher_multipole_qnms}.}
\label{fig:complex_frequency_paths}
\end{figure}

Figure~\ref{fig:complex_frequency_paths} illustrates the effects of rotation and $\qbar$-deformation on the fundamental $(\ell,m, n)=(2,2, 0)$ mode in the complex-frequency plane. The figure contains two sets of trajectories corresponding to variations of different parameters. The dashed curve and black solid line represent the commutative QNM frequencies ($\hat\qbar=0$) obtained by varying the spin parameter $\chi$, using the first-order slow-rotation approximation and the full Kerr spectrum, respectively. In contrast, each colored line shows the $\qbar$uantum-induced displacement of the QNM frequency as $\hat\qbar$ varies at a fixed value of $\chi$. Consequently, the colored trajectories intersect the dashed slow-rotation curve at $\hat\qbar=0$. Their displacement from this curve represents the leading $\qbar$uantum correction, while the separation between the dashed curve and the corresponding full Kerr frequencies illustrates the truncation error of the first-order slow-rotation approximation. The figure therefore provides a complementary visualization of the effects of $\qbar$-deformation and the limitations of the slow-rotation expansion. The corresponding results for $\ell=m=3,4$ are presented in Appendix~\ref{app:higher_multipole_qnms}.

\begin{figure}[H]
    \centering
    \includegraphics[width=0.45\textwidth]{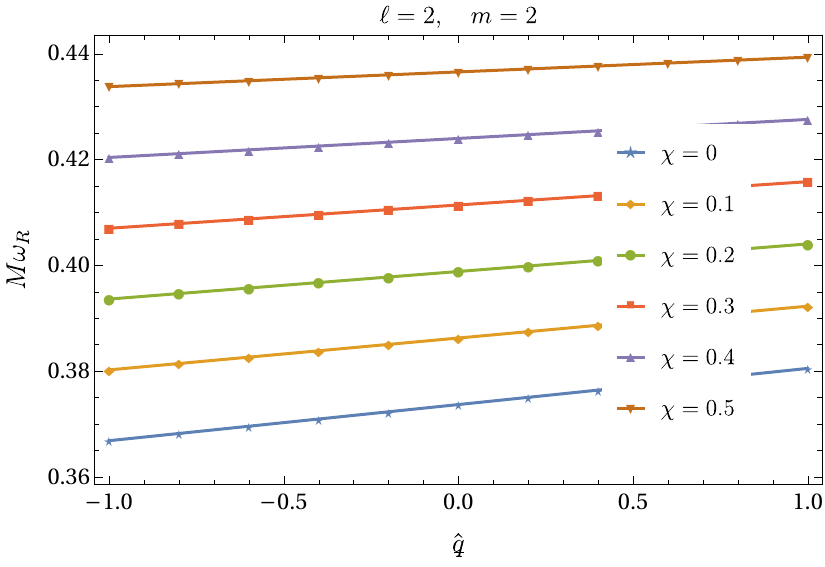}
    \includegraphics[width=0.45\textwidth]{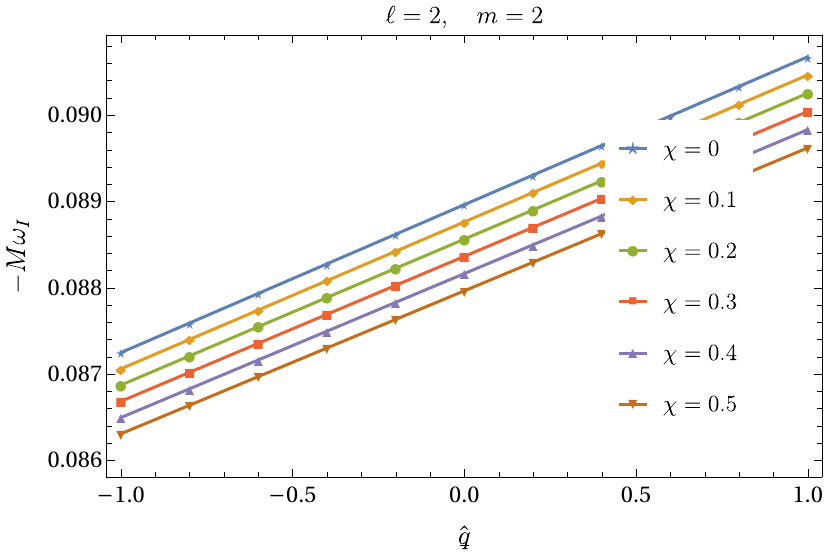}

    \caption{Dependence of the real (left) and imaginary (right)
    parts of the fundamental $(\ell,m, n)=(2,2, 0)$ QNM frequency
    on the deformation parameter $\hat\qbar$ for
    $\chi=0,0.1,0.2,0.3,0.4,$ and $0.5$.
    The plotted spins lie within the adopted $10\%$
    slow-rotation accuracy range $\chi\leq0.6$.
    The corresponding results for $\ell=m=3,4$ are
    presented in Appendix~\ref{app:higher_multipole_qnms}. }
    \label{fig:qnm_nc_parameter_dependence}
\end{figure}

\begin{figure}[H]
    \centering
    \includegraphics[width=0.45\textwidth]{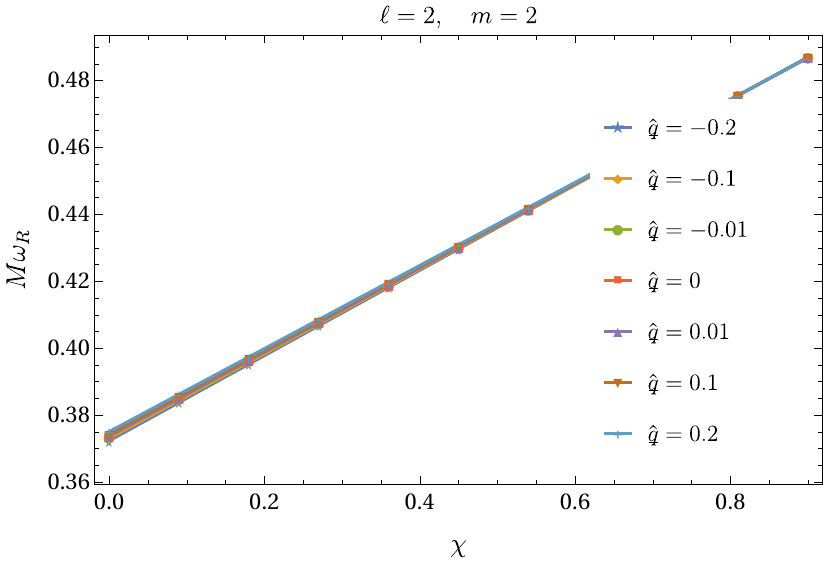}
    \includegraphics[width=0.45\textwidth]{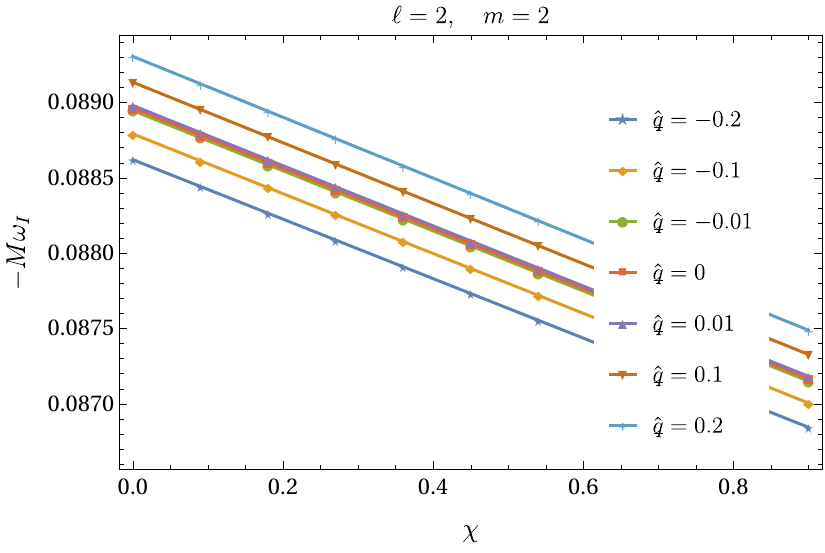}

    \caption{Dependence of the real (left) and imaginary (right)
    parts of the fundamental $(\ell,m, n)=(2,2, 0)$ QNM frequency
    on the dimensionless spin $\chi$ for
    $\hat\qbar=-0.2,-0.1,-0.01,0,0.01,0.1,$ and $0.2$.
    The corresponding results for $\ell=m=3,4$ are
    presented in Appendix~\ref{app:higher_multipole_qnms}. }
    \label{fig:qnm_spin_parameter_dependence}
\end{figure}

Figures~\ref{fig:qnm_nc_parameter_dependence} and \ref{fig:qnm_spin_parameter_dependence} show the dependence of the real and imaginary parts of the fundamental $(\ell,m,n)=(2,2,0)$ QNM frequency on the deformation parameter $\hat\qbar$ and the black hole spin. The variations with either parameter follow the spin- and deformation-dependent slopes in Eqs.~\eqref{eq:nc_deformation_slope} and \eqref{eq:nc_spin_slope}. The present figures include $\chi$ values up to $0.5$, within the working range $\chi\leq0.6$ defined by the adopted $10\%$ criterion. The corresponding results for the higher multipoles $\ell=m=3,4$ are presented in Appendix~\ref{app:higher_multipole_qnms}.

As shown in Fig.~\ref{fig:qnm_nc_parameter_dependence}, for fixed $\chi$, the absolute value of the imaginary part $|\omega_I|$ decreases for negative $\hat\qbar$ and increases for positive $\hat\qbar$. Consequently, negative (positive) deformation leads to weaker (stronger) damping and a longer (shorter) QNM decay time, $\tau_{\rm QNM}=1/|\omega_I|$. While this behaviour characterizes the decay of black hole perturbations rather than the black hole evaporation lifetime, its possible connection to greybody factors and Hawking radiation, particularly for primordial black holes, merits
further investigation.

\section{Ringdown phenomenology}
\label{sec:ringdown_phenomenology}
Black hole ringdown observations provide an opportunity to test the predictions of General Relativity by comparing the measured QNM frequencies and damping times with those predicted for Kerr black holes. Recent studies have investigated the contributions of higher angular modes and overtones~\cite{Carullo:2019flw,Isi:2019aib,Capano:2021etf,Cotesta:2022pci,Siegel:2023lxl,Gennari:2023gmx}, as well as possible deviations from the Kerr QNM spectrum~\cite{Brito:2018rfr,Ghosh:2021mrv,Maggio:2022hre}. More recent gravitational-wave observations have further strengthened these tests ~\cite{LIGOScientific:2026wpt,LIGOScientific:2026qqv, LIGOScientific:2026oim}. In this context, theoretical predictions of QNM corrections beyond classical GR can be related to observable shifts in the ringdown frequency and damping time. We explore this connection for the $\qbar$-deformed axial QNM spectrum obtained in the present work.

In this section, we provide a heuristic discussion of ringdown phenomenology. We estimate the $\qbar$uantum corrections for 11 events with median remnant spins $\chi_f\leq0.6$, selected from GWTC-3 (see O3b, Table IV in \cite{KAGRA:2021vkt}) and GWTC-5.0 (see O4b, Table 3 in \cite{LIGOScientific:2026wfs}). We focus on the fundamental gravitational QNM
\[
(\ell,m,n)=(2,2,0),
\]
and for each event we use the detector frame
remnant mass
\begin{equation}
M_f^{\rm det}=(1+z)M_f
\end{equation}
and the corresponding dimensionless final spin $\chi_f$. The selection is based on the catalog median spins; the 90\% credible intervals need not lie entirely below $\chi_f=0.6$.

Before discussing the $\qbar$uantum correction, it is important to assess the accuracy of the first-order slow-rotation approximation at the spins of the observed remnants. At first order in the dimensionless spin $\chi$, the real and imaginary parts of the QNM frequency take the standard form\footnote{The SR superscript stands for "slow-rotating approximation."}
\cite{Pani:2013wsa}
\begin{align}
M\omega_R^{\rm SR}
&=
M\omega_R^{(0)}
+
m\chi\,M\omega_R^{(1)}
+
\mathcal O(\chi^2),
\\
M\omega_I^{\rm SR}
&=
M\omega_I^{(0)}
+
m\chi\,M\omega_I^{(1)}
+
\mathcal O(\chi^2).
\label{eq:slow_rotation_qnm_pheno}
\end{align}
For the $(2,2,0)$ mode, the coefficients obtained from the continued-fraction calculation (Table~\ref{tab:strict_nc_coefficients}) are
\begin{align}
M\omega_R^{(0)} &= 0.373672,
&
M\omega_I^{(0)} &= -0.088962,
\\
M\omega_R^{(1)} &= 0.062883,
&
M\omega_I^{(1)} &= 0.000998.
\end{align}

For the commutative Kerr reference, we evaluate the gravitational $(\ell,m,n)=(2,2,0)$ QNM frequencies
numerically using the \texttt{qnm} package of Stein~\cite{Stein:2019mop}. The package implements a continued-fraction treatment of the radial Teukolsky equation together with a spectral treatment of the angular equation. We evaluate the dimensionless complex frequency $M\omega_{220}^{\rm Kerr}(\chi)$ at each required spin. These numerical Kerr frequencies serve as the reference for the slow-rotation comparison and the Kerr-anchored noncommutative phenomenology.

We quantify the percentage deviations of the slow-rotation approximation from the full Kerr results, separately for the oscillation frequency and damping rate, as
\begin{align}
\epsilon_R^{\rm SR}
&=
100\,
\frac{
\omega_R^{\rm SR}-\omega_R^{\rm Kerr}
}{
\omega_R^{\rm Kerr}
},
\\
\epsilon_I^{\rm SR}
&=
100\,
\frac{
\omega_I^{\rm SR}-\omega_I^{\rm Kerr}
}{
\omega_I^{\rm Kerr}
}.
\label{eq:slow_rotation_errors}
\end{align}

The 11 selected remnants have median spins in the range
\[
0.14\leq\chi_f\leq0.58.
\]
\begin{table}[H]
\centering
\caption{
Validity of the first-order slow-rotation approximation for the $(\ell,m,n)=(2,2,0)$ mode at the median final spins of the 11 selected events from GWTC-3 (O3b)~\cite{KAGRA:2021vkt} and GWTC-5.0 (O4b)~\cite{LIGOScientific:2026wfs}. The commutative Kerr reference is obtained numerically using the \texttt{qnm} package~\cite{Stein:2019mop}.
}
\label{tab:sr_validity}

\setlength{\tabcolsep}{10pt}

\begin{tabular}{lccccc}
\hline
Event
& $\chi_f$
& $M\omega^{\rm Kerr}$
& $M\omega^{\rm SR}$
& $\epsilon_R^{\rm SR}[\%]$
& $\epsilon_I^{\rm SR}[\%]$
\\
\hline
\multicolumn{6}{l}{\textbf{GWTC-3: O3b}} \\
\hline
GW191219\_163120 & 0.14 & $0.392832-0.088567i$ & $0.391279-0.088683i$ & $-0.40$ & $+0.13$ \\
GW200210\_092254 & 0.34 & $0.427253-0.087428i$ & $0.416432-0.088284i$ & $-2.53$ & $+0.98$ \\
GW200115\_042309 & 0.42 & $0.444339-0.086671i$ & $0.426493-0.088124i$ & $-4.02$ & $+1.68$ \\
GW200105\_162426 & 0.43 & $0.446650-0.086559i$ & $0.427751-0.088104i$ & $-4.23$ & $+1.79$ \\
GW191113\_071753 & 0.45 & $0.451402-0.086321i$ & $0.430266-0.088064i$ & $-4.68$ & $+2.02$ \\

\hline
\multicolumn{6}{l}{\textbf{GWTC-5.0: O4b}} \\
\hline
GW240921\_201835 & 0.52 & $0.469586-0.085325i$ & $0.439070-0.087924i$ & $-6.50$ & $+3.05$ \\
GW241110\_124123 & 0.52 & $0.469586-0.085325i$ & $0.439070-0.087924i$ & $-6.50$ & $+3.05$ \\
GW240507\_041632 & 0.55 & $0.478235-0.084802i$ & $0.442843-0.087865i$ & $-7.40$ & $+3.61$ \\
GW241127\_061008 & 0.56 & $0.481250-0.084613i$ & $0.444101-0.087845i$ & $-7.72$ & $+3.82$ \\
GW240527\_230910 & 0.58 & $0.487494-0.084208i$ & $0.446616-0.087805i$ & $-8.39$ & $+4.27$ \\
GW240930\_035959 & 0.58 & $0.487494-0.084208i$ & $0.446616-0.087805i$ & $-8.39$ & $+4.27$ \\
\hline
\end{tabular}
\end{table}
As shown in Table~\ref{tab:sr_validity}, the relative error in the real part of the QNM frequency, calculated with respect to the numerical Kerr spectrum, ranges from $-0.40\%$ at $\chi_f=0.14$ to $-8.39\%$ at $\chi_f=0.58$. The corresponding relative error in the imaginary part ranges from $+0.13\%$ to $+4.27\%$. For all 11 selected events, the absolute values of both relative errors remain below the adopted $10\%$ threshold (see  Figure \ref{fig:slow_rotation_validity} also). The deviations increase with spin, particularly in the real part, motivating the Kerr-anchored prescription discussed below. This comparison, however, does not establish the accuracy of the noncommutative corrections beyond first order in spin.

\subsection{Kerr-anchored $\qbar$uantum correction}

To the perturbative order considered in this work, the real and imaginary parts of the $\qbar$-deformed QNM frequency may be written as
\begin{align}
M\omega_R^{\hat\qbar,SR}
&=
M\omega_R^{(0)}
+
m\chi\,M\omega_R^{(1)}
+
\hat \qbar\,M\omega_R^{(\hat\qbar)}
+
m\chi\hat \qbar\,M\omega_R^{(\hat\qbar \chi)},
\\
M\omega_I^{\hat\qbar,SR}
&=
M\omega_I^{(0)}
+
m\chi\,M\omega_I^{(1)}
+
\hat \qbar\,M\omega_I^{(\hat\qbar)}
+
m\chi\hat \qbar\,M\omega_I^{(\hat\qbar\chi)}.
\label{eq:nc_qnm_expansion_pheno}
\end{align}
For the fundamental $(2,2,0)$ mode, the $\qbar$-deformed coefficients obtained in Table~\ref{tab:strict_nc_coefficients} are
\begin{align}
M\omega_R^{(\hat\qbar)}
&=
0.006834,
&
M\omega_I^{(\hat\qbar)}
&=
-0.001714,
\\
M\omega_R^{(\hat\qbar\chi)}
&=
-0.004048,
&
M\omega_I^{(\hat\qbar\chi)}
&=
0.000057.
\end{align}

The genuine $\qbar$uantum contribution is isolated by subtracting the commutative slow-rotation result at the same spin,
\begin{equation}
\Delta\omega^{\hat\qbar}
=
\omega^{\hat\qbar,SR}
-
\omega^{\rm SR}.
\label{eq:delta_omega_nc}
\end{equation}
In terms of its real and imaginary parts,
\begin{align}
M\Delta\omega_{R}^{\hat\qbar}
&=
\hat \qbar
\left[
M\omega_R^{(\hat\qbar)}
+
m\chi\,M\omega_R^{(\hat\qbar\chi)}
\right],\label{eq:delta_omega_nc_components1}
\\
M\Delta\omega_{I}^{\hat\qbar}
&=
\hat \qbar
\left[
M\omega_I^{(\hat\qbar)}
+
m\chi\,M\omega_I^{(\hat\qbar\chi)}
\right].
\label{eq:delta_omega_nc_components}
\end{align}

The sizeable commutative slow-rotation error displayed in Table~\ref{tab:sr_validity} would otherwise contaminate the phenomenological $\qbar$uantum shift. We therefore use a Kerr-anchored frequency $\omega^\text{\faAnchor}$,
\begin{equation}
\omega_{R,I}^\text{\faAnchor}(\chi,\hat \qbar)
=
\omega_{R,I}^{\rm Kerr}(\chi)
+
\Delta\omega_{R,I}^{\hat\qbar}(\chi,\hat \qbar)
.
\label{eq:kerr_anchored_frequency}
\end{equation}
This construction is a phenomenological matching prescription rather than a derivation of the complete $\qbar$uantum Kerr spectrum: the commutative part is supplied by the numerical Kerr QNM frequency, while the $\qbar$uantum contribution is retained only through the orders calculated by our Leaver method. The prescription therefore reproduces the numerical full-Kerr frequency at $\hat\qbar=0$, but does not determine the higher-order mixed corrections beginning at $\mathcal O(\hat\qbar\chi^2)$.

\begin{figure}[H]
    \centering
    \includegraphics[width=0.9\textwidth]{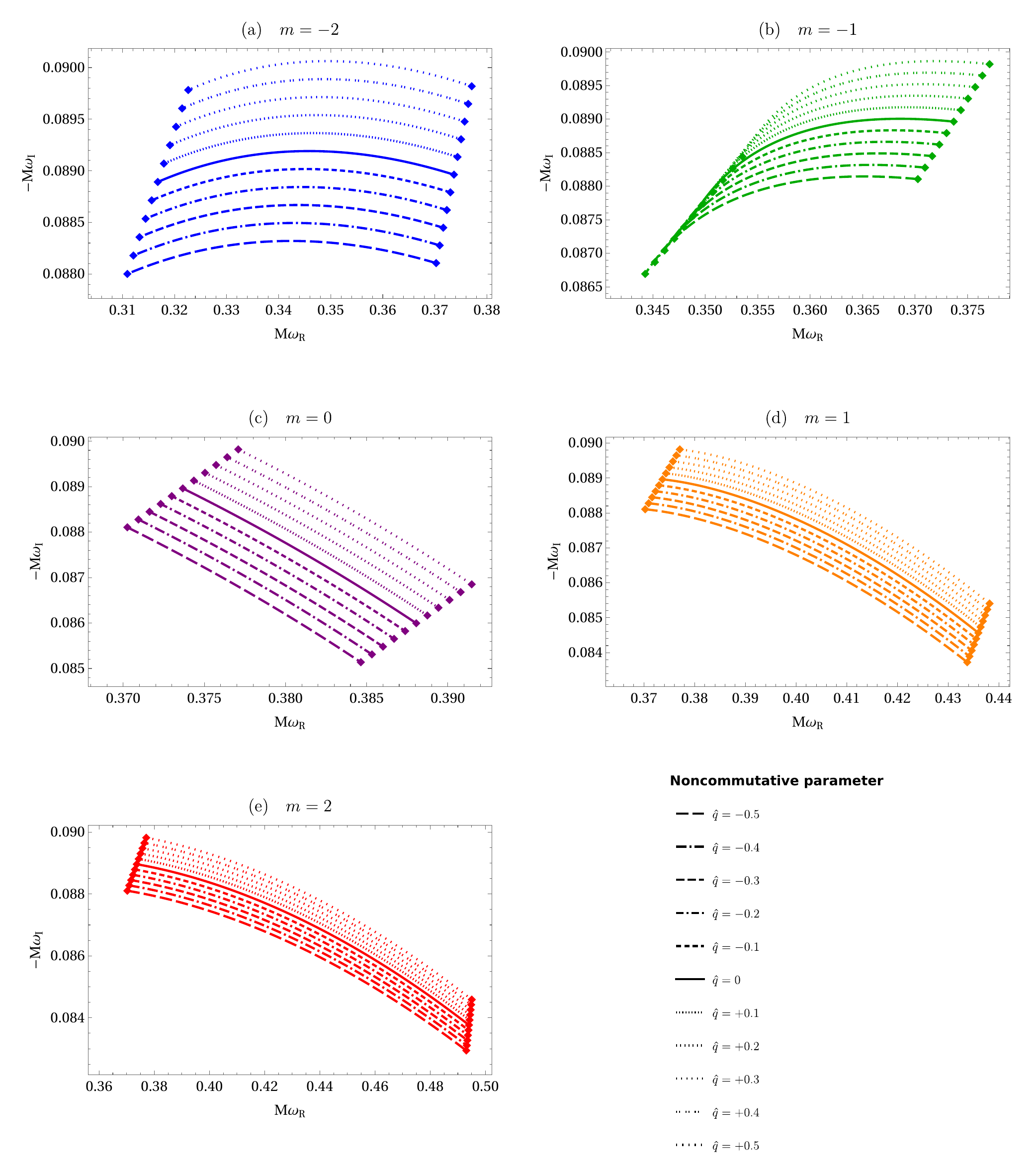}
   \caption{ Kerr-anchored fundamental ($n=0$) QNM trajectories in the complex-frequency plane for $\ell=2$, shown separately for (a) $m=-2$, (b) $m=-1$, (c) $m=0$, (d) $m=1$, and (e) $m=2$. Each curve corresponds to a fixed value of the $\qbar$uantum parameter $\hat{\qbar}\in \{-0.5,-0.4,-0.3,-0.2,-0.1,0,0.1,0.2,0.3,0.4,0.5\}$, with the dimensionless spin varying over $0\leq\chi\leq0.6$. The solid curves represent the full Kerr results ($\hat{\qbar}=0$), whereas dashed and dotted curves correspond to negative and positive values of $\hat{\qbar}$, respectively. Different line patterns distinguish the individual values of $\hat{\qbar}$, as indicated in the common legend. Diamonds mark the endpoints $\chi=0$ and $\chi=0.6$ of each trajectory.  }
    \label{fig:kerr_anchored_fixed_q}
\end{figure}

Equivalently, at the level of the complex frequency\footnote{The Kerr-anchored QNM frequencies are indicated by \faAnchor.},
\begin{equation}
\omega^\text{\faAnchor}
=
\omega^{\rm Kerr}
+
\left(
\omega^{\hat\qbar,SR}-\omega^{\rm SR}
\right).
\end{equation}
This prescription restores the full known spin dependence of the commutative Kerr sector while leaving the calculated $\qbar$uantum correction unchanged. In particular,
\begin{equation}
\omega^\text{\faAnchor}(\chi,0)
=
\omega^{\rm Kerr}(\chi).
\end{equation}
Figure~\ref{fig:kerr_anchored_fixed_q} illustrates the spin dependence of the Kerr-anchored QNM frequencies for different values of the $\qbar$uantum deformation parameter. The trajectories coincide with the full Kerr spectrum at $\hat\qbar=0$ and exhibit systematic shifts for nonzero $\hat\qbar$. For $m=0$, the $\qbar$uantum correction is independent of spin, resulting in a constant displacement of the Kerr trajectory in the complex-frequency plane. For $m\neq0$, the mixed spin-$\qbar$uantum contribution introduces an additional spin dependence into this displacement.

\subsection{Frequency and damping-time shifts}

For a detector-frame remnant mass $M_f^{\rm det}$, the dimensionless QNM frequency is converted to the observed oscillation frequency and damping time according to~\footnote{We use the standard QNM convention $f_{\ell mn}=\omega_{R,\ell mn}/(2\pi)$ and $\tau_{\ell mn}=1/|\omega_{I,\ell mn}|$~\cite{Berti:2005ys,Berti:2009kk}.
}
\begin{align}
f_{220}
&=
\frac{M\omega_R}
{2\pi M_f^{\rm det}t_\odot},
\\
\tau_{220}
&=
\frac{M_f^{\rm det}t_\odot}
{|M\omega_I|},
\label{eq:ringdown_physical_units}
\end{align}
where
\begin{equation}
t_\odot=\frac{GM_\odot}{c^3}
\simeq4.92549\times10^{-6}\ {\rm s}.
\end{equation}
We quantify the effect of the $\qbar$uantum correction relative to Kerr through
\begin{equation}
\delta f_{220}^{\hat\qbar}
=
\frac{
f_{220}^\text{\faAnchor}-f_{220}^{\rm Kerr}
}{
f_{220}^{\rm Kerr}
},
\label{eq:delta_f_nc}
\end{equation}
and
\begin{equation}
\delta \tau_{220}^{\hat\qbar}
=
\frac{
\tau_{220}^\text{\faAnchor}-\tau_{220}^{\rm Kerr}
}{
\tau_{220}^{\rm Kerr}
}.
\label{eq:delta_tau_nc}
\end{equation}
Using Eq.~\eqref{eq:kerr_anchored_frequency}, these may also be written as
\begin{equation}
\delta f_{220}^{\hat\qbar}
=
\frac{
\Delta\omega_R^{\hat\qbar}
}{
\omega_R^{\rm Kerr}
},
\label{eq:delta_f_nc_qnm}
\end{equation}
and
\begin{equation}
\delta \tau_{220}^{\hat\qbar}
=
\frac{
|\omega_I^{\rm Kerr}|
}{
|\omega_I^{\rm Kerr}
+\Delta\omega_I^{\hat\qbar}|
}
-1.
\label{eq:delta_tau_nc_qnm}
\end{equation}

For illustration we consider
\begin{equation}
\hat\qbar\in
\{-0.2,-0.1,-0.01,-0.001,
-10^{-4},-10^{-5},0,
10^{-5},10^{-4},0.001,0.01,0.1,0.2\}.
\end{equation}
For each of the 11 selected events, we use the median remnant mass and dimensionless final spin $\chi_f$ reported in the GWTC-3 (O3b)~\cite{KAGRA:2021vkt} and GWTC-5.0 (O4b)~\cite{LIGOScientific:2026wfs} catalogs. At each catalog median spin $\chi_f$ we calculate the reference Kerr frequency $M\omega_{220}^{\rm Kerr}(\chi_f)$ numerically using the \texttt{qnm} package~\cite{Stein:2019mop}, with $(\ell,m,n)=(2,2,0)$. The NC frequency corrections are obtained from the Leaver coefficients in Table~\ref{tab:strict_nc_coefficients} using \eqref{eq:delta_omega_nc_components1}-\eqref{eq:delta_omega_nc_components} and incorporated through the Kerr-anchored prescription in Eq.~\eqref{eq:kerr_anchored_frequency}. Assigning the same value of $\hat\qbar$ to each event, we calculate the fractional frequency and damping-time deviations using Eqs.~\eqref{eq:delta_f_nc_qnm} and \eqref{eq:delta_tau_nc_qnm}, respectively. The remnant mass sets the conversion to physical frequency and damping time; the fractional deviations at fixed $\hat\qbar$ depend on the remnant spin but not explicitly on the mass.

For $|\hat \qbar|=0.2$, the 11 events give approximately
\begin{equation}
8.77\times10^{-4}
\lesssim
|\delta f_{220}^{\hat\qbar}|
\lesssim
2.90\times10^{-3},
\end{equation}
and
\begin{equation}
3.82\times10^{-3}
\lesssim
|\delta \tau_{220}^{\hat\qbar}|
\lesssim
3.93\times10^{-3}.
\end{equation}
Thus, within the parameter range considered here, the fractional shift of the oscillation frequency remains below approximately $0.3\%$, whereas the damping-time modification is of order $0.4\%$. At fixed $\hat \qbar$, the fractional deviations depend on the median remnant spin, but not explicitly on the remnant mass. The absolute values in the above inequalities characterize the magnitudes of the frequency and damping-time deviations across the 11 selected events for $\hat\qbar=\pm0.2$. In  Table VI we present the full intervals for deviations of the frequencies and damping times for different values of NC deformation parameter $\hat{\qbar}$ across all 11 events. The intervals in Table VI are determined by the minimal and maximal absolute value of the deviation. We also note that negative values of $\hat\qbar$ decrease the oscillation frequency and increase the damping time, whereas positive values produce the opposite behaviour.
\begin{table}[H]
\centering 
\renewcommand{\arraystretch}{1.2}
\setlength{\tabcolsep}{15pt}
\begin{tabular}{cll} 
\hline
$\hat\qbar$ & $|\delta f_{220}^{\hat\qbar}|$ & $|\delta\tau_{220}^{\hat\qbar}|$ \\
\hline
$-0.2$ & $[0.8773,\,2.902]\times10^{-3}$ & $[3.846,\,3.928]\times10^{-3}$ \\
$-0.1$ & $[0.4386,\,1.451]\times10^{-3}$ & $[1.919,\,1.960]\times10^{-3}$ \\
$-10^{-p}$ & $[0.4386,\,1.451]\times10^{-(p+2)}$ & $\simeq[1.9155,\,1.9562]\times10^{-(p+2)}$ \\
$0$ & $0$ & $0$ \\
$10^{-p}$ & $[0.4386,\,1.451]\times10^{-(p+2)}$ & $\simeq[1.9155,\,1.9562]\times10^{-(p+2)}$ \\
$0.1$ & $[0.4386,\,1.451]\times10^{-3}$ & $[1.912,\,1.952]\times10^{-3}$ \\
$0.2$ & $[0.8773,\,2.902]\times10^{-3}$ & $[3.816,\,3.897]\times10^{-3}$ \\
\hline 
\end{tabular}
\caption{Range of the absolute Kerr-anchored $\qbar$uantum fractional deviations in the fundamental $(\ell,m,n)=(2,2,0)$ ringdown frequency and damping time across the 11 selected GWTC-3 (O3b) and GWTC-5.0 (O4b) events. The commutative Kerr QNM frequencies are obtained numerically using the \texttt{qnm} package~\cite{Stein:2019mop}, while the NC frequency corrections are calculated from the Leaver coefficients, including the mixed spin-NC contribution. For each fixed deformation parameter, the intervals represent the minimum and maximum absolute deviations over the 11 catalog median spins; they are not statistical confidence intervals. The absolute values quantify the magnitudes of the frequency and damping-time deviations relative to the corresponding Kerr predictions, irrespective of their signs. For the compact entries $\hat\qbar=\pm10^{-p}$, with $p\in\{2,3,4,5\}$, the damping-time deviations are given to leading order in the deformation parameter.}
\end{table}

To place these theoretical predictions in an observational context, we compare the estimated frequency and damping-time shifts with the uncertainties reported in recent LVK ringdown analyses~\cite{LIGOScientific:2026oim}. For the deformation parameters considered here, the predicted shifts are smaller than the current observational uncertainties. Our estimates use the catalog median remnant parameters and are intended to illustrate the magnitude of the NC corrections rather than to establish observational constraints on the deformation parameter.

An alternative perspective on the phenomenological implications of the present results is provided by the ratios of QNM oscillation frequencies and damping times. As recently discussed in Ref.~\cite{Franchini:2025csk}, these ratios are independent of the black hole mass in classical Kerr spacetime and exhibit a relatively weak dependence on spin for several relevant mode combinations. In the present $\qbar$-deformed framework, however, the dimensionless QNM spectrum depends additionally on $\hat\qbar=\qbar/M^2$. Consequently, if the dimensionful deformation parameter $\qbar$ is assumed to be universal, the ratios generally acquire a residual dependence on the black hole mass through $\qbar/M^2$. Conversely, when $\hat\qbar$ is held fixed, the mass dependence cancels, although the ratios can still deviate from their GR predictions. This distinction provides a possible avenue for investigating the phenomenological signatures of spacetime noncommutativity through multimode black hole spectroscopy.

\subsection{Illustrative comparison with GW250114}
\label{sec:gw250114_comparison}
GW250114 is the loudest gravitational-wave event detected to date, with a network signal-to-noise ratio of approximately 80, providing an opportunity for precision tests of black hole ringdown~\cite{Kumar:2026yhu}. As an illustrative comparison with observations, we consider
the central fractional deviations reported for this event. Using the $\qbar$uantum corrections to the axial QNM frequencies obtained from our Leaver calculation, together with the Kerr-anchored prescription of Eq.~\eqref{eq:kerr_anchored_frequency}, we relate the observational deviations to the deformation parameter $\hat\qbar$.

For the fundamental $(2,2,0)$ mode, the $\qbar$uantum frequency correction in Eq.~\eqref{eq:delta_omega_nc_components} takes the form
\begin{equation}
M\Delta\omega_{220}^{\hat\qbar}
=
\hat\qbar\left(B_2+2\chi C_2\right)
\equiv
\hat\qbar\left[D_R(\chi)+iD_I(\chi)\right],
\end{equation}
where $B_2$ and $C_2$ are the complex Leaver coefficients listed in Table~\ref{tab:strict_nc_coefficients}. Using Eq.~\eqref{eq:delta_f_nc_qnm}, the fractional oscillation-frequency deviation becomes
\begin{equation}
\delta f_{220}^{\hat\qbar}
=
\hat\qbar K_f(\chi),
\qquad
K_f(\chi)
=
\frac{D_R(\chi)}
{M\omega_R^{\rm Kerr}(\chi)}.
\end{equation}
Similarly, expanding Eq.~\eqref{eq:delta_tau_nc_qnm} to first order in $\hat\qbar$, the fractional damping-time deviation is
\begin{equation}
\delta \tau_{220}^{\hat\qbar}
=
\hat\qbar K_\tau(\chi)
+\mathcal O(\hat\qbar^2),
\qquad
K_\tau(\chi)
=
-\frac{D_I(\chi)}
{M\omega_I^{\rm Kerr}(\chi)}.
\end{equation}
Using the numerical Kerr reference from the \texttt{qnm} package~\cite{Stein:2019mop}, which gives $M\omega_{220}^{\rm Kerr}(0.68)=0.523975-0.081513\,i$, we obtain
\begin{equation}
\delta f_{220}^{\hat\qbar}
\simeq 0.002536\,\hat\qbar,
\qquad
\delta \tau_{220}^{\hat\qbar}
\simeq -0.020068\,\hat\qbar.
\end{equation}
The corresponding dimensional deformation parameter can then be obtained from the observed fractional deviations as
\begin{equation}
\begin{aligned}
\qbar_f
&=
\frac{\delta f_{220}^{\rm obs}}{K_f(\chi_f)}
\left(\frac{GM_f^{\rm det}}{c^2}\right)^2,\\
\qbar_\tau
&=
\frac{\delta\tau_{220}^{\rm obs}}{K_\tau(\chi_f)}
\left(\frac{GM_f^{\rm det}}{c^2}\right)^2.
\end{aligned}
\end{equation}
The reported central deviations $\delta f_{220}=0.004$ and $\delta\tau_{220}=0.010$, together with the detector-frame remnant mass $M_f^{\rm det}=68.1M_\odot$, then formally correspond to
\begin{equation}
\begin{aligned}
\qbar_f
&\simeq 1.595\times10^{10}\,\mathrm{m}^2,\\
\qbar_\tau
&\simeq -0.504\times10^{10}\,\mathrm{m}^2.
\end{aligned}
\end{equation}
The frequency and damping-time deviations lead to different estimates of the deformation parameter, including opposite signs for their central values. These estimates should therefore be interpreted separately and do not constitute a joint observational determination of the NC parameter.  Note that the characteristic noncommutative length scale is $\ell_{\rm NC}=\sqrt{|\qbar|}$. The separate mappings of the central frequency and damping-time deviations correspond to $\ell_{\rm NC}^{(f)}\simeq126.3\,\mathrm{km}$ and $\ell_{\rm NC}^{(\tau)}\simeq71.0\,\mathrm{km}$, respectively, using the detector-frame remnant mass. For comparison, the Schwarzschild-radius scale associated with the detector-frame mass is $2GM_f^{\rm det}/c^2\simeq201\,\mathrm{km}$; this is not the Kerr outer-horizon radius. These values are illustrative rather than observational bounds on the deformation scale. In particular, the inferred dimensionless deformation parameters are not necessarily within the perturbative regime, and the remnant spin $\chi_f\simeq0.68$ exceeds the adopted slow-rotation validity range. A reliable constraint on $\qbar$ would therefore require extending the perturbative calculation and performing a dedicated statistical analysis.  Similar illustrative mappings of phenomenological QNM frequency deviations to modified-gravity coupling scales have been considered by the LVK Collaboration for Einstein-dilaton-Gauss-Bonnet and dynamical Chern-Simons gravity (see Table 10 in \cite{LIGOScientific:2026oim} where the corresponding couplings are determined as $\sqrt{\alpha_{EdGB}}=14.3$km and $\sqrt{\alpha_{dCS}}=17$km).

\section{Outlook and Discussion}
\label{sec:discussion}

In this work, we have investigated $\qbar$uantum-spacetime corrections to the axial gravitational quasinormal modes of Kerr black holes. To the best of our knowledge, this is the first systematic investigation of noncommutative Kerr black hole ringdown within the Drinfeld-twist formulation, based on metric perturbations and their quasinormal modes. Using the framework of noncommutative differential geometry and employing a pseudo-semi-Killing Abelian twist generated by $X_1=\partial_t$ and $X_2=\partial_r$, together with the $\qbar$-deformed Einstein-manifold condition, we derived the axial gravitational perturbation equations while preserving the classical Kerr geometry as an exact background solution. In the slow-rotation approximation, we retained terms through first order in the dimensionless spin $\chi$ and deformation parameter $\hat\qbar$, including their mixed contribution $\mathcal O(\chi\hat\qbar)$.

The resulting radial perturbation equation was brought into Schr\"odinger form through a field redefinition and a generalized frequency-dependent tortoise coordinate. The effective potential takes the form $V_{\rm eff}=V_{\rm axial}+\hat\qbar M^2V_{\qbar}$, where the interplay between rotation and noncommutativity introduces additional frequency-dependent contributions, including terms proportional to $\chi\hat\qbar\omega^2$. We calculated the fundamental ($n=0$) axial QNMs for $\ell=2,3,4$ using the WKB approximation and Leaver's continued-fraction method. In the latter approach, a frequency-dependent radial shift brings the radial equation into a form suitable for implementing Leaver's method with the standard QNM boundary conditions. Both methods reproduce consistent trends in the QNM spectrum, with the continued-fraction results serving as the primary numerical reference.

The QNM spectrum was obtained through a perturbative expansion,
\begin{equation}
M\omega_{\ell m}
=
\omega_{00}^{(\ell)}
+m\chi A_\ell
+\hat\qbar B_\ell
+m\chi\hat\qbar C_\ell
+\mathcal O(\chi^2,\hat\qbar^2).
\end{equation}
The complex coefficient $C_\ell$ characterizes the mixed contribution of rotation and spacetime noncommutativity. For the dominant quadrupolar mode, the negative real part of $C_2$ implies that co-rotating spin reduces the deformation-induced increase in the oscillation frequency. For the higher multipoles $\ell=3,4$, the corresponding mixed coefficients have positive real parts, leading to a mild enhancement of this correction. The distinct dependence of the real and imaginary parts of the QNM frequencies on $\chi$ and $\hat\qbar$ is also reflected in their trajectories in the complex-frequency plane. These results demonstrate that rotation and spacetime noncommutativity produce distinct corrections to the real and imaginary parts of the QNM frequencies. Throughout the parameter range examined, the fundamental axial
modes retain negative imaginary parts, indicating the absence of  exponentially growing modes and (super radiant) instabilities, within the sector investigated, in line with the well known result in the field \cite{Whiting:1988vc}.

\begin{figure}[H]
    \centering
    \includegraphics[width=\textwidth]{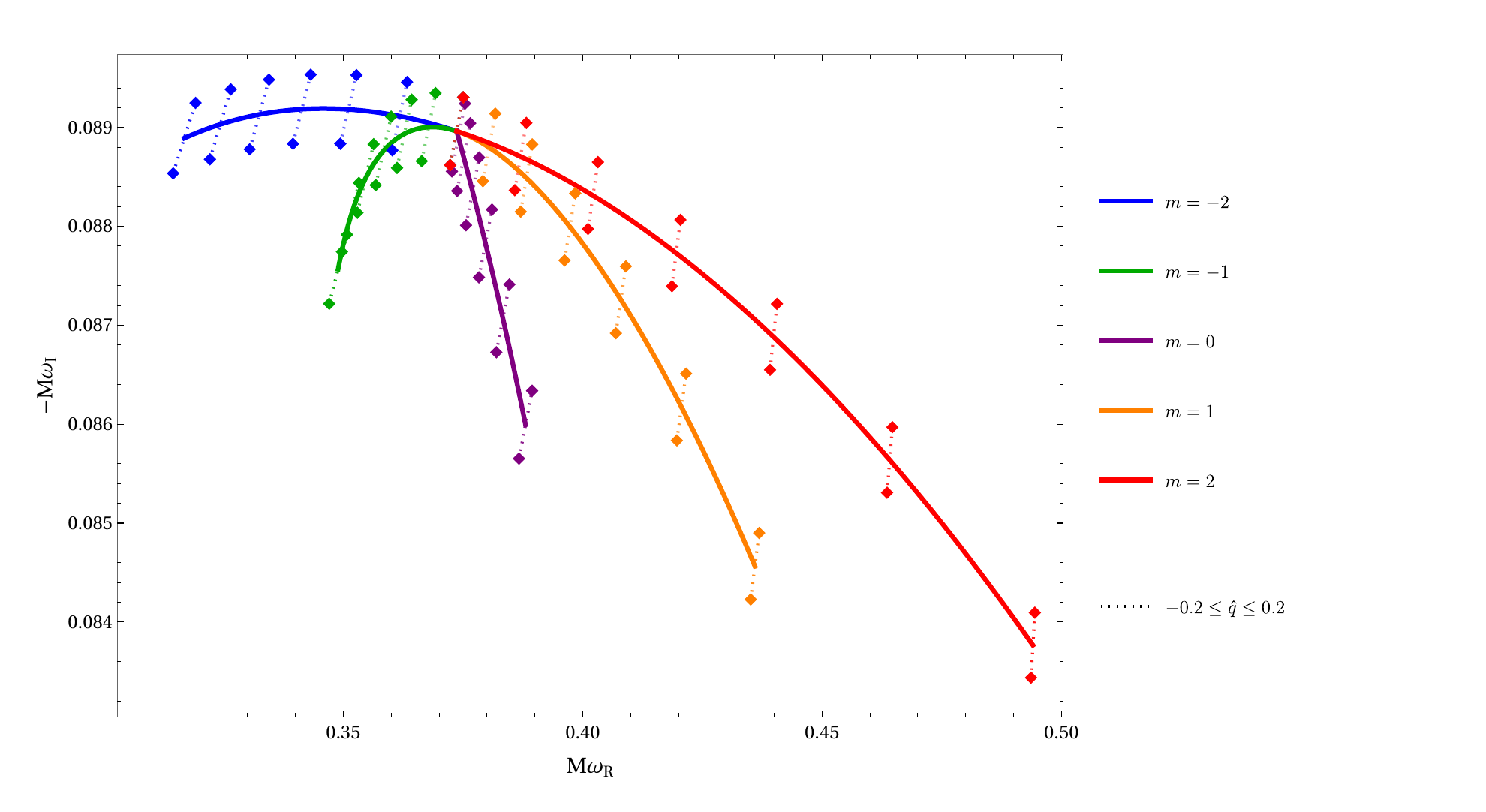}
  \caption{ Kerr-anchored fundamental ($n=0$) QNM frequencies in the complex-frequency plane for $\ell=2$ and all $m$'s. The solid curves represent the full Kerr QNM frequencies, while the dotted lines show the $\qbar$uantum corrections obtained by varying $\hat{\qbar}$ from $-0.2$ to $0.2$ at fixed dimensionless spins $\chi=0,0.1,0.2,0.3,0.4,0.5,0.6$. Different colors distinguish the azimuthal modes, as indicated in the legend. The diamonds mark the endpoints $\hat{\qbar}=\pm0.2$ of the $\qbar$uantum trajectories. The frequencies are constructed using the Kerr-anchored prescription. The colors assigned to the azimuthal modes are identical to those in Fig.~\ref{fig:kerr_anchored_fixed_q}.}
    \label{fig:kerr_anchored_all_m}
\end{figure}

\begin{figure}[H]
    \centering
    \includegraphics[width=0.45\textwidth]{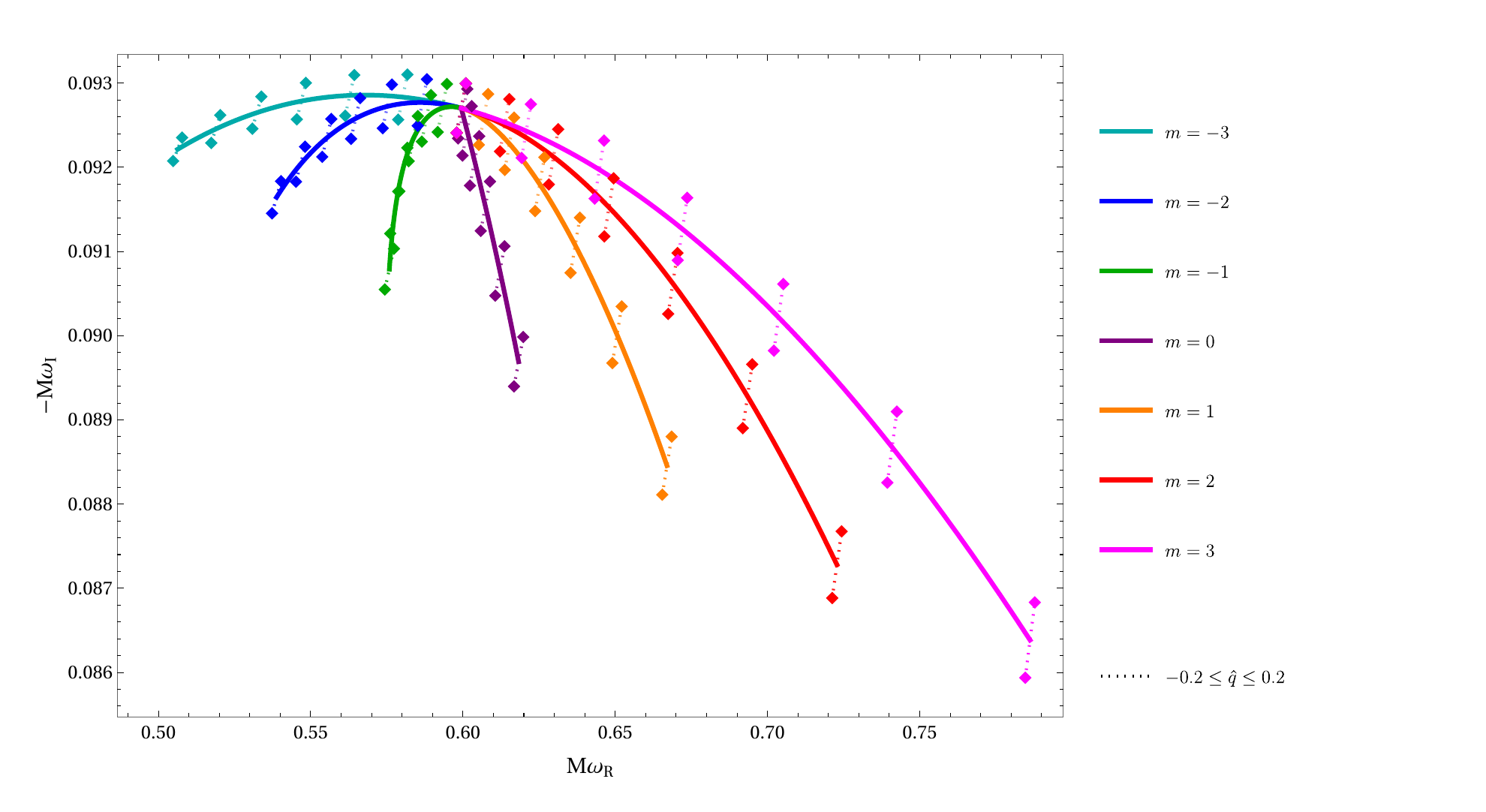}\
        \includegraphics[width=0.45\textwidth]{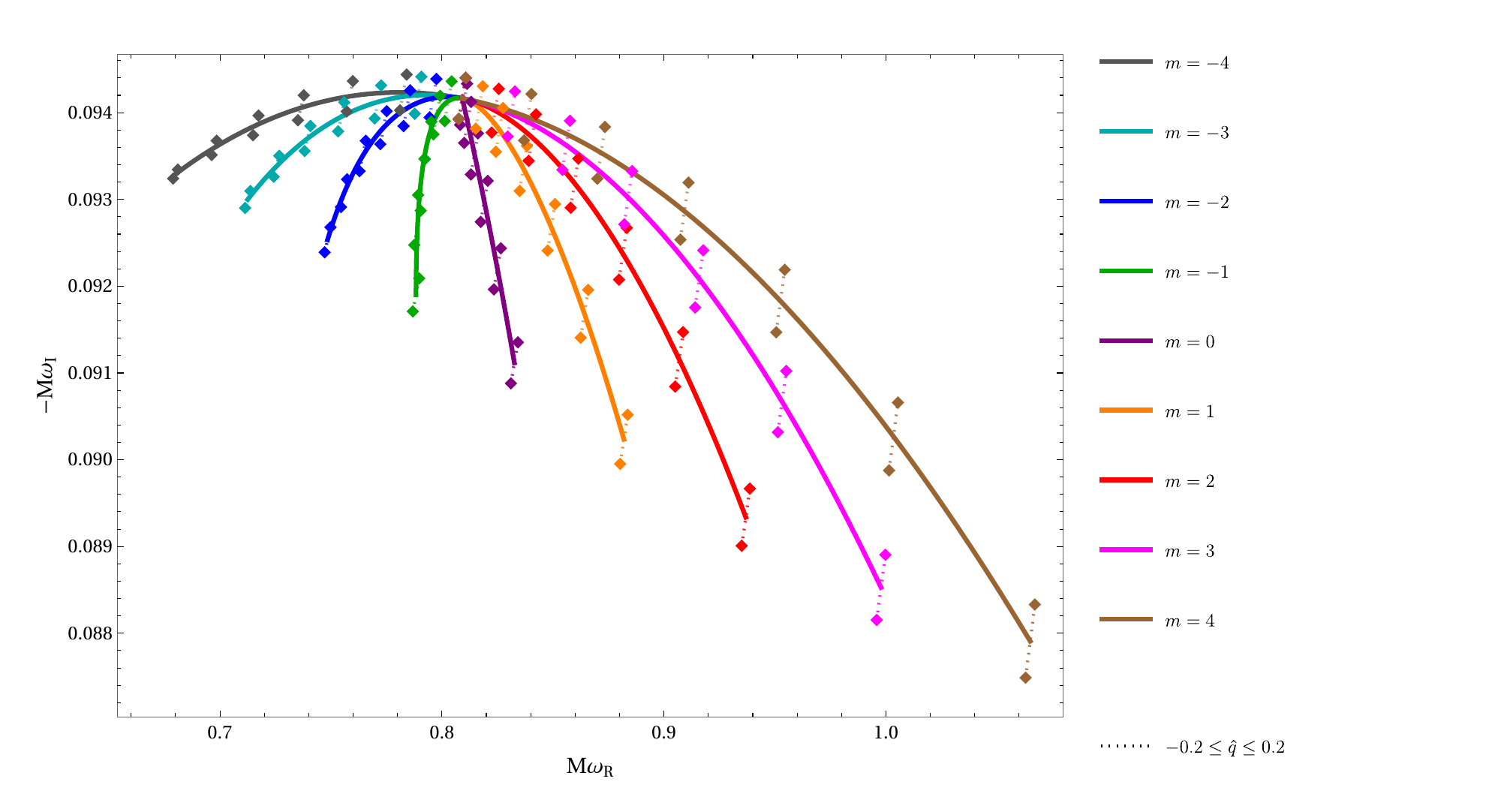}
  \caption{ Kerr-anchored fundamental ($n=0$) QNM frequencies in the complex-frequency plane for $\ell=3$ and $\ell=4$. }
    \label{fig:kerr_anchored_all_m1}
\end{figure}

The combined Kerr-anchored QNM trajectories in Fig.~\ref{fig:kerr_anchored_all_m} and Fig.~\ref{fig:kerr_anchored_all_m1} further illustrate the interplay between rotation and $\qbar$uantum deformation for different azimuthal modes. In the nonrotating limit, the frequencies are independent of $m$, whereas rotation lifts this degeneracy. The $\qbar$uantum corrections shift the trajectories away from their commutative Kerr values, with the magnitude and direction of these shifts determined by the deformation parameter and the mixed spin--$\qbar$uantum contribution.

Similar modifications to gravitational QNM spectra have been investigated in quantum-gravity-inspired black hole models, including loop quantum black holes and self-dual black holes~\cite{Cruz:2015bcj,Yang:2023gas}, as well as covariant effective quantum black holes~\cite{Liu:2025iby}. Noncommutative corrections to scalar QNMs have also been studied for deformed Schwarzschild black holes~\cite{Zhao:2023uam}, while modifications induced by dark matter halos provide another example of changes in gravitational QNM spectra~\cite{Zhang:2021bdr}. More general aspects of QNMs in modified gravity, including possible deviations from the eikonal correspondence and axial-polar isospectrality breaking, have been discussed in Ref.~\cite{Chen:2021cts}. Unlike approaches in which the background geometry or its environment is modified, the present framework preserves the classical Kerr background and introduces noncommutative corrections directly through the gravitational perturbation equations. This provides a complementary approach to investigating possible signatures of quantum spacetime in black hole ringdown.

To assess the applicability of the slow-rotation approximation, we compared the commutative first-order spectrum with the full Kerr QNM frequencies. Under the adopted $10\%$ componentwise accuracy criterion, the approximation remains within this tolerance through $\chi=0.6$ on the sampled spin grid. This provides a working spin range for the present analysis, although the accuracy of the higher-order $\qbar$uantum corrections cannot be established from the commutative comparison alone.

The slow-rotation approximation is particularly suitable for black holes with small spins, such as $\chi\lesssim0.2$, where the commutative QNM frequencies show good agreement with the full Kerr spectrum. The present framework therefore provides theoretical predictions for the $\qbar$uantum corrections to the QNM frequencies and damping times of slowly rotating merger remnants, which may be tested as more precise ringdown observations become available.

We further examined the phenomenological implications of the calculated QNM corrections using a Kerr-anchored prescription that combines the full commutative Kerr spectrum with the perturbative $\qbar$uantum contribution. For 11 merger remnants selected from the GWTC-3 (O3b)~\cite{KAGRA:2021vkt} and GWTC-5.0 (O4b)~\cite{LIGOScientific:2026wfs} catalogs, with median final spins $\chi_f\leq0.6$, we estimated the corresponding shifts in the oscillation frequency and damping time of the fundamental $(2,2,0)$ mode. For $|\hat\qbar|=0.2$, the fractional frequency shifts reach approximately $0.3\%$, while the damping-time corrections are of order $0.4\%$. These deviations are smaller than the current observational uncertainties~\cite{LIGOScientific:2026oim}, but their magnitude motivates further investigation in the context of future high-precision ringdown measurements. The present estimates use catalog median remnant parameters and do not constitute observational constraints on $\hat\qbar$. Moreover, the Kerr-anchored prescription does not determine higher-order mixed corrections beginning at $\mathcal O(\chi^2\hat\qbar)$. We also considered the recent GW250114 ringdown measurements and formally mapped the reported central frequency and damping-time deviations to $\qbar_f\simeq1.595\times10^{10}\,\mathrm{m}^2$ and $\qbar_\tau\simeq-0.504\times10^{10}\,\mathrm{m}^2$, respectively, using the detector-frame remnant mass (this corresponds to the NC scale being  $\ell_{\rm NC}^{(f)}\simeq126.3\,\mathrm{km}$ and $\ell_{\rm NC}^{(\tau)}\simeq71.0\,\mathrm{km}$ which is of the same order of magnitude as the scales reported in \cite{LIGOScientific:2026oim} for the Einstein-dilaton-Gauss-Bonnet and dynamical Chern-Simons models). Also, recently, similar bounds have been obtained from the Solar system test \cite{AraujoFilho:2026rdc}.

\begin{figure}[H]
    \centering
    \includegraphics[width=0.9\linewidth]{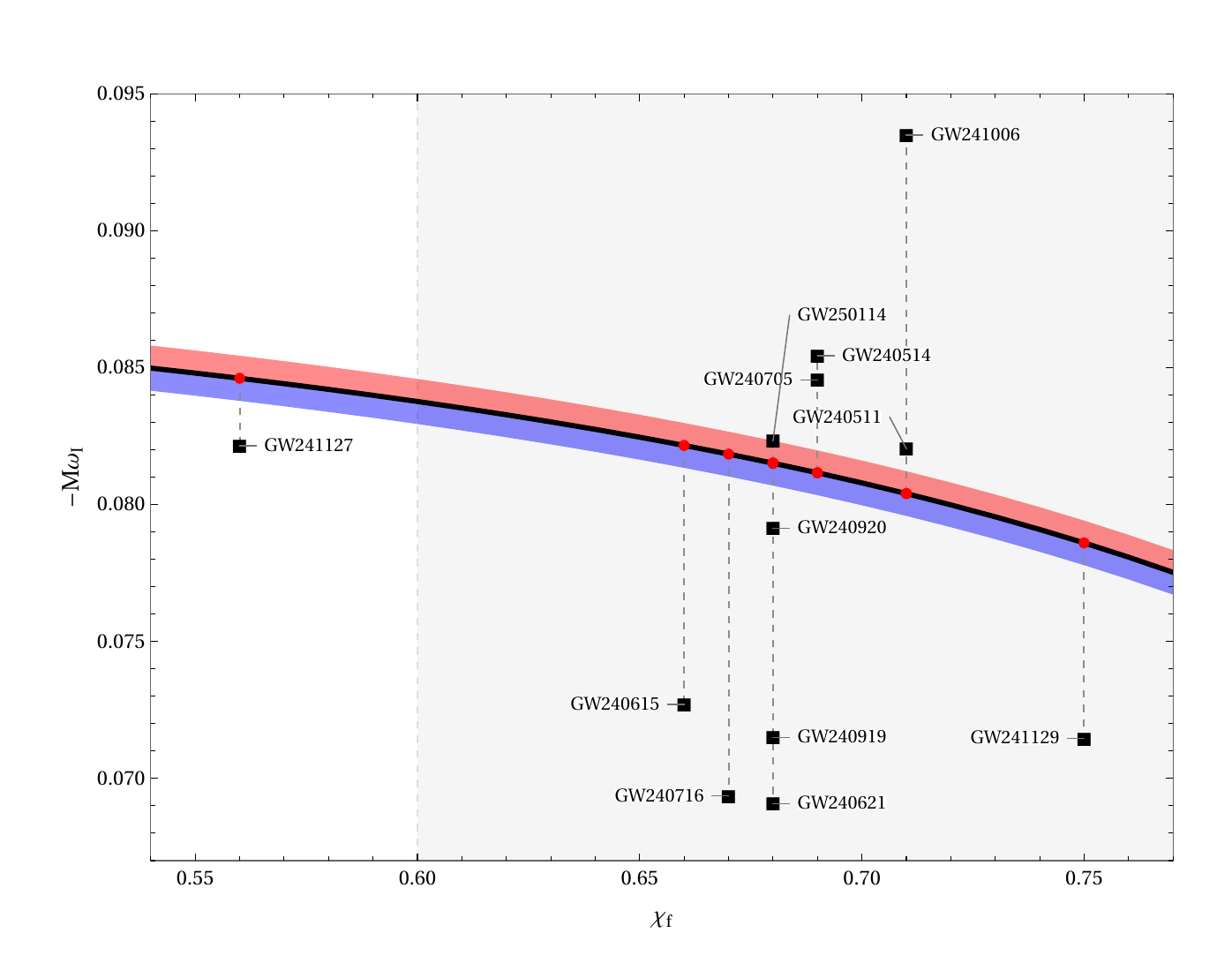}
    \caption{Kerr-anchored dimensionless damping rate $-M\omega_I$ of the fundamental gravitational QNM $(\ell,m,n)=(2,2,0)$ as a function of the dimensionless remnant spin $\chi_f$. The black solid curve represents the full Kerr QNM spectrum calculated using the \texttt{qnm} package~\cite{Stein:2019mop}. The red and blue shaded regions show the $\qbar$uantum corrections obtained by varying the dimensionless deformation parameter over $0\leq\hat\qbar\leq0.5$ and $-0.5\leq\hat\qbar\leq0$, respectively. These corrections are calculated from our Leaver coefficients, including the mixed spin--$\qbar$uantum contribution, and are added to the full Kerr spectrum through the Kerr-anchored prescription. The red circles indicate the Kerr reference damping rates evaluated at the GR median remnant spins of individual events. The black squares represent the corresponding damping rates reconstructed from the median fractional damping-time deviations reported in Table VIII of Ref.~\cite{LIGOScientific:2026oim}. The gray dashed lines connect each Kerr reference value to its corresponding reconstructed observational median. The event names are abbreviated for clarity. For clarity, we focus on twelve events exhibiting comparatively small median-based deviations from the Kerr damping rate. The events GW241225\_082815 and GW250119\_025138, which exhibit larger downward deviations, are not displayed. The light-gray background indicates $\chi_f>0.6$, beyond the adopted validity range of the first-order slow-rotation approximation. The colored regions illustrate theoretical $\qbar$uantum deformations and do not represent observational confidence intervals.}
    \label{fig:nc_kerr_observed_damping}
\end{figure}

To further illustrate the relation between our theoretical predictions and observational ringdown results, we compare the Kerr-anchored damping rates with those reconstructed from the fractional damping-time deviations reported in Table VIII of Ref.~\cite{LIGOScientific:2026oim}. Figure~\ref{fig:nc_kerr_observed_damping} shows the dimensionless damping rate $-M\omega_I$ of the fundamental $(2,2,0)$ mode as a function of the remnant spin. For each event, we evaluate the full Kerr QNM frequency at the corresponding GR median remnant spin and reconstruct the shifted damping rate using $-M\omega_I^{\rm obs}=(-M\omega_I^{\rm Kerr})/(1+\delta\hat\tau_{220})$, where $\delta\hat\tau_{220}$ is the median fractional damping-time deviation reported in Ref.~\cite{LIGOScientific:2026oim}.  On the other side, the NC predictions are obtained using the
Kerr-anchored prescription of Eq.~\eqref{eq:kerr_anchored_frequency}, with the Leaver coefficients in Table~\ref{tab:strict_nc_coefficients} and are represented by the red and blue shaded regions on Fig.\ref{fig:nc_kerr_observed_damping}.

As shown in Fig.~\ref{fig:nc_kerr_observed_damping}, the $\qbar$uantum deformation produces systematic shifts in the damping rate relative to the full Kerr prediction. Positive values of $\hat\qbar$ increase $-M\omega_I$, corresponding to shorter damping times, whereas negative values decrease $-M\omega_I$, leading to longer-lived ringdown oscillations. For the representative deformation range $|\hat\qbar|\leq0.5$, the predicted NC shifts are generally small compared with the median-based deviations reconstructed for several individual events. The observational points illustrate the scale and direction of the reported central deviations, rather than providing direct measurements of the dimensionless QNM frequencies. Moreover, the reconstructed medians do not account for correlations between the remnant parameters and the ringdown deviations. A statistical assessment of the compatibility of the NC predictions with the observations therefore requires the reported posterior uncertainties and, ultimately, a joint parameter-estimation analysis. We also emphasize that most of the events shown have $\chi_f>0.6$, where the NC corrections represent an extrapolation beyond the adopted slow-rotation validity range.

Several extensions of the present analysis merit further investigation. An immediate direction is the calculation of polar gravitational QNMs, which would allow us to examine the possible breaking of axial-polar isospectrality in rotating $\qbar$-deformed black holes, extending the results obtained for the Schwarzschild case~\cite{Herceg:2024vwc}. Although the numerical results presented here focus on the fundamental axial modes with $\ell=2,3,4$, the developed formalism is not restricted to these particular modes and, in principle, allows the calculation of axial QNMs for arbitrary allowed multipole indices $(\ell,m)$ and higher overtones. The continued-fraction framework can therefore be extended to explore a broader QNM spectrum, particularly since enhanced sensitivity of higher overtones to quantum corrections has been reported in other effective quantum black hole models~\cite{Liu:2025iby}. Such predictions could become increasingly relevant as future gravitational-wave observations will enable the identification of additional ringdown modes. Another possibility is to consider the more general pseudo-semi-Killing twist generated by $X_1=\alpha\partial_t+\beta\partial_\varphi$ and $X_2=\partial_r$. An important extension would be to go beyond the slow-rotation approximation and investigate gravitational perturbations of the full Kerr geometry in the noncommutative framework. This would allow us to study rapidly rotating black holes without truncating the spin dependence at first order, potentially through numerical spectral methods, as recently employed for scalar QNM corrections in gravitational effective field theory~\cite{vanderSteen:2026qva}. Rapidly rotating Kerr black holes also exhibit stationary massive-scalar resonances~\cite{Hod:2013zza}, although these describe a different perturbation problem from the gravitational QNMs considered here. Recent studies have further demonstrated the importance of nonlinear, second-order contributions to black hole ringdown~\cite{Cheung:2022rbm,Mitman:2022qdl}. Extending the present framework beyond linear order in metric perturbations would allow us to investigate how spacetime noncommutativity modifies these nonlinear ringdown effects. Finally, investigating greybody factors and Hawking radiation could clarify whether noncommutative corrections have implications for black hole evaporation. In this context, extended primordial black hole lifetimes have been proposed in dark-charged models~\cite{Baker:2025zxm}, although the underlying evaporation mechanism is distinct from the QNM damping effects calculated in the present work. Such extensions would broaden the scope of the framework and provide further theoretical predictions for future black hole spectroscopy.

\section*{Acknowledgement}
This research was supported by the Croatian Science Foundation Project No. IP-2025-02-8625, \emph{Quantum aspects of gravity}.

\appendix
\section{Useful identities}\label{identities}

For completeness, we collect here the spherical-harmonic identities used to reduce the angular dependence of the perturbation equations. The first set follows directly from the scalar spherical-harmonic equation, whereas the second set makes use of the combinations $X^\ell$ and $W^\ell$ defined in Eq.~\eqref{XW}.

\textbf{Set 1:}
\begin{align}
    \partial_\theta^2 Y^\ell \;&=\, -\ell (\ell+1) Y^\ell
    - \frac{\partial_\varphi^2 Y^\ell}{\sin^{2} \theta}
    - \cot \theta \, \partial_\theta Y^\ell, \label{eq:Y20} \\[6pt]
    \partial_\theta^3 Y^\ell \;&=\, -\left(\ell^{2} + \ell + 1\right) \partial_\theta Y^\ell
    + \ell(\ell+1) \cot \theta \, Y^\ell
    + \csc^{2} \theta \left(2 \partial_\theta Y^\ell
    - \partial_\theta \partial_\varphi^2 Y^\ell
    + 3 \cot \theta \, \partial_\varphi^2 Y^\ell\right), \label{eq:Y30} \\[6pt]
    \partial_\theta^4 Y^\ell \;&=\, \csc \theta \Bigg[
        \csc \theta \left(-11 \csc^{2} \theta + \ell^{2} + \ell + 7\right) \partial_\varphi^2 Y^\ell
        + \partial_\theta Y^\ell \left((2 \ell (\ell+1) + 1) \cos \theta
            - 6 \cot \theta \, \csc \theta \right) \notag \\
    &\quad\quad\quad\quad
        + \csc \theta \left(5 \cot \theta \, \partial_\theta \partial_\varphi^2 Y^\ell
            - \partial_\theta^2 \partial_\varphi^2 Y^\ell \right)
    \Bigg]
        + \ell(\ell+1) \left(-3 \csc^{2} \theta + \ell^{2} + \ell + 1\right) Y^\ell,
    \label{eq:Y40} \\[6pt]
    \partial_\theta^2 \partial_\varphi Y^\ell \;&=\, -\ell(\ell+1) \partial_\varphi Y^\ell
    - \cot \theta \, \partial_\theta \partial_\varphi Y^\ell
    - \csc^{2} \theta \, \partial_\varphi^3 Y^\ell, \label{eq:Y21} \\[6pt]
    \partial_\theta^2 \partial_\varphi^2 Y^\ell \;&=\, -\ell(\ell+1) \partial_\varphi^2 Y^\ell
    - \cot \theta \, \partial_\theta \partial_\varphi^2 Y^\ell
    - \csc^{2} \theta \, \partial_\varphi^4 Y^\ell. \label{eq:Y22}
\end{align}
\textbf{Set 2:}
\begin{align}
    \partial_\theta \partial_\varphi Y^\ell \;&=\, \frac{1}{2} \, X^\ell
    + \cot \theta \, \partial_\varphi Y^\ell, \label{eq:Y11} \\[6pt]
    \partial_\varphi^2 Y^\ell \;&=\, -\frac{1}{2} \sin \theta \left(
        \sin \theta \left( \ell(\ell+1) Y^\ell + W^\ell \right)
        + 2 \cos \theta \, \partial_\theta Y^\ell
    \right), \label{eq:Y02} \\[6pt]
    \partial_\theta^2 Y^\ell \;&=\, \frac{1}{2} \left( W^\ell - \ell(\ell+1) Y^\ell \right), \label{eq:Y20b} \\[6pt]
    \partial_\theta^3 Y^\ell \;&=\, \frac{1}{2} \Big(
        -2 \left( \csc^2 \theta + \ell^2 + \ell - 2 \right) \partial_\theta Y^\ell
        - \ell(\ell+1) \cot \theta \, Y^\ell
        - 3 \cot \theta \, W^\ell
        - 2 \csc^2 \theta \, \partial_\theta \partial_\varphi^2 Y^\ell
    \Big), \label{eq:Y30b} \\[6pt]
    \partial_\theta^4 Y^\ell \;&=\, \frac{1}{2} \Bigg[
        -\left( -11 \csc^2 \theta + 2 \ell(\ell+1) + 7 \right) W^\ell
        + \cot \theta \left(
            5 \ell(\ell+1) \cot \theta \, Y^\ell
            + 12 \csc^2 \theta \, \partial_\theta \partial_\varphi^2 Y^\ell
        \right) \notag \\
    &\quad\quad\quad\quad
        + 2 \cot \theta \left( 5 \csc^2 \theta - 6 \right) \partial_\theta Y^\ell
        + 2 \csc^4 \theta \, \partial_\varphi^4 Y^\ell
    \Bigg], \label{eq:Y40b} \\[6pt]
    \partial_\theta^2 \partial_\varphi Y^\ell \;&=\, -\left( \cot^2 \theta + \ell^2 + \ell \right) \partial_\varphi Y^\ell
    - \frac{1}{2} \cot \theta \, X^\ell
    - \csc^2 \theta \, \partial_\varphi^3 Y^\ell, \label{eq:Y21b} \\[6pt]
    \partial_\theta^2 \partial_\varphi^2 Y^\ell \;&=\, \frac{1}{2} \ell(\ell+1) \sin \theta \left(
        \sin \theta \left( \ell(\ell+1) Y^\ell + W^\ell \right)
        + 2 \cos \theta \, \partial_\theta Y^\ell
    \right)
    - \cot \theta \, \partial_\theta \partial_\varphi^2 Y^\ell
    - \csc^2 \theta \, \partial_\varphi^4 Y^\ell, \label{eq:Y22b} \\[6pt]
    \partial_\theta^3 \partial_\varphi Y^\ell \;&=\, \frac{1}{4} \Bigg[
        \csc^3 \theta \left(
            -4 \sin \theta \, \partial_\theta \partial_\varphi^3 Y^\ell
            + \left( 7 \cos \theta + \cos 3\theta \right) \partial_\varphi Y^\ell
            + 12 \cos \theta \, \partial_\varphi^3 Y^\ell
        \right) \notag \\
   &\quad\quad\quad\quad
        - 2 \left( -2 \csc^2 \theta + \ell^2 + \ell + 1 \right) X^\ell
    \Bigg]. \label{eq:Y31}
\end{align}

\section{Expansion coefficients of the Einstein equations}
\label{coefficient_list}

In this appendix we collect the radial coefficients appearing in the angularly separated Einstein equations in both the commutative and $\qbar$uantum calculations. 

\subsection{Commutative coefficients}
\label{app:comm_coefficients}

The GR expansion coefficients appearing in Eq.~\eqref{e13nc} are
\begin{equation}
\begin{aligned}
\alpha^{(t)}_{\ell}
&=0,
\\
\widetilde{\alpha}^{(t)}_{\ell}
&=
\frac{J}{r^{5}}
\left[
2r(r-2M)\,\partial_t h_1
+(16M-7r)\,h_1
\right],
\\
\beta^{(t)}_{\ell}
&=
\frac{1}{2r^{3}}
\left[
r(r-2M)
\left(
r\left(
\partial_t\partial_r h_1
-\partial_t^2 h_0
\right)
+2\,\partial_r h_1
\right)
+\left(\ell(\ell+1)r-4M\right)h_0
\right],
\\
\widetilde{\beta}^{(t)}_{\ell}
&=0,
\\
\eta^{(t)}_{\ell}
&=
-\frac{J\ell(\ell+1)h_1}{2r^{4}},
\\
\xi^{(t)}_{\ell}
&=0,
\\
\zeta^{(t)}_{\ell}
&=
\frac{J\,\partial_r h_0}{r^{2}(r-2M)}
-\frac{J\,h_1}{2r^{4}}.
\end{aligned}
\label{e13comp}
\end{equation}

The GR expansion coefficients appearing in Eq.~\eqref{e23nc} are
\begin{equation}
\begin{aligned}
\alpha^{(r)}_{\ell}
&=0,
\\
\widetilde{\alpha}^{(r)}_{\ell}
&=
\frac{
J\left[
(16M-9r)h_0
-r(2M-r)
\left(
\partial_t h_0+\partial_r h_1
\right)
\right]
}{
r^3(r-2M)^2
},
\\
\beta^{(r)}_{\ell}
&=
-\frac{
r^2
\left[
r\left(
\partial_t\partial_r h_0
-\partial_t^2h_1
\right)
-2\partial_r h_0
\right]
+
(\ell^2+\ell-2)(2M-r)h_1
}{
2r^2(r-2M)
},
\\
\widetilde{\beta}^{(r)}_{\ell}
&=0,
\\
\eta^{(r)}_{\ell}
&=
-\frac{
J\ell(\ell+1)
\left[
3r(2M-r)
\left(
\partial_r h_1-\partial_t h_0
\right)
+(12M-7r)h_0
\right]
}{
2r^3(r-2M)^2
},
\\
\xi^{(r)}_{\ell}
&=0,
\\
\zeta^{(r)}_{\ell}
&=
\frac{
J\left[
(4M-3r)h_0
-r(2M-r)
\left(
\partial_t h_0+\partial_r h_1
\right)
\right]
}{
2r^3(r-2M)^2
}.
\end{aligned}
\label{e23comp}
\end{equation}

The GR expansion coefficients appearing in Eq.~\eqref{e34nc} are
\begin{equation}
\begin{aligned}
f_{\ell}
&=0,
\\
g_{\ell}
&=
\frac{J}{r^3}
\left[
3r\left(
\partial_t h_0-\partial_r h_1
\right)
+
\frac{
\left[
(\ell^2+\ell+6)r-12M
\right]h_0
}{
2M-r
}
\right],
\\
s_{\ell}
&=0,
\\
t_{\ell}
&=
\frac{1}{2r^2}
\left[
\frac{
r\left[
r^2\partial_r h_0
-(r-2M)^2\partial_t h_1
\right]
}{
2M-r
}
+2Mh_1
\right].
\end{aligned}
\label{e34comp}
\end{equation}

\subsection{$\hat\qbar$-dependent corrections}
\label{app:nc_coefficients}

For the $t$-$r$ twist, the radial coefficients appearing in Eqs.~\eqref{e13nc}-\eqref{e34nc} can be written as
\begin{equation}
\mathcal C_{\ell}^{(i)}
=
\mathcal C_{\ell,\mathrm C}^{(i)}
+
\hat\qbar M^2\,\mathcal C_{\ell,\qbar}^{(i)},
\qquad
\mathcal C
\in
\{
\alpha,\widetilde{\alpha},
\beta,\widetilde{\beta},
\eta,\xi,\zeta
\},
\label{eq:appendix_coefficient_split}
\end{equation}
where $\mathcal C_{\ell,\mathrm C}^{(i)}$ denotes the corresponding commutative coefficient listed above.  An analogous decomposition is used for $f_\ell$, $g_\ell$, $s_\ell$, and $t_\ell$.

For the $(t,\theta)$ component, the nonvanishing $\qbar$uantum corrections are
\begin{align}
\widetilde{\alpha}^{(t)}_{\ell,\qbar}
={}&
\frac{iJ}{r^6(r-2M)}
\Bigl[
3r^3\partial_r^2h_0
+
(6M^2-2Mr-r^2)\partial_rh_1
-r(r-2M)(3r-4M)
\partial_t\partial_rh_1
\Bigr],
\label{eq:atnc}
\\
\beta^{(t)}_{\ell,\qbar}
={}&
\frac{i}{2r^4}
\Biggl\{
r
\Bigl[
Mr\left(
\partial_t^2h_0
-\partial_t\partial_r^2h_1
\right)
+
(2r-5M)\partial_t\partial_rh_0
+
(3M-2r)\partial_r^2h_1
\Bigr]
+
\left[\ell(\ell+1)r+M\right]
\partial_rh_0
\Biggr\},
\label{eq:btnc}
\\
\eta^{(t)}_{\ell,\qbar}
={}&
\frac{iJ\ell(\ell+1)M}{
2r^5(r-2M)
}
\,\partial_rh_1,
\label{eq:etanc}
\\
\zeta^{(t)}_{\ell,\qbar}
={}&
-\frac{iJ}{
2r^5(r-2M)^2
}
\Bigl[
r^2(4M-3r)\partial_r^2h_0
+
(2M^2-3Mr+r^2)\partial_rh_1
\Bigr].
\label{eq:zetatnc}
\end{align}
The remaining corrections vanish,
\begin{equation}
\alpha^{(t)}_{\ell,\qbar}
=
\widetilde{\beta}^{(t)}_{\ell,\qbar}
=
\xi^{(t)}_{\ell,\qbar}
=0.
\end{equation}

For the $(r,\theta)$ component, the nonvanishing $\qbar$uantum corrections are
\begin{align}
\widetilde{\alpha}^{(r)}_{\ell,\qbar}
={}&
-\frac{iJ}{
r^4(2M-r)^3
}
\Bigl[
(26M^2-28Mr+7r^2)\partial_rh_0
-Mr(2M-r)
\left(
\partial_t\partial_rh_0
+\partial_r^2h_1
\right)
\Bigr],
\label{eq:arnc}
\\
\beta^{(r)}_{\ell,\qbar}
={}&
\frac{i}{
2r^4(r-2M)^2
}
\Bigl[
-Mr^4\partial_t\partial_r^2h_0
+r^3(4M-r)\partial_r^2h_0
-(r-2M)^2
\left[
9M-(\ell^2+\ell+6)r
\right]
\partial_rh_1
\nonumber\\
&\qquad \qquad \qquad \qquad
+Mr^4\partial_r^3h_1
-r(2M-r)^3
\partial_t\partial_rh_1
\Bigr],
\label{eq:brnc}
\\
\eta^{(r)}_{\ell,\qbar}
={}&
-\frac{
iJ\ell(\ell+1)
}{
4r^4(2M-r)^3
}
\Bigl[
3r(2M-r)(4M-3r)\partial_r^2h_0
+
2(24M^2-29Mr+9r^2)\partial_rh_0
\Bigr],
\label{eq:etarnc}
\\
\zeta^{(r)}_{\ell,\qbar}
={}&
\frac{iJ}{
4r^4(2M-r)^3
}
\Bigl[
2(16M^2-17Mr+5r^2)\partial_rh_0
-r(4M-3r)(2M-r)\partial_rh_1
\Bigr].
\label{eq:zetarnc}
\end{align}
The remaining corrections vanish,
\begin{equation}
\alpha^{(r)}_{\ell,\qbar}
=
\widetilde{\beta}^{(r)}_{\ell,\qbar}
=
\xi^{(r)}_{\ell,\qbar}
=0.
\end{equation}

Finally, for the $(\theta,\varphi)$ component,
\begin{align}
g_{\ell,\qbar}
={}&
-\frac{iJ}{
2r^4(r-2M)^2
}
\Biggl[
3r(r-2M)^2
\left(
\partial_r^2h_1
-\partial_t\partial_rh_0
\right)
+
\left[
3(\ell^2+\ell-1)r^2
-4\ell(\ell+1)Mr
+12M^2
\right]
\partial_rh_0
\Biggr],
\label{eq:gnc}
\\
t_{\ell,\qbar}
={}&
-\frac{i}{
2r^3(r-2M)^2
}
\Bigl[
Mr^3\partial_t\partial_r^2h_0
-3(2M-r)^3\partial_rh_1
+
Mr(r-2M)^2\partial_r^2h_1
\Bigr],
\label{eq:tnc}
\end{align}
while
\begin{equation}
f_{\ell,\qbar}
=
s_{\ell,\qbar}
=0.
\end{equation}

\subsection{Coefficients of the second-order radial equation}
\label{app:nc_radial_coefficients}

After imposing the harmonic time dependence $e^{-i\omega t}$ and eliminating $h_0$, Eq.~\eqref{eq:nc_second_order_h1} is obtained with
\begin{align}
A_{\rm GR}
={}&
-\frac{L(r-2M)}{2r},
\\
A_{\qbar}
={}&
\frac{3\chi M^2m(2M-r)(M-2r)}{r^6}
+
\frac{LM\omega}{2r^2},
\\[1ex]
B_{\rm GR}
={}&
\frac{L(r-5M)}{r^2}
-
\frac{6\chi M^2Lm(2M-r)}{r^5\omega},
\\
B_{\qbar}
={}&
\frac{3\chi M^2m}{2r^7}
\left[
-6(3L-7)Mr
+(7L-16)r^2
+4M^2
\right]
+
\frac{3L\omega(r-4M)}{2r^3},
\end{align}
and
\begin{align}
C_{\rm GR}
={}&
\frac{\chi M^2m}{
2r^8\omega(r-2M)^2}
\left\{
12r^2(r-2M)^2
\left[
(9L-14)M-3(L-2)r
\right]
+
4Lr^6\omega^2(r-2M)
\right\}
\nonumber\\
&+
\frac{L}{
2r^3(r-2M)}
\left[
-2LMr
+(L-2)r^2
-20M^2
+12Mr
-r^4\omega^2
\right],
\label{eq:CGR}
\\[1ex]
C_{\qbar}
={}&
\frac{\chi M^2m}{
2r^8\omega(r-2M)^2}
\left\{
r^5\omega^3
\left[
-8LM+6Lr+6M-3r
\right]
-3\omega(r-2M)^2
\left[
-10(9L+28)Mr
+(25L+86)r^2
+400M^2
\right]
\right\}
\nonumber\\
&+
\frac{L}{
2r^4(r-2M)^2}
\left\{
4(L-32)M^2r\omega
+(L-6)r^3\omega
+
Mr^2
\left[
(49-4L)\omega-r^2\omega^3
\right]
+
104M^3\omega
\right\}.
\label{eq:CNC}
\end{align}


\section{Higher-multipole QNM frequencies}
\label{app:higher_multipole_qnms}

In this appendix, we present the complex-frequency trajectories and the dependence of the fundamental axial QNM frequencies on the deformation parameter $\hat\qbar$ and the dimensionless spin $\chi$ for the higher multipoles $\ell=m=3,4$, complementing the quadrupolar results discussed in Sec.~\ref{sec:results}.

\begin{figure}[H]
\centering
\includegraphics[width=\textwidth]{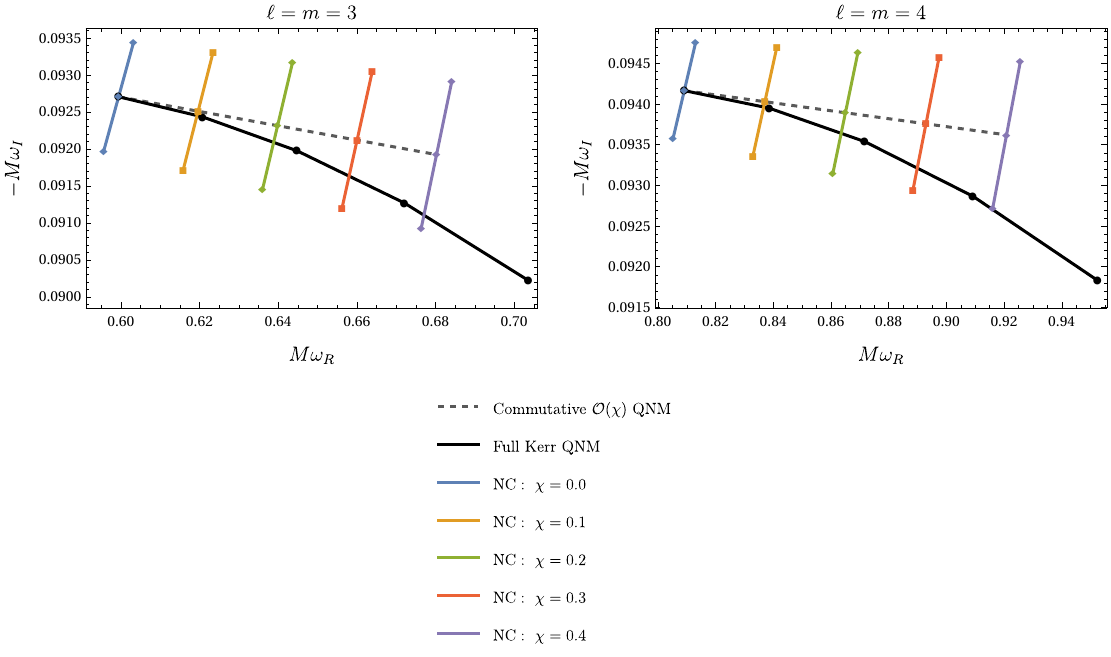}
\caption{Trajectories in the complex-frequency plane for the co-rotating fundamental modes with $\ell=m=3$ (left) and $\ell=m=4$ (right). The dashed curves represent the commutative slow-rotation Kerr results, the black points correspond to the full commutative Kerr spectrum, and the colored lines show the $\qbar$uantum-induced displacement obtained within the slow-rotation formalism by varying $-0.5\leq\hat\qbar\leq0.5$ at fixed $\chi=0,0.1,0.2,0.3$, and $0.4$. The corresponding $\ell=m=2$ result is shown in Fig.~\ref{fig:complex_frequency_paths}.}
\label{fig:complex_frequency_paths_higher}
\end{figure}

\begin{figure}[H]
    \centering
    \includegraphics[width=0.45\textwidth]{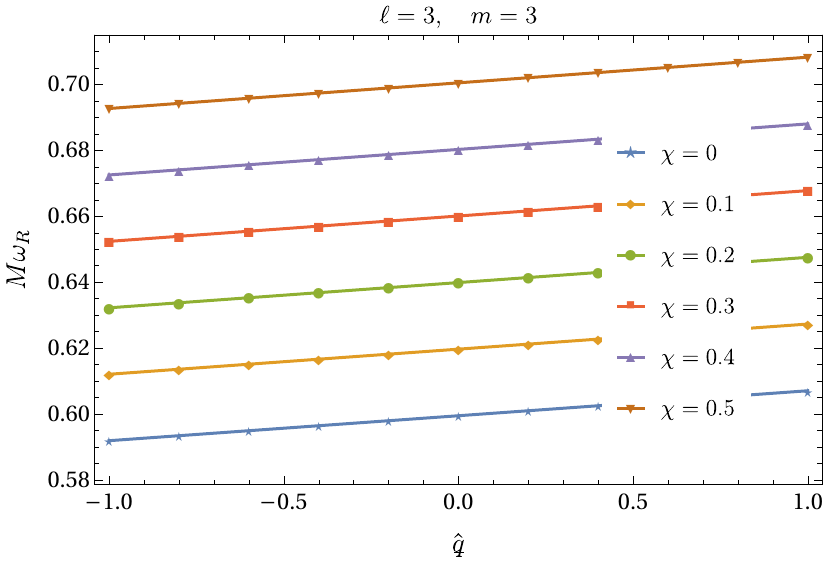}
    \includegraphics[width=0.45\textwidth]{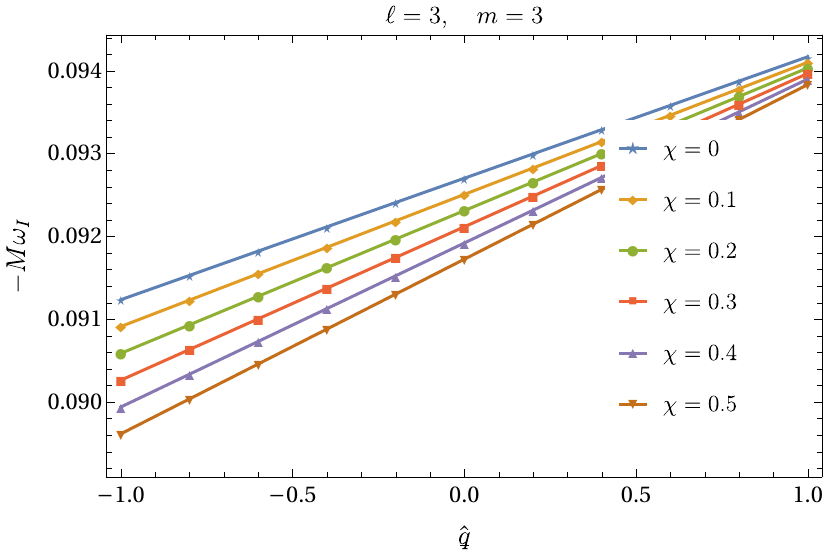}

    \includegraphics[width=0.45\textwidth]{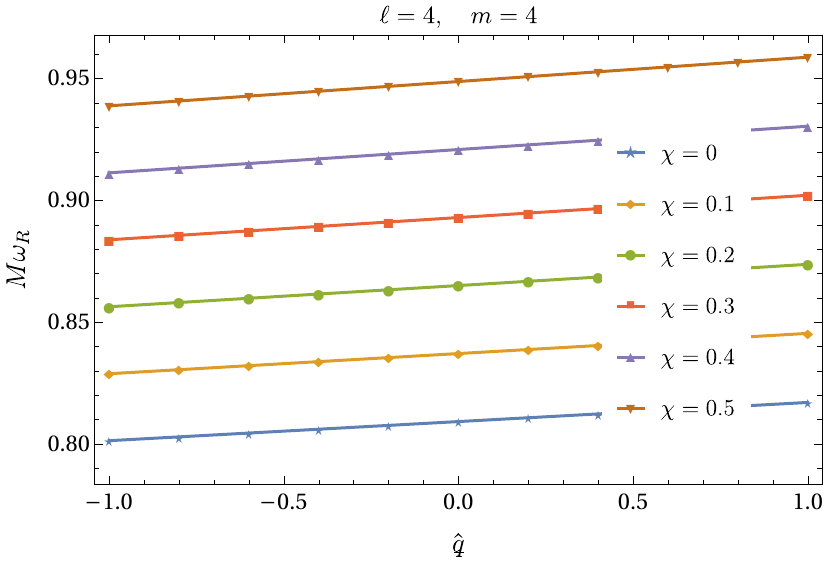}
    \includegraphics[width=0.45\textwidth]{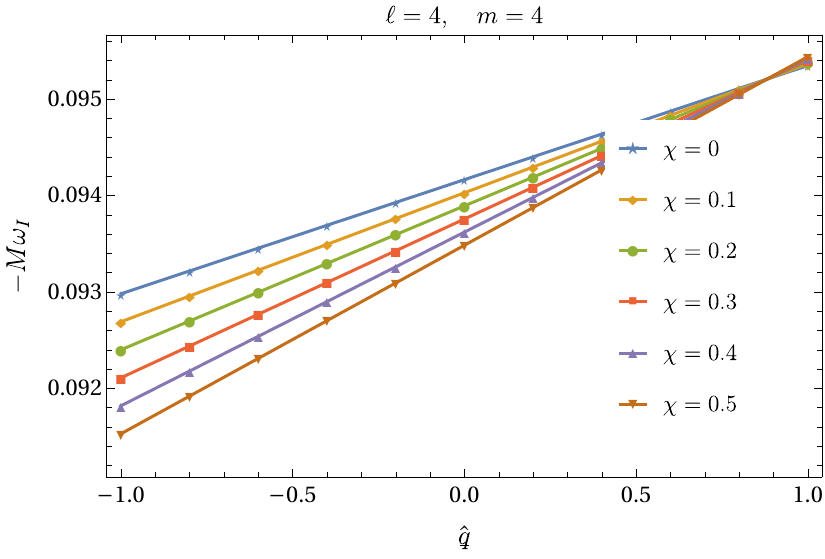}

    \caption{Dependence of the real (left) and imaginary (right)
    parts of the fundamental axial QNM frequencies on the
    deformation parameter $\hat\qbar$ for $\ell=m=3$ (top)
    and $\ell=m=4$ (bottom), with
    $\chi=0,0.1,0.2,0.3,0.4,$ and $0.5$.
    The corresponding $\ell=m=2$ results are shown in
    Fig.~\ref{fig:qnm_nc_parameter_dependence}.}
    \label{fig:qnm_nc_parameter_higher}
\end{figure}

\begin{figure}[H]
    \centering
    \includegraphics[width=0.45\textwidth]{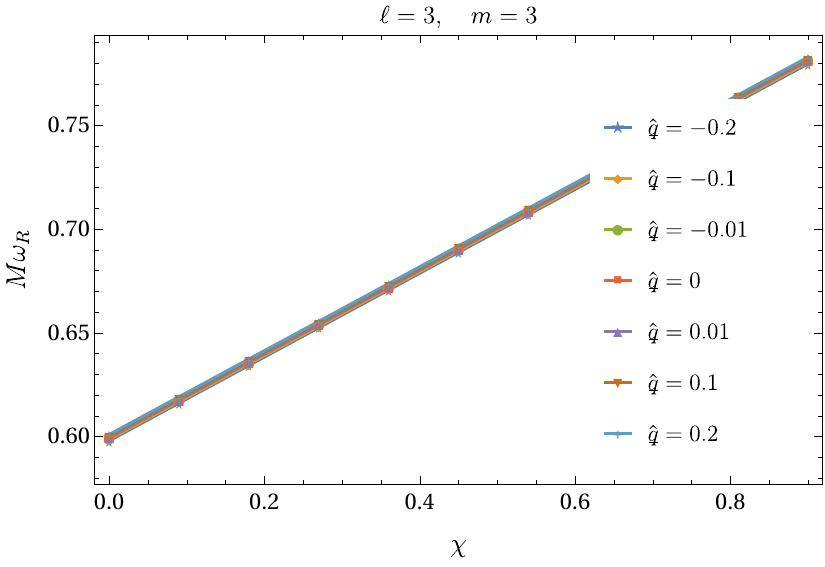}
    \includegraphics[width=0.45\textwidth]{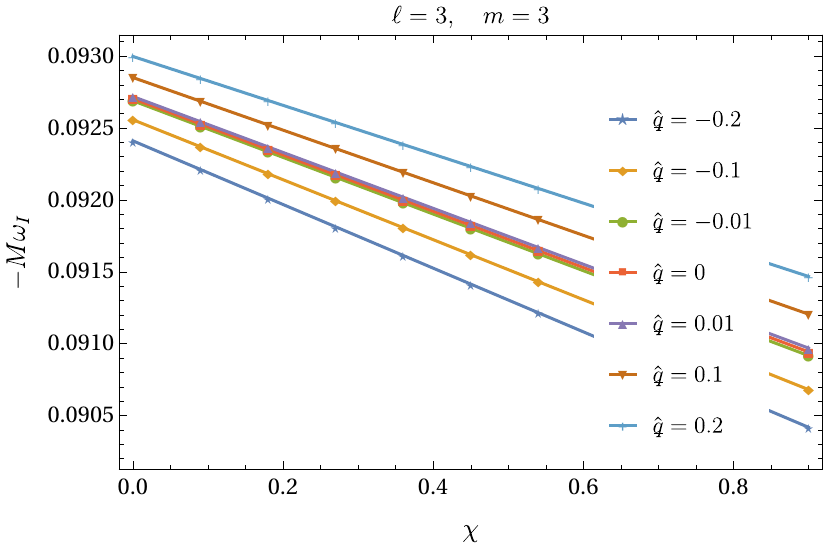}

    \includegraphics[width=0.45\textwidth]{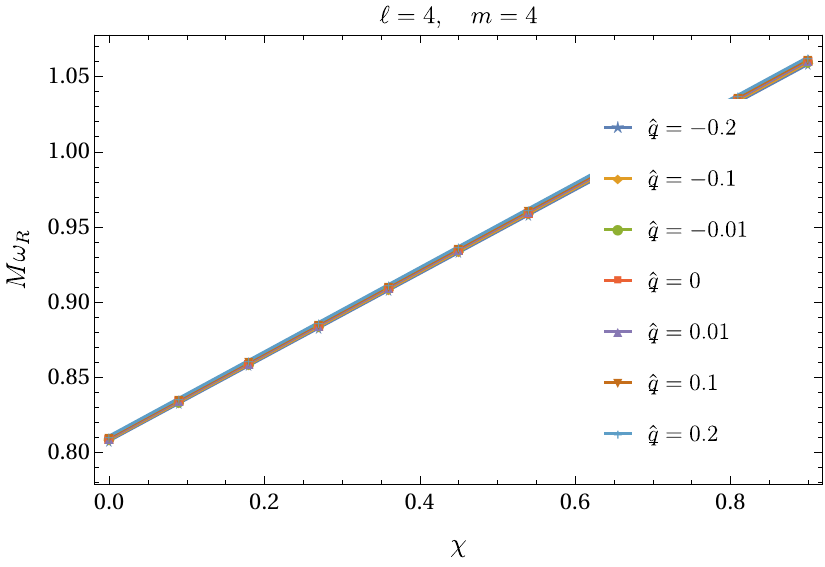}
    \includegraphics[width=0.45\textwidth]{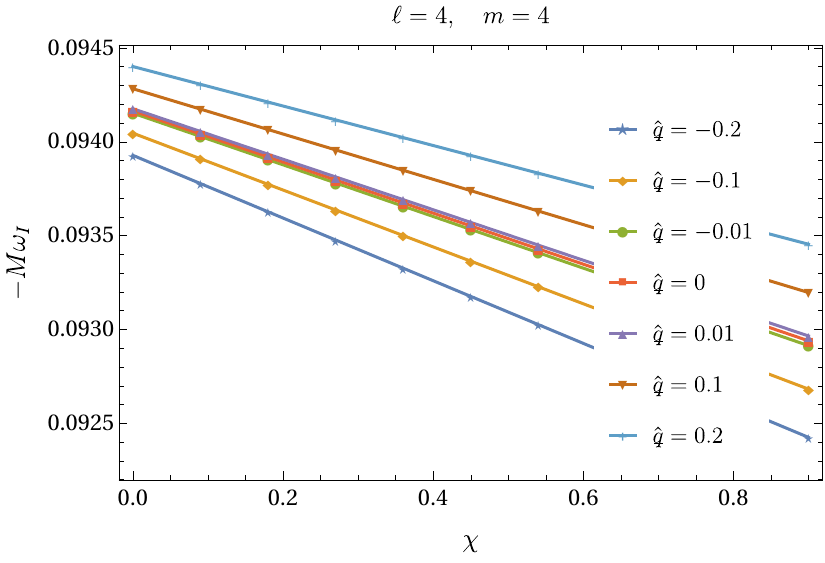}

    \caption{Dependence of the real (left) and imaginary (right)
    parts of the fundamental axial QNM frequencies on the
    dimensionless spin $\chi$ for $\ell=m=3$ (top)
    and $\ell=m=4$ (bottom), with
    $\hat\qbar=-0.2,-0.1,-0.01,0,0.01,0.1,$ and $0.2$.
    The corresponding $\ell=m=2$ results are shown in
    Fig.~\ref{fig:qnm_spin_parameter_dependence}.}
    \label{fig:qnm_spin_parameter_higher}
\end{figure}

\bibliography{BibTex}

\end{document}